\documentclass[11pt,a4paper]{article}
\usepackage{jheppub}       

\usepackage{bm}
\usepackage{xcolor}
\usepackage{tikz}          

\usepackage[T2A]{fontenc}\usepackage[utf8]{inputenc}
\usepackage[english]{babel}

\usepackage{mathbbol}
\usepackage{amsthm}
\usepackage{mathrsfs}
\usepackage{latexsym}
\usepackage{commath} 

\usetikzlibrary{babel,arrows.meta,positioning}
\usepackage{appendix}

\usepackage{subcaption}
\graphicspath{{images/}}

\newcommand{\be}{\begin{equation}}
\newcommand{\ee}{\end{equation}}

\newcommand{\benn}{\begin{equation*}}
\newcommand{\eenn}{\end{equation*}}

\newcommand{\bea}{\begin{equation}\begin{aligned}}
\newcommand{\eea}{\end{aligned}\end{equation}}

\usepackage{mathtools}

\DeclarePairedDelimiterX\braket[2]{\langle}{\rangle}{#1 \delimsize\vert #2}

\newcommand{\cO}{{\cal O}}

\newcommand{\dd}{\mathrm{d}} 
\usepackage{tensor}
\usepackage{cleveref}
\usepackage{lipsum}
\usepackage{comment}
\usepackage{siunitx} 

\newcommand{\AdS}{\mathrm{AdS}}
\newcommand{\RR}{\mathbb{R}}
\newcommand{\EE}{\mathbb{E}}
\newcommand{\cK}{\mathcal{K}}
\newcommand{\cD}{\mathcal{D}}

\newcommand{\cP}{\mathcal{P}}
\newcommand{\cZ}{\mathcal{Z}}

\newcommand{\intp}{\int\frac{\dd^d p}{(2\pi)^d}}
\newcommand{\dV}{\dd V}

\DeclareMathOperator{\Var}{Var}
\DeclareMathOperator{\Cov}{Cov}

\title{Neural Network Field Theory and the AdS/CFT Correspondence}

\author[a,b,1]{Dmitry S. Ageev,}
\author[b,c,d,2]{Yulia A. Ageeva,}
\author[e,3]{and Anna S. Bernakevich}

\affiliation[a]{Department of Mathematical Methods for Quantum Technologies, Steklov Mathematical Institute, Russian Academy of Sciences,
  Gubkin str. 8, 119991 Moscow, Russia}

\affiliation[b]{Institute for Theoretical and Mathematical Physics,
  M.V. Lomonosov Moscow State University,
  Leninskie Gory 1, 119991 Moscow, Russia}

\affiliation[c]{Institute for Nuclear Research of the Russian Academy of
  Sciences, 60th October Anniversary Prospect, 7a, 117312 Moscow, Russia}

\affiliation[d]{Department of Particle Physics and Cosmology, Physics Faculty,
  M.V. Lomonosov Moscow State University,
  Leninskie Gory 1-2, 119991 Moscow, Russia}

\affiliation[e]{Moscow Institute of Physics and Technology,
  Institutskii per. 9, 141702 Dolgoprudny, Russia}

\emailAdd{$^{1}$ageev@mi-ras.ru,
  $^{2}$ageeva@inr.ac.ru,
  $^{3}$bernakevich.as@phystech.edu}

\abstract{
We develop a neural realization of massive scalar field
theory on Euclidean anti--de Sitter space and its
holographic boundary observables within neural network
field theory (NN--FT). We construct Fourier--Bessel
cosine neurons whose covariance reproduces the regulated
massive AdS propagator exactly at finite network width,
with the Gaussian field theory recovered in the
infinite-width limit.

We introduce quartic interactions through two
complementary approaches. The first implements a
prescribed interaction by deforming the neural parameter
measure, while the second generates a local quartic
interaction from the statistics of auxiliary binary
neurons. Applying the BDHM prescription directly to
the neural ensemble, we recover the scalar
bulk-to-boundary propagator, the conformal boundary
two-point function, and the leading connected
four-point contact Witten diagram.

We further show that distinct neural architectures
can reproduce the same leading boundary correlators
while defining inequivalent bulk theories. These
results establish a framework for studying interacting
scalar holography within NN--FT and clarify the role
of neural architecture in encoding bulk dynamics
and boundary observables.
}

\usepackage{url}
\begin{document}


\maketitle

\section{Introduction}
\label{sec:Introduction}

Neural networks have become useful tools throughout physics, from the
analysis of collider data to the study of lattice gauge systems
\cite{BaldiSadowskiWhiteson2014,FavoniEtAl2022,Apte:2024vwn,
JeffersonRamakrishnan2026}. Generative methods, including normalizing
flows, offer ways to sample complicated field-theory measures
\cite{RezendeMohamed2015,PapamakariosEtAl2021,
AlbergoKanwarShanahan2019,AlbergoEtAl2021,KanwarEtAl2020,
BoydaEtAl2021}, while neural architectures have also been used to
investigate holographic geometry and quantum systems
\cite{GanShu2017,HashimotoEtAl2018,Hashimoto2019,
HashimotoEtAl2024QuantumSystems}. These applications generally begin
with a physical problem and use a network to study it. One may instead
reverse the question and ask what physical theory is defined by a neural
network itself. In Euclidean field theory, observables are obtained from
a measure on field configurations. A randomly initialized network likewise
induces a measure on functions through its architecture and the probability
distribution of its parameters.

The link becomes precise for wide networks, which under suitable
assumptions converge to Gaussian processes
\cite{Neal1996,Williams1997,LeeEtAl2018,MatthewsEtAl2018,
NovakEtAl2019,Yang2019,RasmussenWilliams2006}. For a centered Gaussian
field, the covariance is its two-point function and determines all higher
correlations by Wick's theorem. Choosing this covariance to be the Green
function of a free field operator therefore reproduces the corresponding
Gaussian field measure. In neural networks the covariance is set by the
features and their parameter statistics. Kernel methods, random features,
and wide-network limits make this relation explicit
\cite{RahimiRecht2007,ChoSaul2009,JacotEtAl2018,
SchoenholzPenningtonSohlDickstein2017,NavehEtAl2021,LiuEtAl2025KAN}.
Related infinite-network representations have also been studied for
quantum states and entanglement \cite{LuoHalverson2023InfiniteNNQS}.

This idea motivates neural network field theory (NN--FT), in which one
attempts to realize field-theoretic measures through neural ensembles
\cite{HalversonMaitiStoner2021,Halverson2021,Halverson2024TASI}.
Beyond the Gaussian limit, the central difficulty is that finite-width
non-Gaussianity need not describe a specified local interaction.
Proposed approaches include correlated parameter measures, symmetry
constraints, effective actions, and renormalization
\cite{DemirtasHalversonMaitiSchwartzStoner2023,
MaitiStonerHalverson2021,BantaEtAl2023,
ErbinLahocheSamary2022,ErbinLahocheSamary2022b,
HowardEtAl2025BayesianRG,SenVaidya2026}.
Recent work has explored universality, architectural error bounds,
constructive interacting limits, and the relation between neural
statistical laws and quantum mechanics
\cite{FerkoHalversonMutchler2026Universality,
Zhang2026OptimalArchitecture,Frank2026Constructive,
FerkoHalverson2026QuantumMechanics}.

The scope of neural constructions has also expanded to conformal fields,
spinning correlators, Virasoro symmetry, and conformal defects
\cite{HalversonNaskarTian2025,DograHalversonNaskar2026Spinning,
RobinsonVirasoro2025,CapuozzoRobinsonSuzzoni2026}.
Further developments concern anomalies, topology, fermions,
supersymmetry, and transformer-based field measures
\cite{FerkoEtAl2026Anomalies,FerkoEtAl2026,FrankEtAl2025,
AgeevAgeevaTransformer2026}. Neural approaches to string observables,
excited states, and geometric flows provide another line of investigation
\cite{HalversonString2026,AgeevAgeevaStrings2026,
Halverson2026PreStringsLectures,HalversonRuehle2024MetricFlows}.
These works make it natural to ask not only whether an appropriate
network can reproduce a given field theory, but also which features of
the neural architecture have observable consequences.

An especially interesting setting for this question is holography. The AdS/CFT correspondence relates
gravitational theories in anti--de Sitter space to
non-gravitational conformal field theories on its boundary
\cite{Maldacena,GKP,Witten,AharonyEtAl2000,Bousso2002}.
This duality provides a powerful framework for studying
strongly coupled gauge theories and offers a new perspective
on fundamental problems in quantum gravity, including the
physics of black holes and the emergence of bulk spacetime.
Its connection to string theory has also played an important
role in developing our understanding of gravitational
interactions \cite{Polyakov1981,PolchinskiBook,TongString,
Polyakov:1987hqn}. In particular, boundary correlation
functions encode information about bulk dynamics, locality,
and scattering processes
\cite{HeemskerkEtAl2009,Penedones2011,FitzpatrickKaplan2012}.
These features make holography a particularly interesting
setting for neural network field theory. If a bulk field
theory can be represented by a neural ensemble, one may ask
whether its boundary observables can be obtained directly
from the same neural degrees of freedom. Such a construction
would provide a way to investigate how bulk interactions
are encoded in boundary correlation functions and whether
different neural architectures can lead to equivalent
holographic descriptions. More generally, it offers a
new perspective on the relationship between the statistical
structure of neural networks and the emergence of
holographic field theories.

An important direction for this program is the construction
of gauge theories in the NN--FT framework. Gauge theories
play a central role in holography, where large-$N$ gauge
theories admit gravitational descriptions in appropriate
regimes. Developing their neural representations requires
a consistent treatment of gauge symmetry, which introduces
additional structure beyond the choice of neural covariance.
In particular, gauge fixing, ghost fields, and BRST symmetry
must be incorporated consistently
\cite{YangMills1954,FaddeevPopov1967,
BecchiRouetStora1975Higgs,BecchiRouetStora1975,Tyutin1975,
Nakanishi1966,Lautrup1967,Ward1950,Takahashi1957,
Slavnov1972,Taylor1971}. Further difficulties arise from
the Gribov problem and the consistency conditions for
local gauge interactions
\cite{Gribov1978,Neuberger1987,
BarnichHenneaux,BarnichHenneauxTatar,BBH}.
Understanding how these structures can emerge from neural
degrees of freedom is the subject of our forthcoming work,
\emph{Towards Yang--Mills Theory in NN--FT}
\cite{TowardsYangMillsNNFT}.

In the present work, we study a real massive scalar field
on fixed Euclidean $\AdS_{d+1}$ and investigate its
representation within NN--FT. The corresponding free
field theory is well understood. Its stability conditions,
bulk propagators, and possible boundary quantizations
have been extensively studied
\cite{BF,BFPositiveEnergy,KlebanovWitten}.
The associated radial spectral problem admits a precise
description in terms of self-adjoint operators
\cite{IshibashiWald,DerezinskiRichard}, while holographic
renormalization and boundary deformations provide a
framework for understanding its boundary observables
\cite{HenningsonSkenderis1998,deHaroEtAl2001,Skenderis2002,
WittenMultiTrace2001,BerkoozSeverShomer2002}.
These established results provide a useful benchmark
for neural constructions. Our aim is not to derive
the scalar theory anew, but to understand how it can
be realized by different neural architectures, how
local interactions can be introduced into these
representations, and how the resulting bulk dynamics
are reflected in boundary correlation functions.

Our first result is a construction and comparison of neural
representations of the free massive scalar. The main model uses
Fourier--Bessel cosine neurons adapted to the spectral decomposition
of the selected Klein--Gordon Green function,
\begin{equation}
 \phi_N(X)=\frac{1}{\sqrt N}\sum_{i=1}^{N}a_i f_{\eta_i}(X),
 \qquad a_i\sim\mathcal N(0,1),
 \label{eq:intro-neural-field}
\end{equation}
with a joint sampling law for radial and boundary momenta. At fixed
spectral regulator $\rho$, this gives the exact finite-width identity
\begin{equation}
 \mathbb E_0[\phi_N(X)\phi_N(Y)]
 =G_{\Delta,\rho}(X,Y).
 \label{eq:intro-covariance}
\end{equation}
The field is generally non-Gaussian at finite width, but its
finite-dimensional distributions converge to those of the regulated
Gaussian scalar as $N\to\infty$. We also compare deterministic spectral
and heat-kernel factorizations with the Bessel construction. These
representations clarify which properties are fixed by the field
covariance and which depend on the particular neuron family.

Our second result concerns interactions. We first follow the
action-pullback strategy of Demirtas, Halverson and collaborators,
which translates a prescribed local field interaction into a coupling
among neural output weights
\cite{DemirtasHalversonMaitiSchwartzStoner2023}.
We then develop an alternative mechanism in which auxiliary binary
neurons couple to local bulk field readouts. Integrating them out
generates a compensated $\log\cosh$ potential that approaches a positive
local quartic interaction in a controlled population limit. This is
related to Boltzmann-machine and statistical-field constructions
\cite{AckleyHintonSejnowski1985,Hinton2002,
BachtisAartsLucini2021,AartsLuciniPark2024,
GriffithsSimon,EllisNewman1978}.
Unlike direct action pullback, the auxiliary method generates the
quartic term from simpler microscopic variables. We establish
convergence at fixed regulators, identify the relevant stability
conditions, and compare the corrections at finite neural width and
finite auxiliary population.

Our third result is a neural realization of the scalar
bulk-to-boundary dictionary. We obtain boundary neurons by applying
the BDHM limit to the same Fourier--Bessel features that define the
bulk ensemble \cite{BDHM,HKLL2006}. Their mixed averages reproduce
the bulk-to-boundary propagator, and the boundary two-point function
has the expected conformal form. The leading connected four-point
function gives the scalar contact Witten diagram
\cite{FreedmanMathurMatusisRastelli1999,
DHokerFreedman2002,DHokerFreedmanRastelli1999},
\begin{equation}
 \big\langle\mathcal O(x_1)\cdots\mathcal O(x_4)\big\rangle_{\mathrm c}^{(1)}
 =-\lambda A_\Delta^2\,
 \mathcal D_{\Delta\Delta\Delta\Delta}(x_i).
 \label{eq:intro-boundary-four}
\end{equation}
This holds for separated boundary points when the contact integral
converges without further boundary renormalization. The interpretation
of such correlators is informed by conformal blocks and the conformal
bootstrap \cite{DolanOsborn2004,CostaEtAl2011,SimmonsDuffin2016}.

A further result concerns what boundary agreement does \emph{not}
imply. Replacing a coherent Bessel activation by independently sampled
cosine features changes the bulk covariance while preserving selected
leading boundary correlators. The modified bulk covariance contains
additional massive contributions, and the discrepancy survives the
outer infinite-width limit. Thus, different neural architectures can
produce the same boundary two-point function and leading contact
four-point diagram without describing the same bulk field theory.
This explicit example makes neuron design relevant to the question
of which bulk information is accessible holographically.

Several broader issues remain outside the present construction.
Recovering a physical quantum theory from Euclidean correlations also
requires conditions such as reflection positivity
\cite{OsterwalderSchrader1973,OsterwalderSchrader1975}.
Renormalization and continuum limits involve scale dependence,
effective actions, and regulator control
\cite{Callan1970,Symanzik1970,Collins1984,
Wetterich1993,Abbott1981,ReuterWetterich1994,
GrossWilczek1973,Politzer1973}.
Gauge and lattice extensions also bring in loop observables,
Wilson-line renormalization, and improved lattice actions
\cite{Wilson1974,BrandtNeriSato1981,
KorchemskyRadyushkin1987,Symanzik1983,Weisz1983}.
For strings, the corresponding worldsheet questions are tied to
sigma-model renormalization and background-field methods
\cite{Honerkamp1972,Friedan1980,Friedan1985,
AlvarezGaumeEtAl1981,CallanEtAl1985}.
The present claims are instead about regulated Euclidean scalar
measures and the specific large-width, auxiliary-population, and
boundary limits stated in the paper. The construction does not
establish dynamical gravity or a complete AdS/CFT duality.

The paper is organized as follows. Section~\ref{sec:free}
constructs the free Fourier--Bessel neural field and discusses
alternative features and finite-width effects. In
Section~\ref{sec:interactions}, we study the direct and auxiliary
constructions of local quartic interactions. Section~\ref{sec:bdhm}
derives the neural boundary observables and the bulk non-uniqueness
example. Section~\ref{sec:bdhm-summary} discusses the results and
possible extensions. Supporting calculations are collected in the
appendices.

\section{A free massive scalar in AdS and its Fourier--Bessel cosine neural representation}
\label{sec:free}

\subsection{Free massive scalar field theory in AdS: a review}
\label{sec:geometry}

We begin by reviewing the free massive scalar theory on
anti--de Sitter space, focusing on the boundary conditions
and spectral properties that will provide the foundation
for its subsequent neural-network representation.

We begin with a real scalar field on Euclidean anti--de Sitter space
$\AdS_{d+1}$, a maximally symmetric space of constant negative
curvature. The symmetries of AdS strongly constrain bulk correlation
functions, while its conformal boundary allows one to associate bulk
fields with lower-dimensional boundary observables. These properties
underlie the role of AdS in holography \cite{Maldacena,GKP,Witten}.
In this work, however, the AdS geometry is treated as a fixed
background. Our aim is to construct a neural representation of scalar
quantum field theory on AdS and to study the corresponding boundary
observables, without introducing dynamical gravity.

We work in Poincar\'e coordinates, where the bulk is parametrized by
$X=(z,x)$, with $z>0$ and $x\in\mathbb R^d$. The conformal boundary
is approached in the limit $z\to0$, and the metric takes the form
\begin{equation}
 X=(z,x),\qquad z>0,\quad x\in\RR^d,\qquad
 \dd s^2=\frac{L^2}{z^2}(\dd z^2+\dd x^2).
 \label{eq:metric}
\end{equation}
Here $L$ is the AdS curvature radius and the corresponding invariant
volume element and scalar Laplacian are given expressions
\begin{gather}
 \dV_X=\sqrt{g}\,\dd^{d+1}X
 =\frac{L^{d+1}}{z^{d+1}}\,\dd z\,\dd^d x,\\
 \nabla_g^2=\frac{1}{L^2}
 \left[z^2(\partial_z^2+\partial_x^2)
 -(d-1)z\partial_z\right].
 \label{eq:laplacian}
\end{gather}

The free scalar theory is defined by the quadratic action
\begin{equation}
 S_0[\phi]=\frac12\int_{\AdS_{d+1}}\dV_X
 \left(g^{AB}\partial_A\phi\partial_B\phi+m^2\phi^2\right),
 \qquad P_m=-\nabla_g^2+m^2.
 \label{eq:freeaction}
\end{equation}
The effective mass may include a coupling to the background
curvature, $m^2=m_0^2+\xi R$, where
$R=-d(d+1)/L^2$. Unless otherwise stated, we set $L=1$ in
the following calculations.

To specify the free theory, one must fix  the admissible behavior of the field near the conformal
boundary. Indeed, the Klein--Gordon equation admits two
independent asymptotic solutions. Substituting
$\phi\sim z^\Delta$ into $P_m\phi=0$ gives the indicial
equation
\begin{equation}
 \Delta(\Delta-d)=m^2L^2,\qquad
 \Delta_\pm=\frac d2\pm\nu,\qquad
 \nu=\sqrt{\frac{d^2}{4}+m^2L^2}.
 \label{eq:dimensions}
\end{equation}
The requirement that $\nu$ be real yields the
Breitenlohner--Freedman stability bound
\begin{equation}
 m^2L^2\geq-\frac{d^2}{4}.
 \label{eq:bf}
\end{equation}
The standard quantization selects the asymptotic behavior
$z^{\Delta_+}$ and in the mass range $0<\nu<1$, an alternative
quantization with behavior $z^{\Delta_-}$ is also admissible,
provided the corresponding boundary condition and boundary
term are chosen appropriately
\cite{BF,BFPositiveEnergy,KlebanovWitten}.
The two quantizations define distinct scalar theories at the
same bulk mass. The limiting case $\nu=0$ requires separate
treatment, since the two exponents coincide and the second
independent solution develops a logarithmic dependence on $z$.

To treat both quantizations in a unified notation, we introduce
the signed Bessel index
\begin{equation}
 \alpha=\Delta-\frac d2
 =\begin{cases}
 \nu, &\text{standard quantization},\\
 -\nu, &\text{alternative quantization}.
 \end{cases}
 \label{eq:alpha}
\end{equation}
For either admissible choice, $\alpha>-1$, and the selected
near-boundary behavior is $z^{d/2+\alpha}=z^\Delta$.
In what follows, we further restrict to $\Delta>0$, as
required for the positive boundary normalization and the
simple uniform feature bounds used in the neural construction.
This restriction is automatically satisfied for $d\geq2$ in
the quantization ranges considered here. The nonpositive-$\Delta$
part of the $d=1$ alternative-quantization window would
require a separate analysis.

To obtain the radial spectral problem, we absorb the
coordinate-dependent weight in the kinetic term through
the field redefinition
\begin{equation}
 \phi(z,x)=z^{(d-1)/2}\psi(z,x)
 \label{eq:rescale}
\end{equation}
 and integrating the mixed term by parts, the free action
takes the form
\begin{equation}
 S_0[\phi]
 =\frac12\int_0^\infty\dd z\int\dd^dx\,
 \left[
 (\partial_z\psi)^2
 +(\partial_x\psi)^2
 +\frac{\nu^2-\tfrac14}{z^2}\psi^2
 \right]+S_{\partial},
 \label{eq:flat-action}
\end{equation}
where $S_{\partial}$ collects the boundary terms associated
with the chosen quantization. The radial dependence is
thus governed by the inverse-square Schr\"odinger operator
\begin{equation}
 \mathfrak h_\nu
 =-\partial_z^2+\frac{\nu^2-\tfrac14}{z^2},
 \label{eq:radial-differential-expression}
\end{equation}
defined initially as a differential expression on
$L^2(\mathbb R_+,\dd z)$. Its self-adjoint realization,
and hence the radial spectral basis, depends on the
boundary condition at $z=0$.

The choice of boundary condition is encoded in the
self-adjoint realization of the radial operator.
Let $H_{\nu,\min}$ be the closure of $\mathfrak h_\nu$
initially defined on $C_c^\infty(\mathbb R_+)$, and
let $\chi\in C_c^\infty([0,\infty))$ be a cutoff function
equal to unity near $z=0$.
For $0<\nu<1$, both independent near-boundary solutions
are square-integrable, and the two scale-invariant
self-adjoint extensions are specified by the domains
\begin{align}
 \mathcal D\!\left(H_\nu^{\mathrm{std}}\right)
 &=
 \mathcal D(H_{\nu,\min})
 \mathbin{\dotplus}
 \mathbb C\,\chi(z)z^{\frac12+\nu},
 \label{eq:standard-radial-domain}\\
 \mathcal D\!\left(H_\nu^{\mathrm{alt}}\right)
 &=
 \mathcal D(H_{\nu,\min})
 \mathbin{\dotplus}
 \mathbb C\,\chi(z)z^{\frac12-\nu}.
 \label{eq:alternative-radial-domain}
\end{align}
The standard and alternative extensions thus select
the faster and slower radial falloffs, respectively.
Returning to the original field
$\phi=z^{(d-1)/2}\psi$, these correspond to the
asymptotic behaviors $z^{\Delta_+}$ and
$z^{\Delta_-}$ that characterize the two quantizations.
For $\nu\geq1$, the minimal operator is essentially
self-adjoint, leaving no freedom to impose an alternative
boundary condition. The limiting case $\nu=0$
requires separate treatment, since the two independent
solutions behave as $z^{1/2}$ and
$z^{1/2}\log z$ near the boundary.
In all cases, the operator is in the limit-point case
at $z=\infty$, so no boundary condition is needed there.

The choice of boundary condition can be incorporated into a unified
spectral notation. We denote the corresponding self-adjoint
realization of the radial operator by $H_\alpha$, where the signed
Bessel index $\alpha$ distinguishes the two quantizations:
\begin{equation}
 H_\alpha =
 \begin{cases}
  H_\nu^{\mathrm{std}},
   & \alpha=\nu,\\
  H_\nu^{\mathrm{alt}},
   & \alpha=-\nu,\quad 0<\nu<1.
 \end{cases}
 \label{eq:signed-index-domain}
\end{equation}
In the range $0<\nu<1$, these correspond to the two
scale-invariant self-adjoint extensions of the inverse-square
radial operator \cite{IshibashiWald,DerezinskiRichard}.

For either admissible value of $\alpha$, the radial operator
$H_\alpha$ admits a continuous spectral decomposition in terms
of Bessel functions. Its generalized eigenfunctions satisfy
\begin{equation}
 H_\alpha\!\left[\sqrt{kz}\,J_\alpha(kz)\right]
 =
 k^2\sqrt{kz}\,J_\alpha(kz),
 \label{eq:hankelop}
\end{equation}
with the near-boundary behavior
$\sqrt{kz}\,J_\alpha(kz)\propto z^{1/2+\alpha}$,
as required by the corresponding self-adjoint boundary condition.
These eigenfunctions form a complete radial spectral basis,
as expressed by the Hankel completeness relation
\begin{equation}
 \int_0^\infty\dd k\,
 kJ_\alpha(kz)J_\alpha(kw)
 =
 \frac{\delta(z-w)}{z}.
 \label{eq:hankel-completeness}
\end{equation}
Thus, both quantizations admit a common spectral description,
with the choice of boundary condition encoded in the signed
Bessel index $\alpha$.

$\,$

The translational symmetry along the boundary directions
allows us to diagonalize the remaining spatial derivatives
by a Fourier transform. Combining the boundary plane waves
with the radial Bessel eigenfunctions, we obtain the
spectral representation of the massive bulk Green function,
\begin{equation}
 G_\Delta(X,Y)
 =(zw)^{d/2}\intp e^{ip\cdot(x-y)}
 \int_0^\infty\dd k\,
 \frac{kJ_\alpha(kz)J_\alpha(kw)}{p^2+k^2},
 \qquad Y=(w,y).
 \label{eq:spectral-green}
\end{equation}
Here $p^2+k^2$ is the eigenvalue of the rescaled kinetic
operator, whose reciprocal provides the spectral weight.
The signed Bessel index $\alpha$ incorporates both the
scalar mass and the chosen boundary condition.

The normalization of the Green function is fixed by the
completeness of the radial eigenfunctions. Indeed, acting
with the Klein--Gordon operator $P_m$ on the first argument
introduces a factor $z^2(p^2+k^2)$, which cancels the
spectral denominator. Using the Hankel completeness relation
\eqref{eq:hankel-completeness}, we obtain
\begin{equation}
 P_{m,X}G_\Delta(X,Y)
 =z^{d+1}\delta(z-w)\delta^{(d)}(x-y)
 =\frac{\delta^{(d+1)}(X-Y)}{\sqrt{g(X)}}.
 \label{eq:green-delta}
\end{equation}
Thus, $G_\Delta$ is the Green function of the Klein--Gordon
operator with the prescribed self-adjoint boundary condition.

The radial spectral integral can be evaluated explicitly
using the Bessel resolvent identity
\begin{equation}
 \int_0^\infty\dd k\,
 \frac{kJ_\alpha(kz)J_\alpha(kw)}{k^2+p^2}
 =I_\alpha(pz_<)K_\alpha(pz_>),
 \quad z_<=\min(z,w),\quad z_>=\max(z,w),
 \label{eq:IK}
\end{equation}
where $p$ denotes the magnitude of the boundary momentum.
For $\alpha<0$, the appearance of $I_\alpha$ reflects
the alternative boundary condition. Substituting this
identity into the spectral representation gives
\begin{equation}
 G_\Delta(X,Y)=(zw)^{d/2}\intp e^{ip\cdot(x-y)}
 I_\alpha(pz_<)K_\alpha(pz_>).
 \label{eq:mixedgreen}
\end{equation}
This expression makes the dependence on the boundary
momentum and radial coordinates explicit.

Alternatively, the Green function admits a manifestly
AdS-invariant position-space representation. Introducing
the invariant variable $\zeta$ and the normalization
constant $A_\Delta$,
\begin{equation}
 \zeta=\frac{2zw}{z^2+w^2+|x-y|^2},\qquad
 A_\Delta=\frac{\Gamma(\Delta)}{2\pi^{d/2}
 \Gamma(\Delta-d/2+1)},
 \label{eq:normalA}
\end{equation}
the propagator takes the form
\begin{equation}
 G_\Delta(X,Y)=\frac{A_\Delta}{2^\Delta}\zeta^\Delta
 {}_2F_1\!\left(\frac\Delta2,\frac{\Delta+1}{2};
 \Delta-\frac d2+1;\zeta^2\right).
 \label{eq:hypergreen}
\end{equation}
For distinct bulk points, $0<\zeta<1$. In this form,
the dependence on the bulk coordinates is entirely
captured by the AdS-invariant variable $\zeta$, while
the conformal dimension $\Delta$ encodes the scalar
mass and the choice of quantization.

\subsection{Neural field and Bessel-cosine neurons}
\label{sec:bessel-neurons}

We now turn to the construction of a neural field whose
two-point function reproduces the massive AdS Green function
derived above. Our choice of neural features is motivated
by the spectral decomposition of the Klein--Gordon operator.
Translation invariance along the boundary directions
naturally leads to Fourier modes, while the radial dependence
is governed by Bessel functions whose order encodes the
scalar mass and the chosen boundary condition. This suggests
combining a cosine activation in the boundary coordinates
with a Bessel profile in the radial direction, yielding
a neural architecture directly adapted to the spectral
structure of the free bulk theory.

This choice is not unique. More generally, the essential
requirement is that the covariance of the neural field
reproduce the Green function of the Klein--Gordon operator
with the prescribed boundary condition,
\begin{equation}
 G_\Delta(X,Y)=P_m^{-1}(X,Y).
\end{equation}
At fixed regulator, a linear Gaussian representation
amounts to finding a factorization
$G_\Delta=F\Sigma F^T$, where $F$ denotes the feature map
and $\Sigma$ the covariance of the Gaussian coefficients.
Different factorizations can therefore describe the same
free field theory.

The Bessel--cosine architecture is particularly well suited to
the spectral structure of the massive AdS scalar, but it is
by no means the only possible representation. One alternative
is to construct the field using deterministic Fourier--Bessel
modes or, more generally, eigenfunctions of a suitably
regulated kinetic operator. Assigning Gaussian coefficients
with variances determined by the corresponding eigenvalues
then produces an exactly Gaussian field at finite spectral
resolution, with covariance given by the truncated spectral
expansion of the Green function. This approach provides
direct control over the spectral approximation, although
the resulting modes are generally extended throughout
the bulk.

A different possibility is to employ localized basis
functions, heat-kernel profiles, or wavelet-like features.
Such representations can offer greater spatial localization
and may be advantageous for implementing local interactions
or resolving structures at different bulk scales. However,
unlike the Fourier--Bessel modes, these features do not
generally diagonalize the kinetic operator. Reproducing
the desired Green function therefore requires an appropriate
choice of spectral weights, sampling measures, or correlations
among the expansion coefficients. In this sense, the
information carried by the free propagator can be distributed
differently between the feature functions and their
statistical properties.

Randomly sampled neural features introduce an additional
distinction. Even when the output coefficients are Gaussian,
the resulting field is generally not exactly Gaussian at
finite width. Conditional on the sampled features, the field
has a Gaussian distribution, but its covariance depends
on the particular realization of the hidden parameters.
Averaging over these parameters therefore produces a
mixture of Gaussian measures, with non-Gaussian corrections
arising from fluctuations of the empirical covariance.
Under suitable regularity and moment assumptions, these
fluctuations vanish in the infinite-width limit, and the
field converges to the Gaussian theory determined by the
ensemble-averaged covariance.

The choice of activation functions is equally important.
Although the radial Bessel profile can, in the appropriate
range of its index, be represented as a coherent average
of elementary cosine functions, replacing this average
with independently sampled cosine neurons generally
changes the resulting covariance. In particular, such
a replacement need not reproduce the massive AdS
propagator, even when the leading boundary behavior is
preserved. Thus, matching the asymptotic scaling of the
individual features is not sufficient: the complete
spectral structure of the bulk Green function must also
be recovered.

These considerations illustrate the flexibility of neural
representations of free field theories, while emphasizing
the constraints imposed by the underlying kinetic operator.
Different architectures may encode the same Gaussian
measure through distinct combinations of feature profiles
and parameter statistics, but they can differ substantially
in their finite-width corrections, localization properties,
and practical implementation. A more detailed discussion
of these alternatives, together with explicit constructions
and comparisons, is presented in
Appendix~\ref{app:alternative-neurons}.

$\,$

We now specialize to the Bessel--cosine representation, in which
the neural field is constructed from independently weighted
spectral features. We choose the output weights to be Gaussian
and encode the spectral properties of the massive scalar
through the neuron profiles and the probability distribution
of their parameters. These ingredients are chosen consistently
so that the resulting covariance reproduces the regulated
massive AdS Green function.

To separate the prescribed near-boundary scaling from the
remaining radial dependence, we introduce the normalized
Bessel profile
\begin{equation}
 R_\alpha(t)=2^\alpha\Gamma(\alpha+1)t^{-\alpha}J_\alpha(t),
 \qquad R_\alpha(0)=1,
 \qquad C_\alpha=2^\alpha\Gamma(\alpha+1).
 \label{eq:radial-act}
\end{equation}
The normalization $R_\alpha(0)=1$ ensures that the leading
boundary behavior can be extracted as an explicit power
of the radial coordinate. Indeed, using
$\Delta=d/2+\alpha$, we have
\begin{equation}
 z^\Delta R_\alpha(kz)=C_\alpha k^{-\alpha}
 z^{d/2}J_\alpha(kz).
 \label{eq:radial-identity}
\end{equation}
Thus, the factor $z^\Delta$ determines the required
near-boundary scaling, while $R_\alpha(kz)$ captures
the radial dependence of the Fourier--Bessel spectral
modes. This separation is particularly convenient for
implementing the chosen boundary condition and, later,
for extracting the corresponding boundary observables.

Motivated by this structure, we define an individual
real-valued neuron as
\begin{equation}
 f_\eta(z,x)=B_\rho z^\Delta R_\alpha(kz)
 \cos(p\cdot x+b),
 \qquad
 \eta=(p,k,b),\quad k>0,
 \label{eq:neuron}
\end{equation}
where $p\in\mathbb{R}^d$ is the momentum along the boundary
directions, $k$ is the radial spectral parameter, and
$b$ is a phase uniformly distributed on $[0,2\pi)$.
The constant $B_\rho$ fixes the overall normalization
and will be determined together with the sampling
distribution of the spectral parameters.

Although these neurons are built from the spectral modes
of the free theory, they should not be interpreted as
individual solutions of the homogeneous Klein--Gordon
equation. Rather, their Fourier--Bessel structure implies
\begin{equation}
 P_m f_\eta(X)=z^2(p^2+k^2)f_\eta(X),
\end{equation}
where we have set the AdS radius to unity. Each neuron
therefore represents a spectral feature of the rescaled
kinetic operator. The inverse spectral factor
$(p^2+k^2)^{-1}$, required for the Green function,
is incorporated through the distribution of neuron
parameters. The massive bulk propagator is then recovered
from the covariance obtained by averaging over these
features with the appropriate spectral weights.

An additional useful property of the normalized Bessel
profile is its integral representation in terms of
elementary cosine functions. For $\alpha>-1/2$, one has
\begin{equation}
 R_\alpha(t)=
 \int_{-1}^{1}\dd u\,\rho_\alpha(u)\cos(ut),
 \qquad
 \rho_\alpha(u)=
 \frac{\Gamma(\alpha+1)}
 {\sqrt{\pi}\Gamma(\alpha+\tfrac12)}
 (1-u^2)^{\alpha-\frac12}.
 \label{eq:bessel-cosine-average}
\end{equation}
Here $\rho_\alpha(u)$ is an even, positive, normalized
probability density on $[-1,1]$. Substituting this
representation into the neuron profile gives
\begin{equation}
 f_\eta(z,x)
 =
 \int_{-1}^{1}\dd u\,\rho_\alpha(u)\,
 B_\rho z^\Delta
 \cos(p\cdot x+kuz+b).
\end{equation}
Hence, each Bessel--cosine neuron can equivalently be
viewed as a coherent superposition of elementary cosine
features, with the radial dependence generated by
averaging over the auxiliary variable $u$.

The distinction between this coherent average and
independent random sampling is essential. In the
representation above, the integration over $u$ is
performed within each neuron, before assigning its
Gaussian output weight. Consequently, the resulting
covariance involves products of the averaged features.
If the elementary cosines were instead assigned
independent output weights, their covariance would
involve averages of products and would generally differ
from the Fourier--Bessel covariance. The coherent
construction therefore preserves the radial spectral
structure required to reproduce the massive AdS
propagator, whereas independently sampled elementary
cosines need not define the same bulk field theory.

Having specified the neuron profiles, we now construct the
probability distribution of their parameters so that the
ensemble-averaged covariance reproduces the massive AdS Green
function. The sampling measure is determined by the spectral
representation of the propagator, with suitable ultraviolet
and infrared regularization to ensure that all relevant
integrals are well defined.

To this end, we introduce a nonnegative spectral regulator
$\rho(p,k)$ and define the normalization factor
\begin{equation}
 Z_\rho=\int\frac{\dd^d p}{(2\pi)^d}
 \int_0^\infty\dd k\,
 \rho(p,k)\frac{k^{2\alpha+1}}{p^2+k^2}<\infty.
 \label{eq:Zrho}
\end{equation}
We assume $Z_\rho>0$ and that the regulator is even under
$p\mapsto-p$. A convenient choice is a smooth cutoff with
support in a finite spectral annulus, which suppresses both
the infrared and ultraviolet regions. The precise form
of $\rho$ is not essential for the construction, provided
the resulting spectral measure is finite and positive.

We then define the joint probability density of the
boundary momentum $p$ and radial spectral parameter $k$ by
\begin{equation}
 q_\rho(p,k)=\frac{1}{Z_\rho(2\pi)^d}
 \rho(p,k)\frac{k^{2\alpha+1}}{p^2+k^2},
 \qquad
 B_\rho^2=\frac{2Z_\rho}{C_\alpha^2}.
 \label{eq:qneuron}
\end{equation}
The factor $Z_\rho$ ensures that $q_\rho$ is properly
normalized, while $B_\rho$ fixes the overall amplitude of
the neuron features. Importantly, the momenta $p$ and $k$
are sampled jointly rather than independently. Their
distribution contains the factor $(p^2+k^2)^{-1}$ inherited
from the spectral representation of the Klein--Gordon
Green function, together with the radial weight
$k^{2\alpha+1}$. The sampling density and the neuron
normalization are therefore linked: both must be chosen
consistently to reproduce the desired propagator.

The remaining parameter, the phase $b$, is sampled
independently and uniformly on $[0,2\pi)$. We denote
expectation with respect to the complete neuron parameter
distribution by $\mathbb E_\eta$. For any integrable
function $F(\eta)$, this expectation is explicitly
\begin{equation}
 \mathbb{E}_\eta[F(\eta)]
 =
 \int_{\mathbb{R}^d}\dd^d p
 \int_0^\infty\dd k\,q_\rho(p,k)
 \int_0^{2\pi}\frac{\dd b}{2\pi}\,
 F(p,k,b).
 \label{eq:neuron-average}
\end{equation}
This averaging operation refers solely to the statistical
ensemble of hidden neuron parameters. In particular, it
should be distinguished from the integral over the auxiliary
variable $u$ in \eqref{eq:bessel-cosine-average}, which
defines a single Bessel activation through a deterministic
coherent superposition of elementary cosines. The two
operations play different roles: the coherent integral
determines the radial profile of an individual neuron,
whereas the ensemble average determines the statistical
properties of the neural field.

We can now verify that the chosen parameter distribution
reproduces the required covariance. Since the phase is
uniform, the expectation of an individual neuron vanishes,
$\mathbb E_\eta[f_\eta(X)]=0$. The product of two cosine
activations, however, has a nonvanishing phase average,
\begin{equation}
 \mathbb{E}_b[
 \cos(p\cdot x+b)\cos(p\cdot y+b)]
 =
 \frac12\cos\bigl(p\cdot(x-y)\bigr).
\end{equation}
This identity converts the product of boundary activations
into a function of the coordinate difference $x-y$, as
required by translation invariance along the AdS boundary.

Combining the phase average with the radial identity
\eqref{eq:radial-identity} and the sampling density
\eqref{eq:qneuron}, we obtain
\begin{align}
 \mathbb{E}_\eta[f_\eta(X)f_\eta(Y)]
 &=
 \frac{B_\rho^2C_\alpha^2}{2Z_\rho}
 (zw)^{d/2}
 \int\frac{\dd^d p}{(2\pi)^d}
 \int_0^\infty\dd k\,
 \rho(p,k)
 \frac{kJ_\alpha(kz)J_\alpha(kw)}{p^2+k^2}
 \cos\bigl(p\cdot(x-y)\bigr)
 \nonumber\\
 &=G_{\Delta,\rho}(X,Y).
 \label{eq:feature-covariance}
\end{align}
In the second line, we have used the normalization
$B_\rho^2C_\alpha^2/(2Z_\rho)=1$. Moreover, because
$\rho(p,k)$ is even in the boundary momentum, the cosine
integral is equivalent to the Fourier representation of
the regulated Green function.

This establishes the central covariance identity underlying
the Bessel--cosine construction. The neuron profiles encode
the radial eigenfunctions and the prescribed AdS boundary
condition, while the joint sampling distribution supplies
the spectral weights of the inverse Klein--Gordon operator.
Their combination reproduces the regulated massive bulk
propagator directly as an ensemble average of products of
neural features. In particular, the construction fixes
not only the near-boundary scaling of the field but also
its full regulated two-point function throughout the bulk.

Having established the covariance of an individual
Bessel--cosine feature, we now construct the corresponding
finite-width neural field. We introduce independent standard
Gaussian output coefficients $a_i\sim\mathcal N(0,1)$
and independently sampled hidden parameters $\eta_i$,
distributed according to the spectral measure defined
above. The neural field is then given by
\begin{equation}
 \phi_N(X)=\frac{1}{\sqrt{N}}
 \sum_{i=1}^N a_i f_{\eta_i}(X).
 \label{eq:nnfield}
\end{equation}
The normalization $N^{-1/2}$ ensures that the field
covariance remains finite as the number of neurons
increases. We denote by $\mathbb E_0$ expectation over
both the Gaussian output weights and the hidden neuron
parameters, with these two sets of random variables
taken to be independent.

The covariance of the neural field follows directly
from the independence and normalization of the output
coefficients. Since $\mathbb E_0[a_i a_j]=\delta_{ij}$,
all cross terms with $i\neq j$ vanish, giving
\begin{align}
 \mathbb{E}_0[\phi_N(X)\phi_N(Y)]
 &=
 \frac1N\sum_{i,j=1}^N
 \mathbb{E}_0[
 a_i a_j f_{\eta_i}(X)f_{\eta_j}(Y)]
 \nonumber\\
 &=
 \frac1N\sum_{i=1}^N
 \mathbb{E}_{\eta_i}[
 f_{\eta_i}(X)f_{\eta_i}(Y)]
 \nonumber\\
 &=G_{\Delta,\rho}(X,Y).
 \label{eq:twoptexact}
\end{align}
Thus, the ensemble-averaged two-point function agrees
exactly with the regulated massive AdS propagator for
every finite width $N$. This identity is a direct
consequence of the spectral sampling distribution and
does not require taking the infinite-width limit.

Exact agreement of the covariance, however, does not
imply that the finite-width neural field is Gaussian.
For a fixed realization of the hidden parameters,
$\phi_N$ is a linear combination of independent Gaussian
variables and therefore defines a Gaussian field with
conditional covariance
\begin{equation*}
 C_\eta(X,Y)=\frac1N\sum_{i=1}^N
 f_{\eta_i}(X)f_{\eta_i}(Y).
\end{equation*}
This covariance depends on the sampled neurons and
fluctuates between different realizations. Averaging
over the hidden parameters consequently produces a
mixture of Gaussian measures, which is generally
non-Gaussian despite having the exact ensemble-averaged
covariance $G_{\Delta,\rho}$. The distinction disappears
in the large-width limit when the empirical covariance
concentrates around its expectation.

To make this limit precise, consider a finite collection
of bulk points $X_1,\ldots,X_m$ and introduce the
corresponding feature vector
\begin{align}
 \Xi_\eta
 &=
 \bigl(
 f_\eta(X_1),\ldots,f_\eta(X_m)
 \bigr).
 \label{eq:large-width-feature-vector}
\end{align}
At fixed spectral and bulk regulators, we assume that
the feature distribution has a finite fourth moment,
\begin{align}
 \EE_\eta\!\left[\|\Xi_\eta\|^4\right]
 &<\infty.
 \label{eq:large-width-moment-assumption}
\end{align}
This condition controls fluctuations of the sampled
features and is sufficient for the characteristic-function
expansion below.

For arbitrary real coefficients $t_1,\ldots,t_m$,
define
\begin{equation}
 Q_\eta=
 \left(\sum_r t_r f_\eta(X_r)\right)^2.
\end{equation}
Integrating first over the independent Gaussian output
weights, and then averaging over the hidden parameters,
gives the characteristic function
\begin{equation}
 \mathbb{E}_0
 \exp\left(i\sum_r t_r\phi_N(X_r)\right)
 =
 \left[
 \mathbb{E}_\eta
 \exp\left(-\frac{Q_\eta}{2N}\right)
 \right]^N.
\end{equation}
The factorized form follows from the independence of
the neurons. Expanding its logarithm for large $N$,
while keeping the regulators fixed, yields
\begin{align}
 \log\mathbb{E}_0
 \exp\left(i\sum_r t_r\phi_N(X_r)\right)
 &=
 -\frac12\mathbb{E}_\eta[Q_\eta]
 +O(N^{-1})
 \nonumber\\
 &=
 -\frac12\sum_{r,s}
 t_rt_sG_{\Delta,\rho}(X_r,X_s)
 +O(N^{-1}).
 \label{eq:clt}
\end{align}
The leading term is precisely the logarithm of the
characteristic function of a centered multivariate
Gaussian distribution with covariance determined by
the regulated AdS Green function. The remaining terms
describe finite-width deviations from Gaussianity.

More formally, the multivariate central-limit theorem
implies
\begin{align}
 \bigl(
 \phi_N(X_1),\ldots,\phi_N(X_m)
 \bigr)
 &\xrightarrow[N\to\infty]{\mathrm d}
 \mathcal N\!\left(0,G_\rho^{(m)}\right),
 \qquad
 \bigl(G_\rho^{(m)}\bigr)_{rs}
 =
 G_{\Delta,\rho}(X_r,X_s).
 \label{eq:large-width-fdd-limit}
\end{align}
Here $\xrightarrow{\mathrm d}$ denotes convergence in
distribution. Hence, at fixed regulators, the
finite-dimensional distributions of the neural field
converge to those of the Gaussian massive scalar field
with the prescribed AdS boundary condition.

The same limit can be understood directly through the
empirical covariance. Since the hidden parameters are
sampled independently, the variance of $C_\eta(X_r,X_s)$
is suppressed by the number of neurons:
\begin{align}
 &\EE_{\eta_1,\ldots,\eta_N}
 \left[
 \left|
 C_\eta(X_r,X_s)
 -
 G_{\Delta,\rho}(X_r,X_s)
 \right|^2
 \right]
 \notag\\
 &\qquad=
 \frac{1}{N}
 \Var_\eta\!\left[
 f_\eta(X_r)f_\eta(X_s)
 \right]
 \longrightarrow0.
 \label{eq:large-width-covariance-mean-square}
\end{align}
Thus, the conditional covariance converges in mean
square to the deterministic regulated propagator.
This provides a complementary interpretation of the
Gaussian limit: fluctuations of the covariance induced
by random feature sampling vanish as $N\to\infty$.

For the interacting theory developed below, we also
need to control observables involving products of
propagators integrated over the bulk. Pointwise
convergence of the empirical covariance is not by
itself sufficient to justify passing the large-width
limit through such integrals. In particular, the
leading quartic contact interaction involves a product
of four covariance kernels evaluated at a common
bulk vertex.

To formulate an appropriate convergence condition,
consider four fixed external points $X_r$ and define
\begin{align}
 G_r(Y)
 &=
 G_{\Delta,\rho}(X_r,Y).
 \label{eq:contact-limit-kernels}
\end{align}
On a regulated bulk region $\Omega$, we assume
convergence of the empirical covariance in the
corresponding $L^4$ norm, together with a uniform
fourth-moment bound:
\begin{align}
 \EE_{\eta_1,\ldots,\eta_N}
 \left[
 \left\|
 C_\eta(X_r,\cdot)-G_r
 \right\|_{L^4(\Omega,\dV)}^4
 \right]
 &\longrightarrow0,
 \qquad r=1,\ldots,4,
 \label{eq:contact-L4-convergence}\\
 \sup_N
 \EE_{\eta_1,\ldots,\eta_N}
 \left[
 \left\|
 C_\eta(X_r,\cdot)
 \right\|_{L^4(\Omega,\dV)}^4
 \right]
 &<\infty,
 \qquad r=1,\ldots,4.
 \label{eq:contact-L4-bound}
\end{align}
These conditions provide the integrability control
needed for the product of four covariance kernels.
In particular, applying H\"older's inequality to
the difference of the two products gives
\begin{align}
 &\EE_{\eta_1,\ldots,\eta_N}
 \left|
 \int_\Omega\dV_Y\,
 \prod_{r=1}^{4}C_\eta(X_r,Y)
 -
 \int_\Omega\dV_Y\,
 \prod_{r=1}^{4}G_{\Delta,\rho}(X_r,Y)
 \right|
 \longrightarrow0.
 \label{eq:contact-integral-mean-convergence}
\end{align}
The integrated quartic contact expression therefore
converges in mean to the corresponding expression
constructed from the regulated massive propagator.
Under these assumptions, the large-width limit can
be taken through the regulated interaction vertex,
providing the connection between the finite neural
ensemble and the perturbative bulk field theory.

The finite-width corrections encountered in this
construction have a distinct origin from the
interactions introduced later. They arise from
statistical fluctuations of the sampled features,
even though the output coefficients are Gaussian
and no interaction has been imposed. Their leading
contribution to non-Gaussian observables is typically
suppressed by $N^{-1}$ at fixed regulators. It is
therefore important to distinguish this intrinsic
random-feature non-Gaussianity from the correlations
generated by an explicit or dynamically induced
interaction in the neural probability measure.

\subsection{Representation invariance and finite-width cumulants}
\label{sec:representation-invariance}

The preceding construction illustrates a more general
property of Gaussian neural representations: the free
field measure is determined by the full covariance
operator rather than by a particular choice of
activation functions or feature coordinates. Different
neural architectures can therefore describe the same
Gaussian field theory, provided they induce the same
covariance and satisfy the corresponding boundary
conditions.

To make this statement precise, consider a finite
regulated field space and a linear representation
$\phi=Fa$, where the coefficient vector is Gaussian,
$a\sim\mathcal N(0,\Sigma_a)$. The resulting field is
also Gaussian, with covariance
\begin{equation}
 G=F\Sigma_aF^T.
 \label{eq:representation-covariance}
\end{equation}
The feature map $F$ and coefficient covariance
$\Sigma_a$ are not separately unique. Indeed, under
an arbitrary invertible transformation $T$ of the
coefficient coordinates, the redefinitions
\begin{equation}
 F'=FT,\qquad \Sigma_a'=T^{-1}\Sigma_aT^{-T}
 \label{eq:feature-basis-change}
\end{equation}
leave the covariance unchanged,
$F'\Sigma_a'F'^T=G$. Since a centered Gaussian
measure is completely determined by its covariance,
the transformed representation defines exactly the
same free field law.

This equivalence extends to interacting theories
whenever the interaction is specified as a functional
of the field itself. If two representations induce
the same Gaussian measure $\dd\mu_G$ and are
deformed by the same interaction functional
$V[\phi]$, then their normalized expectations agree:
\begin{equation}
 \langle\mathcal A\rangle_{\lambda}
 =\frac{\int\dd\mu_G(\phi)\,\mathcal A[\phi]e^{-\lambda V[\phi]}}
 {\int\dd\mu_G(\phi)\,e^{-\lambda V[\phi]}}.
 \label{eq:representation-interacting-invariance}
\end{equation}
This statement holds whenever the corresponding
reweighted measures are well defined. It expresses
representation invariance at the level of the complete
field probability law, rather than at the level of
individual neurons or their parametrization.

An important qualification arises for randomly sampled
features at finite width. Although their ensemble-averaged
covariance may agree exactly with $G$, the unconditional
field measure is generally a non-Gaussian mixture.
Consequently, equality of two-point functions alone
does not guarantee equality of the complete probability
laws, nor does it imply that their interacting
deformations are equivalent.

The finite-width deviations from Gaussianity can be
characterized explicitly through the connected
cumulants of the neural field. Consider a smeared
observable with test function $J$ and define
\begin{equation}
 \Phi_J=\int\dV_X\,J(X)\phi_N(X),\qquad
 F_{\eta}[J]=\int\dV_X\,J(X)f_\eta(X).
 \label{eq:smeared-feature}
\end{equation}
The neural expansion expresses $\Phi_J$ as a sum
of independent random contributions,
$N^{-1/2}a_iF_{\eta_i}[J]$. Conditioning on the
hidden parameters and integrating over the Gaussian
output weights therefore gives, after averaging
over the features,
\begin{equation}
 \EE_0[e^{t\Phi_J}]
 =\left[\EE_\eta\exp\!\left(\frac{t^2F_\eta[J]^2}{2N}\right)
 \right]^N.
 \label{eq:finite-width-mgf}
\end{equation}
This identity holds wherever the moment-generating
function is finite. It makes the finite-width
statistical structure particularly transparent:
the Gaussian output weights generate the quadratic
dependence on the source $t$, while the fluctuations
of the sampled features enter through the remaining
parameter average.

Expanding the logarithm of this expression and
comparing coefficients in the cumulant-generating
function, we obtain
\begin{equation}
 \kappa_{2r}(\Phi_J)
 =\frac{(2r-1)!!}{N^{r-1}}\,
 \kappa_r\!\left(F_\eta[J]^2\right),\qquad
 \kappa_{2r+1}(\Phi_J)=0.
 \label{eq:finite-width-all-cumulants}
\end{equation}
Here the formula for each cumulant is understood
under the corresponding moment assumptions. All
odd cumulants vanish because the neural field is
symmetric under reversal of the Gaussian output
weights. The even cumulants, on the other hand,
measure fluctuations of powers of the individual
feature contributions and exhibit an explicit
dependence on the neural width.

In particular, the fourth cumulant is
\begin{equation}
 \kappa_4(\Phi_J)
 =\frac{3}{N}\Var_\eta\!\left(F_\eta[J]^2\right)\geq0.
 \label{eq:finite-width-fourth-cumulant}
\end{equation}
This expression shows that the finite-width
non-Gaussianity is controlled by fluctuations of
the feature-induced variance. It is nonnegative
and decreases as $N^{-1}$ at fixed regulators,
provided the corresponding variance is finite.
Thus, although the free two-point function is
reproduced exactly for every width, the connected
four-point function generally approaches its
Gaussian value only in the infinite-width limit.

The same observation identifies the condition under
which an ensemble with Gaussian output weights is
exactly Gaussian even at finite width. In a finite
regulated field space, this requires the conditional
covariance to be independent of the sampled feature
realization. Equivalently, for all regulated
evaluation points,
\begin{align}
 \Var_{\eta_1,\ldots,\eta_N}
 \left[
 C_\eta(X,Y)
 \right]
 &=0,
 \qquad
 \EE_{\eta_1,\ldots,\eta_N}
 \left[
 C_\eta(X,Y)
 \right]
 =
 G(X,Y).
 \label{eq:nonrandom-conditional-covariance}
\end{align}
When this condition holds, the Gaussian mixture
reduces to a single Gaussian measure. Otherwise,
the randomness of the empirical covariance
generates nonzero higher connected cumulants,
even in the absence of any explicit interaction.

For the normalized Bessel--cosine features, the
moment conditions entering these arguments can
be controlled directly in an important range
of parameters. When $\alpha>-1/2$, the positive
integral representation
\eqref{eq:bessel-cosine-average} implies
$|R_\alpha(t)|\leq1$ for real $t$. Consequently,
for $\Delta\geq0$ and $0<z\leq Z$,
\begin{equation}
 |f_\eta(z,x)|\leq B_\rho Z^\Delta.
 \label{eq:regulated-feature-bound}
\end{equation}
At fixed spectral regulator, this provides a
uniform bound on the individual features over
the indicated radial region and therefore
controls their moments. It applies to standard
quantization and to alternative quantization
in the range $0<\nu<1/2$, where the positive
cosine representation is available.

If $\Delta<0$, the same argument requires a
radial cutoff $\epsilon\leq z\leq Z$, with the
bound on $z^\Delta$ taken over that interval.
More generally, these estimates are statements
at fixed regulators and do not establish
uniform control when the ultraviolet cutoff
is removed. The continuum limit may require
additional bounds on the spectral measure and
the corresponding field observables.

Finally, the sign of the fourth cumulant
provides a useful distinction between
random-feature fluctuations and the
non-Gaussianity induced by a positive quartic
interaction. The finite-width contribution
in \eqref{eq:finite-width-fourth-cumulant}
is nonnegative, whereas the leading connected
fourth cumulant of a smeared field generated
by a positive local $\lambda\phi^4$ interaction
is nonpositive. These effects therefore have
different statistical origins and should be
treated separately. The former reflects
fluctuations of the free neural covariance,
while the latter encodes genuine interactions
in the field probability measure. This
distinction will be important when comparing
the interacting neural constructions below;
see also Appendix~\ref{app:gaussian-obstruction}.

\section{Two constructions of a quartically interacting neural field}
\label{sec:interactions}

The free neural construction developed in
Section~\ref{sec:free} provides a Gaussian reference theory
whose covariance is determined by the massive AdS Green
function, with the prescribed boundary condition and
spectral regularization. Although the neural field is
conditionally Gaussian at fixed hidden parameters,
averaging over randomly sampled features generally
produces a non-Gaussian field at finite width. This
non-Gaussianity, however, should not be confused with
the effects of a genuine local interaction. In particular,
finite-width connected correlations arise from fluctuations
of the empirical covariance and do not, in general, exhibit
the sign, locality, or momentum dependence characteristic
of an interacting scalar field theory.

We therefore go beyond the non-Gaussian
corrections intrinsic to random-feature sampling and
construct a neural probability measure that reproduces
a prescribed local interaction in the bulk. Specifically,
we seek a neural ensemble whose induced field measure
agrees, at fixed regulators, with the Gibbs deformation
of the free Gaussian measure,
\begin{align}
 \dd\mu_\lambda(\phi)
 &=
 \frac{1}{\cZ_\lambda}
 \exp\left[
 -S_{\mathrm{int}}[\phi]
 -S_{\mathrm{ct}}[\phi]
 \right]
 \dd\mu_G(\phi),
 \label{eq:interacting-measure-overview}
\end{align}
where $\dd\mu_G$ denotes the Gaussian measure associated
with the selected massive AdS propagator, and
$\cZ_\lambda$ is the corresponding normalization
factor. The functional $S_{\mathrm{int}}$ specifies
the interaction, while $S_{\mathrm{ct}}$ contains
the counterterms required by the chosen renormalization
prescription.

In the present work, we focus on a scalar theory
with a local quartic interaction. The target action
is therefore
\begin{align}
 S[\phi]
 &=
 S_0[\phi]
 +
 \frac{\lambda}{4!}
 \int_\Omega\dV_Y\,\phi(Y)^4
 +
 S_{\mathrm{ct}}[\phi],
 \qquad
 \lambda\geq0.
 \label{eq:target-action}
\end{align}
Here $\lambda$ is the quartic coupling, and the
interaction is integrated over a compact bulk region
$\Omega$ using the invariant AdS volume element.
The restriction $\lambda\geq0$ ensures that the
bare quartic potential is bounded from below.

To formulate the construction precisely, we first
work at fixed spectral or field regulator, with
the bulk region $\Omega$ and the boundary condition
held fixed. This allows the neural and interacting
field measures to be defined within a regulated
finite-dimensional setting, where their normalization
and correlation functions can be analyzed directly.
The removal of regulators and the corresponding
renormalization limits are considered separately.

We realize this interaction \eqref{eq:target-action} in two complementary ways. In the first,
the desired field functional is pulled back directly to the neural
parameters: at fixed hidden features, the Gaussian output-weight measure
is multiplied by
$\exp[-S_{\mathrm{int}}[\phi_N]]$ and normalized. This is the AdS
implementation of the neural action-deformation mechanism of
Ref.~\cite{DemirtasHalversonMaitiSchwartzStoner2023}. It gives an exact
regulated representation once the Gaussian field measure has been fixed,
but it is ``\textit{circular}'' in the sense that the quartic interaction
is prescribed before the neural parameter law is constructed.

The second construction follows a different approach,
in which the quartic interaction is generated rather
than explicitly prescribed. To achieve this, we introduce
additional binary auxiliary neurons that couple linearly
to local values of the Gaussian bulk field. Integrating
out these auxiliary degrees of freedom produces an
effective $\log\cosh$ potential, whose nonlinear structure
is determined by the statistics of the binary variables.
A suitable quadratic compensator cancels the induced
quadratic contribution, leaving a positive quartic term
as the leading nontrivial interaction. By introducing
replicated auxiliary neurons with appropriately scaled
couplings and taking a controlled weak-unit limit,
the higher-order contributions are suppressed, and
the effective theory approaches a local
$\lambda\phi^4$ interaction.
In this formulation, the quartic dependence arises
from the non-Gaussian statistics of the auxiliary
neurons rather than from a polynomial interaction
inserted into the original neural probability measure.
The locality of the resulting interaction, however,
is not generated by these statistics alone. It is
built into the coupling structure, since each auxiliary
neuron interacts with a point-local readout of the
bulk field. The construction therefore separates
the statistical origin of the quartic interaction
from the architectural choice that determines its
spatial locality.

Both constructions use the same quenched convention for randomly
sampled hidden features: the interacting output-weight measure is
normalized at fixed features, and only then averaged over the original
spectral sampling law. When their complete Gaussian field laws, quartic
couplings, counterterms, and feature-averaging prescriptions are matched,
the two constructions yield the same regulated quartic field theory and
the same leading bulk contact correlator. Their finite-population and
higher-point corrections, however, need not agree.

\subsection{Quartic interactions through direct action pullback}
\label{sec:pullback}

We begin with the direct pullback construction. A realization of the
hidden features is first sampled from the free product measure and then
held fixed. The interacting output-weight measure is formed and normalized
for this fixed realization; the quenched average over hidden features is
taken only afterward. The interaction therefore changes the conditional
law of the output weights without reweighting the spectral sampling law
$q_\rho$, see \eqref{eq:qneuron}.

At fixed features, the construction is simply the ordinary Gibbs
deformation of the induced Gaussian field measure. Its nontrivial neural
content is that the local field interaction becomes a generally dense
correlation among the output coefficients, with the corresponding
coupling tensor determined by the bulk geometry and the neuron
functions. Below we make this correspondence explicit.

We now express the quartic interaction in terms of the neural output coefficients. Recall the Bessel--cosine neural field
\begin{align}
 \phi_N(X)
 &=\frac{1}{\sqrt N}\sum_{i=1}^{N}a_i f_{\eta_i}(X).
 \label{eq:forward-field}
\end{align}
The output weights are distributed according to the normalized Gaussian
measure, while the hidden parameters are sampled independently from the
spectral law,
\begin{align}
 &\dd\mu_0(a)
 =\prod_{i=1}^{N}\frac{e^{-a_i^2/2}}{\sqrt{2\pi}}\,\dd a_i,
 \label{eq:forward-free-weights}\\
 &\dd Q_N(\eta)
 =\prod_{i=1}^{N}
 \left[
 q_\rho(p_i,k_i)\,\dd^d p_i\,\dd k_i\,
 \frac{\dd b_i}{2\pi}
 \right],
 \qquad \eta=(\eta_1,\ldots,\eta_N).
 \label{eq:forward-hidden-law}
\end{align}
We write $\EE_{0,a}$ for expectation with respect to $\mu_0$ at fixed
$\eta$, and $\EE_\eta$ for expectation with respect to $Q_N$.

At fixed hidden features, the field is a linear function of the Gaussian
output weights and therefore has conditional covariance
\begin{align}
 C_\eta(X,Y)
 &=\frac{1}{N}\sum_{i=1}^{N}
 f_{\eta_i}(X)f_{\eta_i}(Y).
 \label{eq:forward-conditional-cov}
\end{align}
For a test source $J$, Gaussian integration gives
\begin{align}
 &\EE_{0,a}\left[
 \exp\left(i\int\dV_X\,J(X)\phi_N(X)\right)
 \biggm|\eta\right]\nonumber\\
 &=\exp\left[
 -\frac12\int\dV_X\dV_Y\,
 J(X)C_\eta(X,Y)J(Y)
 \right].
 \label{eq:pushfree}
\end{align}

The covariance of the neural field at fixed hidden parameters
depends on the particular realization of the sampled features.
Nevertheless, the sampling distribution was constructed so that
its ensemble average reproduces the regulated massive AdS
Green function exactly. Since the hidden parameters are sampled
independently, fluctuations of the empirical covariance around
this target are suppressed as the network width increases.

More precisely, at fixed spectral regulator, and assuming
finite second moments of the feature products, we have
\begin{align}
 \EE_{\eta_1,\ldots,\eta_N}
 \left[
 C_\eta(X,Y)
 \right]
 &=
 G_{\Delta,\rho}(X,Y),
 \label{eq:forward-cov-average}\\
 \EE_{\eta_1,\ldots,\eta_N}
 \left[
 \left|
 C_\eta(X,Y)-G_{\Delta,\rho}(X,Y)
 \right|^2
 \right]
 &=
 \frac{1}{N}
 \Var_\eta\!\left[
 f_\eta(X)f_\eta(Y)
 \right]
 \longrightarrow0,
 \label{eq:forward-cov-mean-square}
\end{align}
as $N\to\infty$. The first identity shows that the
empirical covariance is an unbiased representation of
the regulated propagator at every finite width, while
the second establishes its mean-square convergence
to the target covariance for fixed bulk points $X$ and
$Y$. In particular, the mean-square deviation decreases
as $N^{-1}$, reflecting the suppression of statistical
fluctuations through independent feature sampling.

Thus, although individual finite-width realizations
generally define Gaussian fields with different
conditional covariances, these fluctuations vanish
in the infinite-width limit, and the covariance
approaches the prescribed massive AdS propagator.
This statistical convergence distinguishes the
random-feature construction from deterministic
spectral representations, where the feature functions
and Gaussian coefficient variances can instead be
chosen to reproduce a prescribed finite-dimensional
Green matrix exactly at finite resolution.

We now introduce the interaction at fixed hidden features. Define the
conditional partition function and the normalized output-weight measure by
\begin{align}
 \cZ_\lambda(\eta)
 &=\int_{\RR^N}\dd\mu_0(a)\,
 \exp\left[-S_{\mathrm{int}}[\phi_N(a,\eta)]\right],
 \label{eq:forward-conditional-partition}\\
 \dd\cP_\lambda(a\mid\eta)
 &=\frac{1}{\cZ_\lambda(\eta)}
 \exp\left[-S_{\mathrm{int}}[\phi_N(a,\eta)]\right]
 \dd\mu_0(a).
 \label{eq:pullbacklaw}
\end{align}
The full interacting neural ensemble is defined by the quenched joint law
\begin{align}
 \dd\cP_\lambda^{\mathrm q}(a,\eta)
 &=\dd Q_N(\eta)\,\dd\cP_\lambda(a\mid\eta)
 \notag\\
 &=\dd Q_N(\eta)\,
 \frac{\exp\left[-S_{\mathrm{int}}[\phi_N(a,\eta)]\right]}
 {\cZ_\lambda(\eta)}\,\dd\mu_0(a).
 \label{eq:quenched-joint-law}
\end{align}
The normalization is performed separately for every feature realization.
Consequently, integrating out the output weights returns the original
hidden-feature law,
\begin{align}
 \int_{\RR^N}\dd\cP_\lambda^{\mathrm q}(a,\eta)
 &=\dd Q_N(\eta)\,
 \frac{1}{\cZ_\lambda(\eta)}
 \int_{\RR^N}\dd\mu_0(a)\,
 e^{-S_{\mathrm{int}}[\phi_N(a,\eta)]}
 \notag\\
 &=\dd Q_N(\eta).
 \label{eq:quenched-hidden-marginal}
\end{align}
Thus the interaction changes the conditional distribution of the output
weights but does not change the marginal distribution of the hidden
features.

For any field observable $\mathcal A[\phi]$, the fixed-feature and quenched
expectations are
\begin{align}
 \EE_\lambda\!\left[\mathcal A[\phi_N]\mid\eta\right]
 &=\frac{
 \displaystyle\int\dd\mu_0(a)\,
 \mathcal A[\phi_N(a,\eta)]
 e^{-S_{\mathrm{int}}[\phi_N(a,\eta)]}}
 {\displaystyle\int\dd\mu_0(a)\,
 e^{-S_{\mathrm{int}}[\phi_N(a,\eta)]}},
 \label{eq:pullbackidentity}\\
 \EE_\lambda^{\mathrm q}\!\left[\mathcal A[\phi_N]\right]
 &=\int\dd Q_N(\eta)\,
 \EE_\lambda\!\left[\mathcal A[\phi_N]\mid\eta\right].
 \label{eq:quenched-observable}
\end{align}
Since the free conditional field law is Gaussian with covariance $C_\eta$,
the first line may equivalently be written as
\begin{align}
 \EE_\lambda\!\left[\mathcal A[\phi_N]\mid\eta\right]
 &=\frac{
 \displaystyle\int\dd\mu_{C_\eta}(\phi)\,
 \mathcal A[\phi]e^{-S_{\mathrm{int}}[\phi]}}
 {\displaystyle\int\dd\mu_{C_\eta}(\phi)\,
 e^{-S_{\mathrm{int}}[\phi]}}.
 \label{eq:forward-measure-identity}
\end{align}

Here $\dd\mu_{C_\eta}$ denotes the Gaussian field measure
induced by the neural map for a fixed realization of the
hidden parameters, with covariance $C_\eta$. The identity
above is exact at finite resolution and establishes a
direct correspondence between the interacting neural
parameter measure and the Gibbs deformation of the
associated Gaussian field theory.

If the feature representation is chosen such that
$C_\eta$ coincides exactly with the prescribed regulated
Green function $G$, the induced interacting measure
reproduces the target scalar field theory at that
regulator. For randomly sampled features, however,
the conditional covariance fluctuates between
realizations. At finite width, the resulting theory
is therefore a quenched mixture of interacting field
measures, each defined by its own covariance $C_\eta$.

Recovering the target interacting theory in the
infinite-width limit consequently requires more than
the convergence of the free covariance alone. One
must also establish that the corresponding normalized
interacting measures converge as
$C_\eta\to G_{\Delta,\rho}$, with sufficient control over
the interaction functional and its normalization.
Under these conditions, the quenched ensemble approaches
the Gibbs deformation of the prescribed regulated
Gaussian measure.

Throughout the paper, interacting
expectations without an explicit conditioning on
the hidden parameters are understood in this
quenched sense: the interacting measure is first
normalized at fixed features, and the resulting
expectation is then averaged over their original
sampling distribution.

Substituting the neural field \eqref{eq:forward-field} into the
quartic interaction gives
\begin{align}
 S_{\mathrm{int}}[\phi_N]
 &=
 \frac{\lambda}{4!}
 \int_\Omega\dV_Y
 \left[
 \frac{1}{\sqrt N}
 \sum_{i=1}^{N}a_i f_{\eta_i}(Y)
 \right]^4
 \nonumber\\
 &=
 \frac{\lambda}{4!N^2}
 \sum_{i,j,k,l=1}^{N}
 a_i a_j a_k a_l
 \int_\Omega\dV_Y\,
 f_{\eta_i}(Y)f_{\eta_j}(Y)
 f_{\eta_k}(Y)f_{\eta_l}(Y).
 \label{eq:forward-quartic-expansion}
\end{align}
Introduce the four-neuron overlap tensor
\begin{equation}
 V_{ijkl}
 \equiv
 \int_\Omega\dV_Y\,
 f_{\eta_i}(Y)f_{\eta_j}(Y)
 f_{\eta_k}(Y)f_{\eta_l}(Y).
 \label{eq:Voverlap}
\end{equation}
It measures the overlap of four neural features over the AdS bulk.
The integration uses the invariant volume element, so the geometry
enters directly into the interaction tensor. The interaction becomes
\begin{equation}
 S_{\mathrm{int}}[\phi_N]
 =
 \frac{\lambda}{4!N^2}
 \sum_{i,j,k,l=1}^{N}
 V_{ijkl}a_i a_j a_k a_l.
 \label{eq:bare-pullback}
\end{equation}
Consequently, at fixed hidden features the interacting output-weight
measure is
\begin{align}
 \dd\cP_\lambda(a\mid\eta)
 &=\frac{1}{\cZ_\lambda(\eta)}
 \exp\left[
 -\frac{\lambda}{4!N^2}
 \sum_{i,j,k,l}V_{ijkl}a_i a_j a_k a_l
 \right]\dd\mu_0(a).
 \label{eq:forward-parameter-density}
\end{align}
This measure correlates the output weights. The hidden parameters remain
independently distributed according to $Q_N$.

This establishes the central result of the direct
action-pullback construction: a local quartic interaction
of the bulk scalar field is mapped into a quartic
coupling among the neural output weights. The resulting
interaction is not introduced through an independently
chosen parameter-space tensor. Instead, its coefficients
are determined by the four-neuron overlap tensor
$V_{ijkl}$, which is fixed entirely by the feature
functions and their integration over the AdS geometry.
The neural interaction therefore inherits its structure
directly from the prescribed local bulk field theory.

The overall factor $N^{-2}$ follows from the
normalization of the neural field, where each neuron
contributes a factor $N^{-1/2}$. Substituting four
such factors into the quartic interaction produces
the corresponding $N^{-2}$ scaling. This normalization
is naturally associated with the dense sum over four
neuron indices and is compatible with a nontrivial
interaction in the large-width limit. Its precise form,
however, depends on the chosen neural parameterization
and should not be interpreted as a universal scaling
law for arbitrary network architectures.

An important consequence of this construction is
the emergence of correlations among neural parameters
that were independent in the free theory. Before
introducing the interaction, the Gaussian probability
measure factorizes over the individual output weights.
The quartic deformation breaks this factorization by
coupling different weights through the overlap tensor
$V_{ijkl}$. Although the original interaction is local
in the bulk field, its representation in neural
parameter space is generally nonlocal and involves
many distinct combinations of neuron indices.

This distinction reflects the extended spatial
structure of the Bessel--cosine features. Since each
feature typically has support throughout the bulk,
the overlap of four such functions is generally
nonvanishing, giving rise to a dense interaction
tensor. Consequently, locality of the field-theoretic
interaction does not translate into \textit{sparsity} of
the neural parameter couplings. Rather, the local
bulk dynamics is encoded collectively through
correlations among the output weights, with the
geometry and spectral properties of the neural
features determining the detailed structure
of these correlations.

\paragraph{Wick ordering and interacting bulk correlators.}
The quartic interaction introduces nontrivial correlations
among the neural output weights, which were statistically
independent in the free Gaussian ensemble. We now examine
how these interaction-induced correlations arise in neural
parameter space and how they are transmitted to the bulk
field through the neural representation. In particular,
we derive the leading connected correlators of the output
weights and show that, when combined with the corresponding
neuron profiles, they reproduce the perturbative correlation
functions of the interacting massive scalar field on AdS.

It is convenient to use a Wick-ordered interaction,
\begin{equation}
 S_{\mathrm{int},N}^{\mathrm{Wick}}
 =
 \frac{\lambda}{4!}
 \int_\Omega \dV_Y\,:\!\phi_N(Y)^4\!:\,_{C_\eta},
 \label{eq:forward-wick-action}
\end{equation}
where $C_\eta$ is the conditional covariance defined in
\eqref{eq:forward-conditional-cov}, and
\begin{equation}
 :\!\phi_N^4\!:\,_{C_\eta}
 =
 \phi_N^4-6C_\eta(Y,Y)\phi_N^2+3C_\eta(Y,Y)^2.
 \label{eq:forward-wick}
\end{equation}

To compare neural ensembles with different hidden-feature
realizations, it is useful to distinguish this prescription from
normal ordering relative to the deterministic target covariance
$G=G_{\Delta,\rho}$. Set
$\delta C_\eta(Y)=C_\eta(Y,Y)-G(Y,Y)$. Directly from the Wick
polynomials,
\begin{equation}
 :\!\phi_N(Y)^4\!:_{C_\eta}
 =:\!\phi_N(Y)^4\!:_{G}
 -6\delta C_\eta(Y):\!\phi_N(Y)^2\!:_{G}
 +3\delta C_\eta(Y)^2.
 \label{eq:conditional-vs-fixed-wick}
\end{equation}
Accordingly, the corresponding interaction actions differ by
\begin{equation}
 S_{\mathrm{int},N}^{\mathrm{Wick}}-S_{\mathrm{int},N}^{G}
 =-\frac{\lambda}{4}\int_\Omega\dV_Y\,
 \delta C_\eta(Y):\!\phi_N(Y)^2\!:_{G}
 +\frac{\lambda}{8}\int_\Omega\dV_Y\,\delta C_\eta(Y)^2.
 \label{eq:wick-action-difference}
\end{equation}
Thus conditional ordering adds a realization-dependent quadratic term
and a constant relative to a fixed-$G$ scheme. These terms disappear
only under appropriate covariance-concentration and integrability
conditions. For $\lambda\geq0$, fixed-$G$ ordering gives a
feature-independent lower bound at fixed regulators, while
\eqref{eq:regulated-feature-bound} also bounds the conditional
covariance for the regulated Bessel features. The present
conditional scheme remains convenient for the explicit parameter
cumulants derived below, but its finite-width law is not identical
to a fixed-$G$ interaction.

This subtraction removes the first-order tadpole contribution and
keeps the free two-point function unchanged through order $\lambda$.
It corresponds to a particular choice of mass renormalization scheme.

Since the neural field is linear in the Gaussian output weights, the
Wick-ordered interaction can be written in parameter space as
\begin{equation}
 S_{\mathrm{int},N}^{\mathrm{Wick}}
 =
 \frac{\lambda}{4!N^2}
 \sum_{i,j,k,l=1}^{N}V_{ijkl}H_{ijkl}(a),
 \label{eq:forward-hermite-action}
\end{equation}
where
\begin{equation}
 H_{ijkl}(a)
 \equiv :\!a_i a_j a_k a_l\!:
\end{equation}
is the fourth multivariate Hermite polynomial with respect to the
independent standard Gaussian weights. The overlap tensor
$V_{ijkl}$ was defined in \eqref{eq:Voverlap}.

The conditional partition function is generally feature dependent. For the
Wick-ordered interaction, conditional normal ordering removes the term
linear in $\lambda$, while Wick's theorem gives
\begin{align}
 \cZ_\lambda(\eta)
 &=\int\dd\mu_0(a)\,
 e^{-S_{\mathrm{int},N}^{\mathrm{Wick}}[a;\eta]}
 \notag\\
 &=1+\frac{\lambda^2}{48}
 \int_\Omega\dV_Y\dV_Z\,C_\eta(Y,Z)^4
 +O(\lambda^3).
 \label{eq:conditional-partition-expansion}
\end{align}
In the quenched construction, this dependence does not alter
the marginal distribution of the hidden features. The factor
$\cZ_\lambda(\eta)^{-1}$ normalizes the interacting output-weight
measure for each fixed realization $\eta$, ensuring that the
feature distribution remains exactly $Q_N$, as shown in
Eq.~\eqref{eq:quenched-hidden-marginal}. Thus, the expansion
above characterizes fluctuations of the conditional partition
function rather than an interaction-induced modification of
the spectral sampling law. These fluctuations are distinct
from the $1/N$ corrections associated with finite-width
random-feature sampling.

To calculate the induced parameter correlations, introduce sources
$j_i$ coupled to the output weights and define
\begin{equation}
 Z_\eta[j]
 =
 \frac{
 \displaystyle\int \dd\mu_0(a)\,
 \exp\left(j\cdot a-S_{\mathrm{int},N}^{\mathrm{Wick}}[a]\right)
 }{
 \displaystyle\int \dd\mu_0(a)\,
 \exp\left(-S_{\mathrm{int},N}^{\mathrm{Wick}}[a]\right)
 }.
 \label{eq:forward-weight-generating}
\end{equation}
The connected parameter correlators are derivatives of
$W_\eta[j]=\log Z_\eta[j]$. Expanding to first order in $\lambda$
and using the Gaussian identity
\begin{equation}
 \EE_0\left[
 H_{ijkl}(a)e^{j\cdot a}
 \right]
 =
 j_i j_j j_k j_l\,e^{j^2/2},
\end{equation}
we obtain
\begin{equation}
 W_\eta[j]
 =
 \frac12\sum_i j_i^2
 -
 \frac{\lambda}{4!N^2}
 \sum_{i,j,k,l}V_{ijkl}j_i j_j j_k j_l
 +O(\lambda^2).
 \label{eq:forward-weight-functional}
\end{equation}
Taking derivatives with respect to the sources yields
the leading connected cumulants of the output weights,
\begin{equation}
 \kappa_2(a_i,a_j\mid\eta)
 =
 \delta_{ij}+O(\lambda^2),
 \qquad
 \kappa_4(a_i,a_j,a_k,a_l\mid\eta)
 =
 -\frac{\lambda}{N^2}V_{ijkl}
 +O(\lambda^2).
 \label{eq:forward-weight-cumulants}
\end{equation}
Thus the quartic interaction generates a nonzero connected
four-point cumulant of the neural output weights. Its magnitude is
determined by the overlap of four neuron features, while its negative
sign agrees with the leading connected fourth cumulant of a
positive-coupling $\phi^4$ theory.

We can now reconstruct the bulk four-point function by substituting
the neural expansion \eqref{eq:forward-field}:
\begin{align}
 G_{\mathrm{c},N}^{(4)}(X_1,\ldots,X_4\mid\eta)
 &=
 \frac{1}{N^2}
 \sum_{i,j,k,l}
 \prod_{r=1}^{4}f_{\eta_{i_r}}(X_r)\,
 \kappa_4(a_i,a_j,a_k,a_l\mid\eta)
 \nonumber\\
 &=
 -\lambda\int_\Omega\dV_Y
 \prod_{r=1}^{4}C_\eta(X_r,Y)
 +O(\lambda^2),
 \label{eq:forward-four-finite}
\end{align}
where $(i_1,i_2,i_3,i_4)=(i,j,k,l)$.
In the second line, the sums over neuron indices have been
recombined into the conditional covariance $C_\eta$.

Equation~\eqref{eq:forward-four-finite} is the connected four-point
function at fixed hidden features. For the full quenched ensemble, the law
of total cumulance gives, for the centered field,
\begin{align}
 G_{\mathrm c,N}^{(4),\mathrm q}(X_1,X_2,X_3,X_4)
 &=\EE_\eta\!\left[
 G_{\mathrm c,N}^{(4)}(X_1,X_2,X_3,X_4\mid\eta)
 \right]
 \notag\\
 &\quad+
 \sum_{(ab\mid cd)}
 \Cov_\eta\!\left(
 G_{\lambda,\eta}^{(2)}(X_a,X_b),
 G_{\lambda,\eta}^{(2)}(X_c,X_d)
 \right),
 \label{eq:quenched-total-fourth-cumulant}
\end{align}
where the sum runs over the three pairings
$(12\mid34)$, $(13\mid24)$, and $(14\mid23)$. In the conditional
Wick-ordering scheme,
$G_{\lambda,\eta}^{(2)}=C_\eta+O(\lambda^2)$, so the covariance terms are
the finite-width random-feature contribution already present in the free
mixture. They are distinct from the order-$\lambda$ contact vertex and
vanish only after the required covariance concentration in the
large-width limit.

Taking the infinite-width limit at fixed regulators gives
$C_\eta\to G_{\Delta,\rho}$, provided the corresponding feature
moments are controlled. After removing the regulators in a
well-defined perturbative prescription, we recover
\begin{equation}
 \left\langle
 \phi(X_1)\phi(X_2)\phi(X_3)\phi(X_4)
 \right\rangle_{\mathrm c}
 =
 -\lambda\int_{\AdS_{d+1}}\dV_Y
 \prod_{r=1}^{4}G_\Delta(X_r,Y)
 +O(\lambda^2).
 \label{eq:forward-bulk-contact}
\end{equation}
This is precisely the leading connected quartic contact diagram
of the interacting massive scalar field. The four external points
are connected to a common bulk vertex $Y$ by the free massive
propagator $G_\Delta$.

The same result follows from the source generating functional
of the interacting field. For the Wick-ordered theory, its
first-order expansion is
\begin{equation}
 W_\lambda[J]
 =
 \frac12(J,G_\Delta J)
 -
 \frac{\lambda}{4!}
 \int_{\AdS_{d+1}}\dV_Y\,
 \bigl[(G_\Delta J)(Y)\bigr]^4
 +O(\lambda^2),
 \label{eq:forward-field-functional}
\end{equation}
where
\begin{equation}
 (G_\Delta J)(Y)
 =
 \int_{\AdS_{d+1}}\dV_X\,G_\Delta(Y,X)J(X).
\end{equation}
Four functional derivatives with respect to $J$ reproduce
\eqref{eq:forward-bulk-contact}.

If we instead use the bare quartic interaction, the first-order
two-point function receives the usual tadpole correction,
\begin{equation}
 \delta G^{(2)}(X_1,X_2)
 =
 -\frac{\lambda}{2}
 \int_{\AdS_{d+1}}\dV_Y\,
 G_\Delta(X_1,Y)G_\Delta(Y,Y)G_\Delta(Y,X_2)
 +O(\lambda^2).
 \label{eq:forward-tadpole}
\end{equation}
The coincident propagator $G_\Delta(Y,Y)$ requires ultraviolet
regularization and an appropriate mass counterterm. Normal ordering
removes this contribution at first order, but it does not eliminate
the need for higher-order renormalization.

At order $\lambda^2$, the same neural measure generates the usual
bulk loop diagrams, including the connected four-point bubble,
\begin{align}
 \delta^{(2)}G_{\mathrm c}^{(4)}(X_1,\ldots,X_4)
 &=
 \frac{\lambda^2}{2}
 \sum_{(ab\mid cd)}
 \int\dV_Y\,\dV_Z\,
 G_\Delta(X_a,Y)G_\Delta(X_b,Y)
 \nonumber\\
 &\qquad\times
 G_\Delta(Y,Z)^2
 G_\Delta(X_c,Z)G_\Delta(X_d,Z)
 +\text{counterterms},
 \label{eq:forward-bubble}
\end{align}
where the sum runs over the three pairings of the external points.
Higher-order diagrams follow from the same interacting probability
measure, with the necessary renormalization conditions imposed.

The calculation shows that the neural representation of the free
propagator does not affect the form of the resulting interaction
diagrams once the Gaussian field law and the pulled-back interaction
are fixed. Different feature families may produce different
finite-width corrections, but their limiting interacting bulk
correlators agree when they represent the same regulated covariance
and field functional. The BDHM prescription can subsequently be
applied to the external bulk points to obtain the corresponding
boundary correlation functions, as discussed in
Section~\ref{sec:bdhm}.

\subsection{Quartic interactions from auxiliary neurons}
\label{sec:aux}

The direct action-pullback construction developed in the
previous subsection introduces the desired quartic interaction
explicitly into the neural parameter distribution. Although
this approach reproduces the interacting scalar field theory
when the underlying regulated Gaussian measure is matched,
the interaction itself is \textit{prescribed} rather than generated
by the neural degrees of freedom. We now explore an
alternative construction in which the quartic interaction
emerges from the non-Gaussian statistics of additional
auxiliary neurons.

The central idea is to extend the Gaussian neural field
by introducing binary auxiliary variables that couple
linearly to local bulk field values. Integrating out these
variables generates an effective ``$\log\cosh$'' potential,
as familiar from Gaussian--Bernoulli restricted Boltzmann
machines; see, for example, Ref.~\cite{AartsLuciniPark2024}.
By including a suitable quadratic compensator, the induced
quadratic contribution is canceled, leaving a positive
quartic term as the leading nontrivial interaction.
With an appropriate scaling of the auxiliary couplings,
a replicated weak-unit limit suppresses higher-order
contributions and yields the desired local
$\lambda\phi^4$ interaction.

The essential distinction from the direct construction
lies in the origin of the interaction. Rather than
prescribing a quartic polynomial or introducing a
four-neuron overlap tensor from the outset, we obtain
the quartic structure and its positive coefficient
from the statistical properties of the auxiliary
variables. The locality of the resulting interaction,
however, remains an architectural input: each auxiliary
neuron couples to a point-local readout of the bulk field.
Thus, the construction generates the nonlinear interaction
statistically, while its spatial structure is determined
by the chosen coupling architecture.

\paragraph{Local auxiliary neurons and the induced interaction.}

We begin with the Gaussian neural field constructed in Section~\ref{sec:free},
which can be written as
\begin{equation}
 \phi(X)=\sum_i F_i(X)a_i,
 \qquad
 a_i\sim\mathcal N(0,1).
 \label{eq:aux-free-field}
\end{equation}
Here $F_i$ denotes a regulated set of neural features, which may be
obtained from the Bessel--cosine construction. At fixed features, the
field is Gaussian with covariance
\begin{equation}
 C(X,Y)=\sum_iF_i(X)F_i(Y).
 \label{eq:aux-free-covariance}
\end{equation}
For a correctly chosen feature representation, this covariance
approximates the regulated massive AdS Green function $G_{\Delta,\rho}$.

When the features $F_i$ are obtained from the randomly sampled
Bessel--cosine construction, all measures in this subsection are understood
at fixed $\eta$. The same quenched convention is used as in
Eq.~\eqref{eq:quenched-joint-law}:
\begin{align}
 \dd\cP_{\mathrm{aux}}^{\mathrm q}(a,\sigma,\eta)
 &=\dd Q_N(\eta)\,
 \dd\cP_{\mathrm{aux}}(a,\sigma\mid\eta),
 \label{eq:aux-quenched-law}\\
 \sum_{\{\sigma\}}\int_{\RR^N}
 \dd\cP_{\mathrm{aux}}^{\mathrm q}(a,\sigma,\eta)
 &=\dd Q_N(\eta).
 \label{eq:aux-quenched-marginal}
\end{align}
For a deterministic feature representation the conditioning label may, of
course, be omitted.

Divide the regulated bulk region $\Omega$ into cells $C_A$ with
reference points $Y_A$ and invariant volumes $v_A>0$. At each
cell define a local scalar readout
\begin{equation}
 S_A(a)=\phi(Y_A)=\sum_iF_{Ai}a_i,
 \qquad F_{Ai}=F_i(Y_A).
 \label{eq:local-readout}
\end{equation}
The covariance between two readouts is
\begin{equation}
 C_{AB}=\EE_0[S_AS_B]
 =\sum_iF_{Ai}F_{Bi}=C(Y_A,Y_B).
 \label{eq:aux-readout-cov}
\end{equation}
We now introduce one \textit{auxiliary binary neuron} at each bulk cell,
\begin{equation}
 \sigma_A\in\{-1,+1\},
\end{equation}
with equal reference probabilities for the two values.
Its interaction with the local field is controlled by a real
coupling $g_A$. Conditional on the fixed features, we define the joint law of the Gaussian
coefficients and the auxiliary binary variables by
\begin{align}
 \dd\cP_{\mathrm{aux}}(a,\sigma\mid\eta)
 &=\frac{1}{\cZ_{\mathrm{aux}}(\eta)}
 \prod_A\left[
 \frac12\exp\left(
 g_A\sigma_A S_A(a)
 -\frac{g_A^2}{2}S_A(a)^2
 \right)
 \right]\dd\mu_0(a).
 \label{eq:joint-binary}
\end{align}
The first term in the exponent is a linear coupling between a binary
neuron and the scalar readout. The second is a quadratic compensator
whose purpose will become clear after the binary variable is summed
over. Here $\cZ_{\mathrm{aux}}(\eta)$ is the conditional partition function obtained by summing over all binary configurations and integrating over the Gaussian coefficients, and is chosen so that the joint measure at fixed $\eta$ is normalized.

Since each $\sigma_A$ takes only two values, it can be integrated
out exactly
\begin{align}
 \frac12\sum_{\sigma_A=\pm1}
 \exp\left(
 g_A\sigma_AS_A-\frac{g_A^2}{2}S_A^2
 \right)
 &=
 e^{-g_A^2S_A^2/2}\cosh(g_AS_A)=
 e^{-W(g_AS_A)},
 \label{eq:aux-spin-sum}
\end{align}
where we have defined the effective potential
\begin{equation}
 W(t)\equiv \frac{t^2}{2}-\log\cosh t.
 \label{eq:Wpotential}
\end{equation}
Consequently, after summing the binary variables, the conditional measure
of the Gaussian coefficients becomes
\begin{align}
 \dd\cP_{\mathrm{aux}}(a\mid\eta)
 &=\frac{1}{\cZ_{\mathrm{aux}}(\eta)}
 \exp\left[-\sum_AW(g_AS_A(a))\right]\dd\mu_0(a).
 \label{eq:auxmarginal}
\end{align}
The interaction between output coefficients is generated by the auxiliary
variables, whereas the hidden-feature law remains $Q_N$.
The interaction between neural parameters has therefore appeared
automatically after the auxiliary variables have been eliminated.

To identify its leading form, expand the effective potential around
$t=0$
\begin{equation}
 \log\cosh t
 =
 \frac{t^2}{2}
 -\frac{t^4}{12}
 +\frac{t^6}{45}
 -\frac{17t^8}{2520}
 +O(t^{10}),
\end{equation}
so that
\begin{equation}
 W(t)
 =
 \frac{t^4}{12}
 -\frac{t^6}{45}
 +\frac{17t^8}{2520}
 +O(t^{10}).
 \label{eq:Wseries}
\end{equation}
The quadratic term cancels exactly because of the compensator in
\eqref{eq:joint-binary}. The first surviving term is quartic and
has a positive coefficient. Its origin can also be understood from
the fourth cumulant of the symmetric binary distribution,
\begin{equation}
 \kappa_4(\sigma)
 =
 \EE[\sigma^4]-3\EE[\sigma^2]^2
 =-2.
 \label{eq:aux-fourth-cumulant}
\end{equation}
The negative fourth cumulant of the auxiliary variable produces a
positive quartic contribution to the effective scalar action.

The complete potential is stable, not merely its leading
quartic approximation. Indeed,
\begin{equation}
 W(0)=W'(0)=0,
 \qquad
 W''(t)=\tanh^2t\geq0.
\end{equation}
Thus, $W$ is convex and
\begin{equation}
 W(t)\geq0
 \label{eq:Wconvex}
\end{equation}
for all real $t$. Thus the \emph{bare compensated} auxiliary-neuron probability law
\eqref{eq:auxmarginal} is normalizable at every fixed
finite-dimensional Gaussian regulator, even though the truncated
Taylor expansion contains a negative sextic term. This statement uses
only $W(t)\geq0$ and does not include a subsequently added negative
quadratic mass counterterm. At finite auxiliary population, a
Wick-subtracted family requires the separate large-field stability condition
\eqref{eq:aux-wick-positive-definite}, derived in Appendix~\ref{sec:aux-wick-stability}.

We first state the cell-refinement limit at fixed ultraviolet regulator.
Let $\mathcal P_n=\{C_A^{(n)}\}$ be partitions of the compact regulated
region $\Omega$, with invariant volumes $v_A^{(n)}$, reference points
$Y_A^{(n)}$, and
\begin{align}
 \delta_n
 &=
 \max_A\operatorname{diam}_g\!\left(C_A^{(n)}\right)
 \longrightarrow0,
 \qquad
 v_{\max}^{(n)}
 =
 \max_Av_A^{(n)}
 \longrightarrow0.
 \label{eq:cell-refinement-assumptions}
\end{align}
At fixed finite-dimensional field regulator, $\phi(a,\cdot)$ is continuous
on $\overline\Omega$. Therefore
\begin{align}
 \sum_Av_A^{(n)}
 \phi\!\left(a,Y_A^{(n)}\right)^4
 &\longrightarrow
 \int_\Omega\dV_Y\,\phi(a,Y)^4,
 \label{eq:quartic-riemann-limit}\\
 \sum_A
 \left(v_A^{(n)}\right)^{3/2}
 \left|
 \phi\!\left(a,Y_A^{(n)}\right)
 \right|^6
 &\leq
 \operatorname{Vol}(\Omega)
 \left(v_{\max}^{(n)}\right)^{1/2}
 \|\phi(a,\cdot)\|_\infty^6
 \longrightarrow0.
 \label{eq:cell-remainder-limit}
\end{align}
With
$g_A^{(n)}=\gamma\left(v_A^{(n)}\right)^{1/4}$, the expansion of
$W$ consequently gives
\begin{align}
 \sum_A
 W\!\left(
 g_A^{(n)}
 \phi\!\left(a,Y_A^{(n)}\right)
 \right)
 &\longrightarrow
 \frac{\gamma^4}{12}
 \int_\Omega\dV_Y\,\phi(a,Y)^4.
 \label{eq:cell-weak}
\end{align}
Comparing with the conventional interaction
\begin{equation}
 S_{\mathrm{int}}[\phi]
 =
 \frac{\lambda}{4!}\int_\Omega\dV_Y\,\phi(Y)^4
\end{equation}
gives
\begin{equation}
 \lambda=2\gamma^4.
 \label{eq:lambdagamma}
\end{equation}
A more controlled construction keeps the field regulator and the
bulk cells fixed and introduces $K$ auxiliary binary neurons per
cell. Their couplings are chosen as
\begin{equation}
 g_{A,r}=\gamma v_A^{1/4}K^{-1/4},
 \qquad r=1,\ldots,K.
 \label{eq:aux-replicated-coupling}
\end{equation}
Since all $K$ neurons couple to the same local readout, integrating
them out produces
\begin{equation}
 V_K(a)
 =
 \sum_AK\,W\left(
 \gamma v_A^{1/4}K^{-1/4}S_A(a)
 \right).
 \label{eq:Kaux}
\end{equation}
Using the expansion \eqref{eq:Wseries}, we obtain
\begin{align}
 V_K(a)
 &=
 \frac{\gamma^4}{12}\sum_Av_AS_A^4
 -\frac{\gamma^6}{45\sqrt K}
 \sum_Av_A^{3/2}S_A^6
 +O(K^{-1})\longrightarrow
 V_4(a)
 =
 \frac{\lambda}{4!}\sum_Av_AS_A^4,
 \label{eq:Klimit}
\end{align}
for $K\to\infty.$ This limit is taken with the features, bulk cells, and finite-dimensional
field regulator fixed. For every fixed coefficient vector $a$,
Eq.~\eqref{eq:Klimit} gives
\begin{align}
 V_K(a)
 &\longrightarrow
 V_4(a).
 \label{eq:K-pointwise-limit}
\end{align}
Since $W(t)\geq0$,
\begin{align}
 0
 \leq
 e^{-V_K(a)}
 \leq1.
 \label{eq:K-domination}
\end{align}
Writing
\begin{align}
 \cZ_K
 &=
 \int\dd\mu_0(a)\,e^{-V_K(a)},
 \qquad
 \cZ_4
 =
 \int\dd\mu_0(a)\,e^{-V_4(a)},
 \label{eq:K-partition-functions}
\end{align}
dominated convergence gives
\begin{align}
 \cZ_K
 &\longrightarrow
 \cZ_4,\\
 \frac{
 \displaystyle
 \int\dd\mu_0(a)\,
 \mathcal A(a)e^{-V_K(a)}
 }{
 \cZ_K
 }
 &\longrightarrow
 \frac{
 \displaystyle
 \int\dd\mu_0(a)\,
 \mathcal A(a)e^{-V_4(a)}
 }{
 \cZ_4
 }
 \label{eq:K-measure-limit}
\end{align}
for every polynomial observable $\mathcal A$. Thus the normalized
auxiliary-neuron measure converges to the regulated positive quartic Gibbs
measure. This statement does not remove the field or ultraviolet
regulator.

The convergence argument above applies to the bare
finite-$K$ auxiliary measure. Imposing Wick subtraction at
finite $K$ requires an additional stability condition,
since the exact auxiliary potential grows quadratically
rather than quartically at large field amplitudes.
Normalizability holds when the quadratic matrix $Q_K$
is positive definite; a sufficient condition is
$K\geq K_{\mathrm{stab}}$.
The derivation and the cutoff dependence of this bound
are discussed in Appendix~\ref{sec:aux-wick-stability}.

\paragraph{Derivation of the interacting four-point function.}
We now demonstrate that the auxiliary-neuron construction
reproduces the leading connected four-point function of
the massive scalar theory with a local $\lambda\phi^4$
interaction.
After integrating out the binary neurons, the interacting neural
measure is
\begin{equation}
 \dd\mu_K(a)
 =
 \frac{1}{\cZ_K}
 e^{-V_K(a)}\dd\mu_0(a),
 \label{eq:aux-interacting-measure}
\end{equation}
where $\dd\mu_0$ is the free Gaussian parameter measure and
$V_K$ is defined in \eqref{eq:Kaux}. In the limit $K\to\infty$,
the interaction becomes
\begin{equation}
 V_4[\phi]
 =
 \frac{\lambda}{4!}\sum_Av_A\phi(Y_A)^4.
 \label{eq:aux-quartic-limit}
\end{equation}
Introduce a source $J(X)$ coupled to the scalar field and define
the normalized generating functional
\begin{equation}
 Z_\lambda[J]
 =
 \frac{
 \displaystyle
 \EE_0\left[
 \exp\left(
 \int\dV_X\,J(X)\phi(X)-V_4[\phi]
 \right)
 \right]
 }{
 \displaystyle
 \EE_0\left[e^{-V_4[\phi]}\right]
 }.
 \label{eq:aux-generating}
\end{equation}
The connected correlation functions are obtained from
\begin{equation}
 W_\lambda[J]=\log Z_\lambda[J].
 \label{eq:aux-connected-functional}
\end{equation}
We first expand the generating functional to first order in
$\lambda$. For the Gaussian reference field,
\begin{equation}
 \EE_0\left[
 e^{(J,\phi)}
 \right]
 =
 \exp\left[
 \frac12(J,CJ)
 \right],
 \label{eq:aux-gaussian-source}
\end{equation}
where
\begin{equation}
 (J,\phi)\equiv \int\dV_X\,J(X)\phi(X),
 \qquad
 (J,CJ)\equiv \int\dV_X\dV_Y\,J(X)C(X,Y)J(Y).
\end{equation}
Expanding the exponent, 
we find
\begin{equation}
 Z_\lambda[J]
 =
 \frac{
 \EE_0[e^{(J,\phi)}]
 -
 \frac{\lambda}{4!}\sum_Av_A
 \EE_0[\phi(Y_A)^4e^{(J,\phi)}]
 +O(\lambda^2)
 }{
 1-\frac{\lambda}{4!}\sum_Av_A
 \EE_0[\phi(Y_A)^4]+O(\lambda^2)
 }.
 \label{eq:aux-generating-expansion}
\end{equation}
The Gaussian expectation containing four fields can be evaluated
using Wick's theorem. It is useful to introduce
\begin{equation}
 m_A[J]\equiv (CJ)(Y_A)
 =\int\dV_X\,C(Y_A,X)J(X).
 \label{eq:aux-source-shift}
\end{equation}
Under the Gaussian measure weighted by $e^{(J,\phi)}$, the field
$\phi(Y_A)$ has mean $m_A[J]$ and variance $C_{AA}$.
Its fourth moment is therefore
\begin{equation}
 \EE_{0,J}[\phi(Y_A)^4]
 =
 m_A[J]^4+6C_{AA}m_A[J]^2+3C_{AA}^2.
 \label{eq:aux-shifted-fourth}
\end{equation}
Equivalently,
\begin{align}
 \EE_0[\phi(Y_A)^4e^{(J,\phi)}]
 &=
 e^{\frac12(J,CJ)}
 \left[
 m_A[J]^4
 +6C_{AA}m_A[J]^2
 +3C_{AA}^2
 \right].
 \label{eq:aux-gaussian-fourth}
\end{align}
Substituting this result into \eqref{eq:aux-generating-expansion}
and expanding the denominator, the source-independent terms
proportional to $3C_{AA}^2$ cancel. Taking the logarithm gives
\begin{equation}
 W_\lambda[J]
 =
 \frac12(J,CJ)
 -
 \frac{\lambda}{4!}\sum_Av_A
 \left[
 m_A[J]^4+6C_{AA}m_A[J]^2
 \right]
 +O(\lambda^2).
 \label{eq:aux-Wtree}
\end{equation}
The first term describes the free Gaussian field. The term
quadratic in $m_A[J]$ produces the first-order correction to the
two-point function, while the fourth power generates the
connected four-point function.

To obtain the latter, take four functional derivatives:
\begin{equation}
 G_{\mathrm c}^{(4)}(X_1,X_2,X_3,X_4)
 =
 \left.
 \frac{\delta^4W_\lambda[J]}
 {\delta J(X_1)\delta J(X_2)
  \delta J(X_3)\delta J(X_4)}
 \right|_{J=0}.
 \label{eq:aux-fourpoint-definition}
\end{equation}
Since $m_A[J]$ is linear in the source,
\begin{equation}
 \frac{\delta m_A[J]}{\delta J(X)}
 =
 C(Y_A,X),
 \label{eq:aux-source-derivative}
\end{equation}
and therefore
\begin{align}
 \left.
 \frac{\delta^4m_A[J]^4}
 {\delta J(X_1)\delta J(X_2)
  \delta J(X_3)\delta J(X_4)}
 \right|_{J=0}
 =
 4!\prod_{r=1}^{4}C(Y_A,X_r).
 \label{eq:aux-four-derivatives}
\end{align}
The terms containing at most two powers of $J$ do not
contribute to the fourth derivative. Thus, the connected four-point function is
\begin{equation}
 G_{\mathrm c}^{(4)}(X_1,X_2,X_3,X_4)
 =
 -\lambda\sum_Av_A
 \prod_{r=1}^{4}C(X_r,Y_A)
 +O(\lambda^2).
 \label{eq:aux-fourpoint-finite}
\end{equation}
This expression has a simple interpretation. Each of the four
external points is connected by a free propagator to the same
bulk cell $Y_A$. The sum over cells represents the position
of the local interaction vertex. The factor $4!$ produced
by differentiating the generating functional cancels the
factor $1/4!$ in the interaction action.

We now consider the relevant limits, beginning with the
convergence of the Gaussian neural covariance to the
regulated massive AdS Green function,
\begin{equation}
 C(X,Y)\longrightarrow G_{\Delta,\rho}(X,Y).
\end{equation}
Then, as the bulk quadrature is refined,
\begin{equation}
 \sum_Av_AF(Y_A)\longrightarrow
 \int_\Omega\dV_Y\,F(Y).
\end{equation}
Consequently,
\begin{equation}
 G_{\mathrm c}^{(4)}(X_1,X_2,X_3,X_4)
 =
 -\lambda\int_\Omega\dV_Y\,
 \prod_{r=1}^{4}G_{\Delta,\rho}(X_r,Y)
 +O(\lambda^2).
 \label{eq:aux-fourpoint-regulated}
\end{equation}
After the controlled removal of the spectral regulator and
any necessary renormalization, the leading result becomes
\begin{equation}
 \left\langle
 \phi(X_1)\phi(X_2)\phi(X_3)\phi(X_4)
 \right\rangle_{\mathrm c}
 =
 -\lambda\int_{\AdS_{d+1}}\dV_Y\,
 \prod_{r=1}^{4}G_\Delta(X_r,Y)
 +O(\lambda^2).
 \label{eq:aux-bulk-contact}
\end{equation}
This is exactly the leading connected quartic contact diagram
obtained in the circular forward construction. The two approaches
therefore produce the same \emph{leading connected bulk four-point
function} when their Gaussian covariances and quartic couplings are
matched. This does not imply equality of their unrenormalized
finite-width measures, because the forward cumulant calculation uses
conditional Wick ordering while the auxiliary construction produces
the bare quartic interaction. For a comparison of complete measures,
the normal-ordering convention, mass counterterms, and conditional
versus joint normalization must also be aligned.

The difference lies in the origin of the interaction. In the
forward construction, the quartic term is inserted directly
into the probability distribution of the original neural
parameters. In the auxiliary-neuron construction, it emerges
after summing elementary binary variables and taking the
weak-unit limit.

For completeness, the quadratic term in
\eqref{eq:aux-Wtree} gives the familiar first-order tadpole
correction,
\begin{equation}
 \delta G^{(2)}(X_1,X_2)
 =
 -\frac{\lambda}{2}
 \int_{\AdS_{d+1}}\dV_Y\,
 G_\Delta(X_1,Y)G_\Delta(Y,Y)G_\Delta(Y,X_2)
 +O(\lambda^2).
 \label{eq:aux-tadpole}
\end{equation}
This correction is present because the auxiliary construction
generates the ordinary, un-normal-ordered quartic interaction. To cancel the first-order tadpole correction to the two-point
function, one may choose a regulated covariance $\mathcal C$ and use
the Wick polynomial
\begin{equation}
 :\!\phi(Y)^4\!:\,_{\mathcal C}
 =
 \phi(Y)^4
 -6\mathcal C(Y,Y)\phi(Y)^2
 +3\mathcal C(Y,Y)^2.
 \label{eq:auxwick}
\end{equation}
For the exact finite-$K$ auxiliary family, adding the corresponding
negative quadratic counterterm nonperturbatively requires the
stability condition
\eqref{eq:aux-wick-positive-definite}. Alternatively, one may keep the
finite-$K$ auxiliary measure bare and impose the Wick subtraction only
after taking the $K\to\infty$ quartic limit. Under either prescription,
the subtraction removes the order-$\lambda$ tadpole but does not
change the leading connected four-point function
\eqref{eq:aux-bulk-contact}. It does not remove the need for
higher-order mass and coupling renormalization.

The result demonstrates that elementary local auxiliary neurons
can produce the same leading perturbative bulk contact interaction as
a directly prescribed quartic action. When the complete free Gaussian field law, interaction functional,
renormalization scheme, and quenched feature-averaging prescription are
matched, the resulting regulated interacting observables are independent
of the particular covariance factorization
\eqref{eq:representation-interacting-invariance}. At finite random-feature
width, equal ensemble-averaged covariance alone is insufficient.
The resulting bulk correlators can therefore be related to
boundary observables through the BDHM prescription developed
in Section~\ref{sec:bdhm}. The remaining challenge is to
control the ultraviolet and boundary limits so that this
agreement extends beyond the regulated theory.

\paragraph{Comparison of neural interaction mechanisms and finite-width effects.}

The neural realization of an interacting field theory
is not unique, as different choices of feature functions
and parameter distributions can represent the same
underlying field measure. To make this freedom explicit,
consider a linear Gaussian representation at fixed
finite-dimensional regulator,
\begin{align}
 \phi
 &=
 Fa,
 \qquad
 a\sim\mathcal N(0,\Sigma_a),
 \qquad
 G
 =
 F\Sigma_aF^T.
 \label{eq:interaction-representation-summary}
\end{align}
Here $F$ denotes the feature map, $\Sigma_a$ is the
covariance matrix of the Gaussian coefficients, and
$G$ is the resulting field covariance.

Different feature maps and coefficient covariances may therefore induce
the same Gaussian field measure. Once the complete Gaussian measure
$\dd\mu_G$, the interaction functional $V[\phi]$, and the renormalization
prescription have been fixed, normalized field observables are independent
of this factorization:
\begin{align}
 \left\langle\mathcal A\right\rangle_V
 &=
 \frac{
 \displaystyle
 \int\dd\mu_G(\phi)\,
 \mathcal A[\phi]e^{-V[\phi]}
 }{
 \displaystyle
 \int\dd\mu_G(\phi)\,
 e^{-V[\phi]}
 }.
 \label{eq:interaction-universality-summary}
\end{align}
This is the precise sense in which the direct and auxiliary
constructions become equivalent after their regulated measures have been
matched. It does not imply equivalence of arbitrary finite-width neural
ensembles that happen to have the same ensemble-averaged two-point
function.

This qualification is essential because non-Gaussianity by itself does
not determine the interaction. For a Gaussian mixture generated by a
random conditional covariance, the fourth cumulant of a smeared field
$\Phi_J$ has the form
\begin{align}
 \kappa_4^{\mathrm{mix}}(\Phi_J)
 &=
 3\Var_\eta\!\left(V_\eta[J]\right)
 \geq0,
 \label{eq:interaction-mixture-sign}
\end{align}
where $V_\eta[J]$ is the conditional variance. By contrast, a positive
local quartic interaction gives, to first order,
\begin{align}
 \kappa_4^{\lambda\phi^4}(\Phi_J)
 &=
 -\lambda
 \int_\Omega\dV_Y\,
 \bigl[(GJ)(Y)\bigr]^4
 +
 O(\lambda^2)
 \leq0.
 \label{eq:interaction-quartic-sign}
\end{align}
Finite-width random-feature corrections must therefore be kept separate
from the prescribed or auxiliary-generated quartic vertex. This
comparison also gives a useful estimate of how large the neural width
must be before the interaction-induced non-Gaussianity dominates over
the non-Gaussianity already present from random feature sampling.

Indeed, for the smeared observable $\Phi_J$ defined in
\eqref{eq:smeared-feature}, the free finite-width contribution is
\begin{equation}
 \kappa_4^{\mathrm{width}}(\Phi_J)
 =
 \frac{3}{N}
 \Var_\eta\!\left(F_\eta[J]^2\right),
\end{equation}
whereas the desired positive-coupling quartic interaction gives, at
leading order,
\begin{equation}
 \kappa_4^{\mathrm{int}}(\Phi_J)
 =
 -\lambda
 \int_\Omega\dV_Y\,[(GJ)(Y)]^4
 +O(\lambda^2).
\end{equation}
Thus, at fixed regulators and to leading order in both the finite-width
and interaction corrections, their combined behavior is schematically
\begin{equation}
 \kappa_4(\Phi_J)
 =
 \frac{3}{N}
 \Var_\eta\!\left(F_\eta[J]^2\right)
 -
 \lambda
 \int_\Omega\dV_Y\,[(GJ)(Y)]^4
 +O\!\left(\frac{\lambda}{N},\lambda^2\right).
 \label{eq:width-interaction-competition}
\end{equation}

For a probe $J$ with $\lambda>0$ and for which
$\int_\Omega\dV_Y\,[(GJ)(Y)]^4>0$, it is therefore natural to define
the crossover width
\begin{equation}
 N_{\mathrm{cross}}(J)
 :=
 \frac{
 3\,\Var_\eta\!\left(F_\eta[J]^2\right)
 }{
 \lambda
 \displaystyle\int_\Omega\dV_Y\,[(GJ)(Y)]^4
 }.
 \label{eq:finite-width-crossover}
\end{equation}
When $N\gg N_{\mathrm{cross}}(J)$, the unwanted fourth cumulant caused
by finite random-feature width is much smaller than the leading
quartic-interaction signal for this observable. For
$N\sim N_{\mathrm{cross}}(J)$ the two effects are comparable and can
partially cancel because they have opposite signs, while for
$N\ll N_{\mathrm{cross}}(J)$ the apparent non-Gaussianity may be
dominated by finite-width sampling rather than by the intended
interaction. The quantity $N_{\mathrm{cross}}(J)$ depends on the probe
$J$ and on the fixed regulators, so it is a practical leading-order
diagnostic rather than a universal width bound.
It compares only the leading fourth-cumulant contributions; it is
not a bound ensuring convergence of the full interacting law. The
$O(\lambda/N,\lambda^2)$ remainder presumes sufficient regulated
moments and controlled small-$\lambda$, large-$N$ expansions. If the
numerator vanishes (as for an exact deterministic covariance), the
finite-width mixture contribution is absent.

A more general diagnostic is to amputate the connected four-point
function and test whether the resulting vertex approaches a point-local
kernel. Amputation should first be defined on a regulated field
space where the relevant covariance matrix is invertible (or on its
specified nondegenerate case).

Further microscopic mechanisms for generating scalar
interactions are explored in
Appendix~\ref{app:alternative-interactions},
including critical binary populations, generalized
auxiliary-neuron distributions, and nonlocal
interaction architectures.
 In
the critical binary-population construction, the scalar field itself is
a collective magnetization,
\begin{align}
 x_A
 &=
 \frac{M_A}{n^{3/4}},
\end{align}
and the quartic potential arises from the entropy of a critically tuned
Ising-type population
\cite{GriffithsSimon,EllisNewman1978}. This construction does not derive
the AdS geometry or kinetic operator: the quadratic form and the critical
tuning remain prescribed. More general auxiliary probability laws can
generate other even interactions, while smeared or extended field
readouts produce finite-range, derivative, or nonlocal kernels. In this
broader class, the order of the interaction is controlled by auxiliary
cumulants, whereas its spatial locality is controlled separately by the
readout architecture. Correlated auxiliary units must additionally
satisfy positivity of the complete induced quartic form; negativity of
each individual fourth cumulant is not sufficient.

There is a second, independent non-uniqueness at the level of the bulk
feature representation. Distinct feature architectures may reproduce a
small set of boundary observables without defining the same bulk theory.
In particular, replacing the coherent Bessel activation by independently
sampled elementary cosines can preserve the leading BDHM boundary
two-point function and contact diagram while changing the bulk covariance
and introducing additional massive components. Agreement of a few
boundary correlators is therefore weaker than equality of the regulated
bulk field measures. The main results of this work use the
Fourier--Bessel covariance together with the direct pullback or the
compensated auxiliary construction; the detailed alternative
factorizations, collective-spin limits, stability conditions, and
nonlocal generalizations are retained in
Appendices~\ref{app:alternative-neurons}
and~\ref{app:alternative-interactions}.

\section{Boundary observables from neural field theory}
\label{sec:boundary}
\label{sec:bdhm}

The AdS/CFT correspondence establishes a relation between
a gravitational or quantum field theory in $(d+1)$-dimensional
anti-de Sitter space and a conformal field theory defined
on its $d$-dimensional boundary
\cite{Maldacena,GKP,Witten}. A central aspect of this
correspondence is the identification of boundary correlation
functions with quantities computed from the bulk theory,
providing a holographic dictionary between bulk fields
and boundary operators.

In the Gubser--Klebanov--Polyakov--Witten (GKPW)
prescription, this relation is formulated as an equality
between the generating functional of boundary correlation
functions and the renormalized bulk partition function
evaluated with prescribed boundary sources:
\begin{equation}
 Z_{\mathrm{CFT}}[J]
 =
 \left\langle
 \exp\left(\int \dd^d x\,J(x)\cO(x)\right)
 \right\rangle_{\mathrm{CFT}}
 =
 \frac{Z_{\mathrm{bulk}}^{\mathrm{ren}}[\phi_{(0)}=J]}
 {Z_{\mathrm{bulk}}^{\mathrm{ren}}[\phi_{(0)}=0]}.
 \label{eq:AdSCFT-partition}
\end{equation}
Here $J(x)$ is the source coupled to a boundary operator
$\cO(x)$, while $\phi_{(0)}(x)$ denotes the corresponding
source coefficient in the near-boundary expansion of the
bulk field. For a scalar field in standard quantization,
this coefficient appears in the asymptotic behavior
$\phi(z,x)\sim z^{d-\Delta}J(x)$ as $z\to0$,
where $\Delta$ is the conformal dimension of the
associated boundary operator.

In the semiclassical bulk approximation, the renormalized
partition function is determined by the on-shell action,
$Z_{\mathrm{bulk}}^{\mathrm{ren}}[J]
\sim e^{-S_{\mathrm{bulk,ren}}^{\mathrm{on-shell}}[J]}$
\cite{Skenderis2002}. Boundary correlation functions
can then be obtained by taking functional derivatives
with respect to the source $J$. This correspondence
provides a systematic relation between the dynamics
of bulk fields and the correlation functions of their
associated boundary operators.

The holographic interpretation also assigns a natural
role to the radial coordinate of AdS. The radial direction
is associated with the energy scale of the boundary theory,
with the limit $z\to0$ corresponding to its ultraviolet
regime. Thus, the near-boundary behavior of bulk fields
encodes the scaling properties of the corresponding
boundary observables.

For a scalar field of mass $m$, the corresponding boundary operator
has a conformal dimension determined by
\begin{equation}
 m^2L^2=\Delta(\Delta-d).
 \label{eq:holographic-mass}
\end{equation}
As discussed in Section~\ref{sec:geometry}, standard quantization
assigns $\Delta=\Delta_+$, while alternative quantization with
$\Delta=\Delta_-$ is possible in the appropriate mass window.
Thus the scalar mass and the chosen boundary condition determine
the scaling behavior of the dual operator.

Our purpose is to apply this correspondence to the neural field
theories constructed in the previous sections. We have obtained
a neural representation of a free massive scalar field on AdS
and introduced a local quartic interaction through different
neural constructions. We now want to understand how these bulk
neural fields give rise to boundary observables. In particular,
we aim to reproduce the connected four-point function of the
boundary scalar operator from the correlation functions of
the interacting neural network.

There are several related formulations of the holographic dictionary.
The mentioned GKPW prescription
\cite{GKP,Witten} identifies the generating functional of boundary
correlators with the renormalized bulk partition function evaluated
with prescribed boundary sources. At the classical level, this
amounts to differentiating the renormalized on-shell bulk action
with respect to the boundary value of the field.

The Banks--Douglas--Horowitz--Martinec (BDHM) prescription
\cite{BDHM} provides an alternative formulation of the
holographic dictionary by identifying boundary operators
directly with the appropriate near-boundary limits of
fluctuating bulk quantum fields. A complementary approach
is the Hamilton--Kabat--Lifschytz--Lowe (HKLL)
reconstruction, which expresses bulk fields in terms of
boundary operators within a suitable perturbative regime
\cite{HKLL2006}. Although these prescriptions describe
closely related aspects of the bulk--boundary
correspondence, they differ in their starting points
and in how the relation between bulk and boundary
observables is implemented. In particular, a consistent
comparison requires careful treatment of the boundary
conditions and normalization conventions for the
corresponding operators.

The BDHM prescription is particularly natural for NN--FT. A neural
ensemble already defines a fluctuating bulk field, and its correlation
functions are obtained by averaging over neural parameters. We can
therefore define a boundary neural operator by taking the appropriate
near-boundary limit of each bulk neuron. The resulting boundary
correlators can then be calculated from the same probability
distribution that defines the bulk neural field.

\subsection{The BDHM dictionary and boundary correlation functions}
\label{sec:bdhm-basics}

The BDHM prescription provides a natural framework for
relating neural field theories (NN--FT) in the
AdS bulk to their corresponding boundary observables.
In the neural formulation, the bulk field is represented
by an ensemble of neurons whose statistical correlations
are determined by the underlying probability distribution
of neural parameters. This structure allows the BDHM
prescription to be implemented directly at the level of
individual neuron profiles, by taking their appropriately
normalized near-boundary limits.

The resulting boundary neural operators are therefore
constructed from the same degrees of freedom that define
the bulk field, rather than introduced as independent
objects. Their correlation functions can subsequently
be obtained by averaging over the original neural
parameter distribution, providing a direct connection
between the statistical structure of the bulk neural
theory and its holographic boundary observables.

For a scalar with conformal dimension $\Delta$, we define the
unnormalized operator
\begin{equation}
 \cO_{\mathrm{raw}}(x)
 =
 \lim_{z\to0}z^{-\Delta}\phi(z,x).
 \label{eq:bdhm-raw}
\end{equation}
This limit is understood at the level of suitably renormalized
correlation functions. In particular, it need not define an ordinary
random variable at an individual boundary point.

More precisely, let $f\in C_c^\infty(\mathbb R^d)$ and, at fixed spectral
cutoff $\Lambda$, define
\begin{align}
 \mathcal O_{\Lambda,z}(f)
 &=
 A_\Delta^{-1/2}
 \int_{\mathbb R^d}\dd^dx\,
 f(x)z^{-\Delta}\phi_\Lambda(z,x).
 \label{eq:smeared-BDHM-field}
\end{align}
We assume that the products of smeared fields entering each regulated
correlator are uniformly integrable for $0<z<z_0$. Since the regulated
features have a pointwise boundary limit, this assumption permits the
limit to pass through the regulated expectations, and we define
\begin{align}
 \mathcal O_\Lambda(f)
 &=
 \lim_{z\to0}\mathcal O_{\Lambda,z}(f)
 \label{eq:regulated-BDHM-limit}
\end{align}
in the sense of moments. The spectral cutoff is removed only after this
fixed-cutoff boundary limit and after subtraction of any required local
contact terms. At mutually separated boundary points, the formulas below
are the ordinary representatives of these correlation distributions.

The normalization of the boundary operator can be fixed from the
free massive bulk propagator. Recall the mixed position-space
Green function obtained in Section~\ref{sec:free}. If
$X=(z,x)$ approaches the boundary while $Y=(w,y)$ remains
in the bulk, then
\begin{equation}
 G_\Delta((z,x),(w,y))
 \underset{z\to0}{\sim}
 A_\Delta z^\Delta
 \left(
 \frac{w}{w^2+|x-y|^2}
 \right)^\Delta,
 \label{eq:bdhm-bulk-asymptotic}
\end{equation}
where
\begin{equation}
 A_\Delta
 =
 \frac{\Gamma(\Delta)}
 {2\pi^{d/2}\Gamma(\Delta-\frac d2+1)}.
 \label{eq:bdhm-A}
\end{equation}
Taking both points to the boundary, for $x\neq y$, gives
\begin{equation}
 \lim_{z,w\to0}
 z^{-\Delta}w^{-\Delta}
 G_\Delta((z,x),(w,y))
 =
 \frac{A_\Delta}{|x-y|^{2\Delta}}.
 \label{eq:bdhm-free-boundary}
\end{equation}
Consequently, the free two-point function of the unnormalized
BDHM operator is
\begin{equation}
 \langle\cO_{\mathrm{raw}}(x)
 \cO_{\mathrm{raw}}(y)\rangle_0
 =
 \frac{A_\Delta}{|x-y|^{2\Delta}}.
 \label{eq:bdhm-raw-two}
\end{equation}
We choose the unit-normalized boundary operator
\begin{equation}
 \cO(x)
 =
 A_\Delta^{-1/2}\cO_{\mathrm{raw}}(x)
 =
 A_\Delta^{-1/2}
 \lim_{z\to0}z^{-\Delta}\phi(z,x).
 \label{eq:bdhmoperator}
\end{equation}
Its free two-point function has the standard conformal form
\begin{equation}
 \langle\cO(x)\cO(y)\rangle_0
 =
 \frac{1}{|x-y|^{2\Delta}}.
 \label{eq:unitboundary}
\end{equation}
An important quantity for the interacting theory is the
bulk-to-boundary propagator. In the BDHM prescription, it is
obtained by taking the boundary limit of one external point
of the bulk Green function:
\begin{equation}
 \cK_\Delta(x;Y)
 =
 A_\Delta^{-1/2}
 \lim_{z\to0}z^{-\Delta}
 G_\Delta((z,x),Y).
 \label{eq:bdhm-K-definition}
\end{equation}
Using \eqref{eq:bdhm-bulk-asymptotic}, we obtain
\begin{equation}
 \cK_\Delta(x;w,y)
 =
 \sqrt{A_\Delta}
 \left(
 \frac{w}{w^2+|x-y|^2}
 \right)^\Delta.
 \label{eq:unitK}
\end{equation}
This kernel connects a boundary insertion at $x$ to a bulk
point $Y=(w,y)$; standard AdS propagator conventions and their
normalizations are reviewed in
Refs.~\cite{FreedmanMathurMatusisRastelli1999,DHokerFreedman2002}.

The GKPW prescription uses a closely related kernel, but its
normalization is chosen to reproduce a prescribed boundary
source. For standard quantization with $\nu>0$, the
source-normalized kernel is
\begin{equation}
 K_{\mathrm{src}}(x;w,y)
 =
 c_\Delta
 \left(
 \frac{w}{w^2+|x-y|^2}
 \right)^\Delta,
 \qquad
 c_\Delta=2\nu A_\Delta.
 \label{eq:gkpw-kernel}
\end{equation}
The source-normalized boundary operator and the raw BDHM
operator are related by a factor $2\nu$ in their nonlocal
correlation functions, with possible local contact terms.
This distinction is important when comparing calculations
performed using different holographic prescriptions.
Alternative quantization and the BF-bound case require
their corresponding boundary prescriptions.

We now consider the interacting bulk scalar field with action
\begin{equation}
 S[\phi]
 =
 S_0[\phi]
 +
 \frac{\lambda}{4!}
 \int_{\AdS_{d+1}}\dV_Y\,\phi(Y)^4
 +
 S_{\mathrm{ct}}[\phi].
 \label{eq:bdhm-interacting-action}
\end{equation}
At first order in $\lambda$, its connected four-point function is
\begin{equation}
 \begin{aligned}
 &\langle
 \phi(X_1)\phi(X_2)\phi(X_3)\phi(X_4)
 \rangle_{\mathrm c}^{(1)} =
 -\lambda\int_{\AdS_{d+1}}\dV_Y
 \prod_{a=1}^{4}G_\Delta(X_a,Y).
 \end{aligned}
 \label{eq:bdhm-bulk-four}
\end{equation}
The four bulk propagators connect the external points to a
single interaction vertex $Y$. Applying the BDHM prescription to all four external fields
gives
\begin{equation}
 \langle
 \cO(x_1)\cO(x_2)\cO(x_3)\cO(x_4)
 \rangle_{\mathrm c}^{(1)}
 =
 -\lambda\int_{\AdS_{d+1}}\dV_Y
 \prod_{a=1}^{4}\cK_\Delta(x_a;Y).
 \label{eq:bdhmcontact}
\end{equation}
This is the leading connected boundary four-point function
generated by the bulk quartic interaction. It is represented
by a contact Witten diagram \cite{Witten,FreedmanMathurMatusisRastelli1999}.

In Poincar\'e coordinates, where
\begin{equation}
 \dV_Y=\frac{\dd w\,\dd^dy}{w^{d+1}},
\end{equation}
the result becomes
\begin{equation}
 \langle
 \cO(x_1)\cdots\cO(x_4)
 \rangle_{\mathrm c}^{(1)}
 =
 -\lambda
 \int_0^\infty\frac{\dd w}{w^{d+1}}
 \int_{\RR^d}\dd^dy
 \prod_{a=1}^{4}\cK_\Delta(x_a;w,y).
 \label{eq:bdhm-boundary-contact}
\end{equation}
For $\AdS_4$, this reduces to
\begin{equation}
 \langle
 \cO(x_1)\cdots\cO(x_4)
 \rangle_{\mathrm c}^{(1)}
 =
 -\lambda
 \int_0^\infty\frac{\dd w}{w^4}
 \int_{\RR^3}\dd^3y
 \prod_{a=1}^{4}\cK_\Delta(x_a;w,y).
 \label{eq:bdhm-AdS4-contact}
\end{equation}
It is useful to introduce the standard contact integral
\cite{FreedmanMathurMatusisRastelli1999,DHokerFreedman2002}
\begin{equation}
 \cD_{\Delta\Delta\Delta\Delta}(x_i)
 =
 \int_0^\infty\frac{\dd w}{w^{d+1}}
 \int_{\RR^d}\dd^dy
 \prod_{a=1}^{4}
 \left(
 \frac{w}{w^2+|x_a-y|^2}
 \right)^\Delta.
 \label{eq:boundary-D}
\end{equation}
In the normalization \eqref{eq:bdhmoperator}, the boundary
four-point function is
\begin{equation}
 \langle
 \cO(x_1)\cdots\cO(x_4)
 \rangle_{\mathrm c}^{(1)}
 =
 -\lambda A_\Delta^2
 \cD_{\Delta\Delta\Delta\Delta}(x_i).
 \label{eq:unitD}
\end{equation}
The contact integral has the conformal transformation
properties expected for four identical scalar operators.
For mutually separated boundary points, the ordinary unrenormalized
integral in \eqref{eq:boundary-D} is convergent only in the range
\begin{equation}
 4\Delta>d.
 \label{eq:boundary-contact-convergence}
\end{equation}
Indeed, in the near-boundary region $w\to0$, with $y$ kept away from
all external points $x_a$, the radial dependence of the integrand is
\begin{equation}
 \frac{\dd w}{w^{d+1}}
 \prod_{a=1}^{4}
 \left(
 \frac{w}{w^2+|x_a-y|^2}
 \right)^\Delta
 \sim
 \dd w\,w^{4\Delta-d-1}.
\end{equation}
The radial integral therefore converges at $w=0$ precisely when
$4\Delta>d$. This condition is automatic in standard quantization.
In alternative quantization, where
$\Delta=\Delta_-=d/2-\nu$, it becomes
\begin{equation}
 0<\nu<\min\left\{1,\frac d4\right\}.
 \label{eq:alternative-contact-range}
\end{equation}
This restriction concerns the unrenormalized quartic contact diagram,
not the existence of the free alternative quantization itself. At
$4\Delta=d$ the contact integral has a logarithmic boundary
divergence, while for $4\Delta<d$ the divergence is power-like. In
these cases the diagram must be defined with a boundary regulator and
an additional boundary renormalization prescription.
For completeness, the remaining integration regions do not impose
stronger restrictions for separated $x_a$ and $\Delta>0$: when
$|y-x_a|=O(w)$, rescaling $y-x_a=wu$ gives a contribution proportional
to $\int_0 w^{2\Delta-1}\,\dd w$, which converges; the large-distance
region is likewise integrable for positive $\Delta$ once the
near-boundary condition $4\Delta>d$ holds. The resulting boundary
subtractions need not be restricted to contact terms at coincident
external positions: boundary interactions or multitrace counterterms
may be required, and their effect on boundary conditions must be fixed
in the renormalization prescription.

We now turn to the derivation of these boundary
observables directly from the neural representation
of the bulk field. In particular, we show that the
bulk-to-boundary propagator emerges from the statistical
correlations of the neural features, without being
introduced as an independent component of the
neural architecture.

\subsection{Boundary neurons, free correlators, and ultraviolet behavior}
\label{sec:bdhm-neural-free}

We return to the free Bessel--cosine network constructed in
Section~\ref{sec:bessel-neurons}:
\begin{equation}
 \phi_N(z,x)
 =
 \frac{1}{\sqrt N}
 \sum_{i=1}^{N}a_i f_{\eta_i}(z,x),
 \label{eq:bdhm-neural-bulk}
\end{equation}
where
\begin{equation}
 f_\eta(z,x)
 =
 B_\rho z^\Delta R_\alpha(kz)
 \cos(p\cdot x+b),
 \qquad
 \alpha=\Delta-\frac d2.
 \label{eq:bdhm-bulk-neuron}
\end{equation}
The output weights $a_i$ are independent standard Gaussian
variables, while the hidden parameters $\eta_i=(p_i,k_i,b_i)$
are sampled from the spectral distribution $q_\rho$ defined
in \eqref{eq:qneuron}.

Since the BDHM prescription extracts the leading normalizable
boundary behavior of the bulk field, it is natural to apply
the same limit to every neuron. We define
\begin{equation}
 h_\eta(x)
 =
 A_\Delta^{-1/2}
 \lim_{z\to0}z^{-\Delta}f_\eta(z,x).
 \label{eq:bdhm-boundary-neuron-limit}
\end{equation}
The radial activation was normalized so that
$R_\alpha(0)=1$. Therefore,
\begin{equation}
 h_\eta(x)
 =
 \frac{B_\rho}{\sqrt{A_\Delta}}
 \cos(p\cdot x+b).
 \label{eq:boundary-neuron}
\end{equation}
The boundary neuron has a simpler form than the bulk
neuron: its explicit Bessel dependence disappears, and
only the cosine activation along the boundary remains.
Nevertheless, the radial momentum $k$ is still part of
the hidden parameter distribution. The mass dependence
is therefore retained in the spectral statistics even
though it is not visible in the boundary activation itself.

We define the boundary neural operator by
\begin{equation}
 \cO_N(x)
 =
 \frac{1}{\sqrt N}
 \sum_{i=1}^{N}a_i h_{\eta_i}(x).
 \label{eq:boundary-neural-field}
\end{equation}
The bulk and boundary neural fields use the same output weights and the
same hidden parameters. The boundary network is therefore not an
independently initialized model, but the extrapolated readout of the
original bulk network. Their fixed-feature correlators are computed with
the same conditional interacting measure, and their unconditioned
correlators are obtained from the same quenched average over $Q_N$.

We first calculate the free boundary two-point function.
Using the independence of the Gaussian output weights,
\begin{align}
 \EE_0[\cO_N(x)\cO_N(y)]
 &=
 \frac1N\sum_{i,j}
 \EE_0[a_i a_j]\,
 \mathbb E_\eta[
 h_{\eta_i}(x)h_{\eta_j}(y)]=
 \mathbb E_\eta[h_\eta(x)h_\eta(y)].
 \label{eq:boundary-neural-two}
\end{align}
Since the phase average gives
\begin{equation}
 \mathbb E_b[
 \cos(p\cdot x+b)\cos(p\cdot y+b)]
 =
 \frac12\cos\bigl(p\cdot(x-y)\bigr),
 \label{eq:boundary-phase-average}
\end{equation}
we have
\begin{equation}
 \mathbb E_\eta[h_\eta(x)h_\eta(y)]
 =
 \frac{B_\rho^2}{2A_\Delta}
 \int_{\RR^d}\dd^dp
 \int_0^\infty\dd k\,
 q_\rho(p,k)
 \cos\bigl(p\cdot(x-y)\bigr).
 \label{eq:boundary-neural-spectral-two}
\end{equation}
We can identify this expression using the free bulk covariance.
From the definition of the boundary neuron,
\begin{align}
 \mathbb E_\eta[h_\eta(x)h_\eta(y)]
 &=
 \frac1{A_\Delta}
 \lim_{z,w\to0}
 z^{-\Delta}w^{-\Delta}
 \mathbb E_\eta[
 f_\eta(z,x)f_\eta(w,y)]
 \nonumber\\
 &=
 \frac1{A_\Delta}
 \lim_{z,w\to0}
 z^{-\Delta}w^{-\Delta}
 G_{\Delta,\rho}((z,x),(w,y)).
 \label{eq:boundary-neural-G-limit}
\end{align}
At a fixed spectral cutoff, these limits are well defined
for the regulated features. After the appropriate removal
of the regulator, the massive propagator has the separated-point
boundary behavior
\begin{equation}
 G_\Delta((z,x),(w,y))
 \underset{z,w\to0}{\sim}
 A_\Delta
 \frac{(zw)^\Delta}{|x-y|^{2\Delta}},
 \qquad x\neq y.
 \label{eq:boundary-G-asymptotic}
\end{equation}
It follows that
\begin{equation}
 \lim_{\rho\to1}
 \lim_{N\to\infty}
 \EE_0[\cO_N(x)\cO_N(y)]
 =
 \frac1{|x-y|^{2\Delta}},
 \qquad x\neq y,
 \label{eq:neural-boundary-two-result}
\end{equation}
where $\rho\to1$ denotes removal of the spectral regulator
in the prescribed distributional sense.

Thus, the boundary neurons reproduce the free two-point
function of a conformal scalar primary of dimension
$\Delta$. The result agrees with the BDHM prediction
\eqref{eq:unitboundary}.
The ensemble-averaged two-point identity holds at every finite
$N$ at fixed spectral regulator; the $N\to\infty$ limit in
\eqref{eq:neural-boundary-two-result} is needed to obtain the Gaussian
field law, not to make its covariance exact.

An important qualification is that
\eqref{eq:neural-boundary-two-result} is an ordinary function only
for separated points, $x\neq y$. The coincidence limit $x\to y$ is
singular and must be interpreted with a regulator and, when required,
subtraction of local contact terms. Boundary smearing improves the
short-distance behavior by suppressing large boundary momenta, but it
does not automatically remove all ultraviolet growth inherited from
the radial spectral integral. Thus, although
$\mathcal O_\Lambda(f)$ is an ordinary random variable at every fixed
cutoff $\Lambda$, its convergence to a cutoff-independent random
variable is an additional statement that is not established here.
The results below are therefore interpreted as separated-point
correlators or as suitably renormalized boundary correlation
distributions.

The radial spectral integral introduces an additional
ultraviolet subtlety in the boundary limit. In standard
quantization, at fixed boundary momentum $p$, the
large-$k$ contribution to the boundary covariance
behaves as
$\int^\Lambda k^{2\nu+1}(p^2+k^2)^{-1}\,\dd k$,
which grows as $\Lambda^{2\nu}$ for $\nu>0$.
Consequently, smearing the field along the boundary
directions does not, by itself, eliminate ultraviolet
divergences associated with the radial momentum.
This behavior can be examined explicitly through
the pointwise statistics of the free boundary neurons.

For an arbitrary normalized spectral density $q_\rho$,
averaging over the uniformly distributed phase yields
\begin{equation}
 \sigma_\rho^2
 :=\EE_0[\cO_N(x)^2]
 =\frac{B_\rho^2}{2A_\Delta},\qquad
 \kappa_4(\cO_N(x))
 =\frac{3}{2N}(\sigma_\rho^2)^2.
 \label{eq:boundary-variance-and-kurtosis}
\end{equation}
The second identity follows from
\eqref{eq:finite-width-fourth-cumulant} and the fourth moment of
a uniformly phased cosine. For the specific Gaussian spectral cutoff
$\rho_\Lambda(p,k)=\exp[-(p^2+k^2)/\Lambda^2]$ in \emph{standard}
quantization, normalization of \eqref{eq:qneuron} gives
\begin{equation}
 \sigma_\Lambda^2
 =\frac{\Lambda^{2\Delta}}
 {2^{2\Delta}\Gamma(\Delta+1)},\qquad
 \kappa_4(\cO_{N,\Lambda}(x))
 =\frac{3}{2N}(\sigma_\Lambda^2)^2
 \propto\frac{\Lambda^{4\Delta}}{N}.
 \label{eq:boundary-gaussian-cutoff-growth}
\end{equation}
Therefore a relative $1/N$ suppression does not by itself
control the absolute ultraviolet growth. These are coincident-point
statistics of a regulated field, not statements that its separated,
renormalized CFT correlations diverge. Consistently, the exact
unregulated mixed propagator in standard quantization satisfies
$z^{-2\Delta}G_\Delta(p;z,z)\sim z^{-2\nu}/(2\nu)$ for $\nu>0$;
at the BF point the divergence is logarithmic. The extrapolate
operator must consequently be interpreted at the level of suitable
renormalized distributions and observables, with the boundary limit
and ultraviolet limit controlled separately.

\subsection{The bulk-to-boundary propagator from neural averaging}
\label{sec:bdhm-neural-kernel}

We next show that the bulk-to-boundary propagator
\eqref{eq:unitK} can also be obtained directly from
the neural ensemble. Consider one boundary neuron $h_\eta(x)$ and one bulk
neuron $f_\eta(w,y)$ with the same hidden parameters.
Their average is
\begin{equation}
 \cK_{\Delta,\rho}^{\mathrm{NN}}(x;w,y)
 =
 \mathbb E_\eta[
 h_\eta(x)f_\eta(w,y)].
 \label{eq:neural-K-definition}
\end{equation}
Using the explicit neuron functions,
\begin{align}
 \cK_{\Delta,\rho}^{\mathrm{NN}}(x;w,y)
 &=
 \frac{B_\rho^2}{\sqrt{A_\Delta}}
 w^\Delta
 \mathbb E_\eta\left[
 R_\alpha(kw)
 \cos(p\cdot x+b)\cos(p\cdot y+b)
 \right]
 \nonumber\\
 &=
 \frac{B_\rho^2}{2\sqrt{A_\Delta}}
 w^\Delta
 \int_{\RR^d}\dd^dp
 \int_0^\infty\dd k\,
 q_\rho(p,k)R_\alpha(kw)
 \cos\bigl(p\cdot(x-y)\bigr).
 \label{eq:neural-K-phase}
\end{align}
The second line follows from the phase average
\eqref{eq:boundary-phase-average}.

Recall the spectral probability density
\begin{equation}
 q_\rho(p,k)
 =
 \frac{\rho(p,k)}
 {Z_\rho(2\pi)^d}
 \frac{k^{2\alpha+1}}{p^2+k^2},
 \qquad
 B_\rho^2
 =
 \frac{2Z_\rho}{C_\alpha^2},
 \qquad
 C_\alpha=2^\alpha\Gamma(\alpha+1).
 \label{eq:neural-K-spectral-data}
\end{equation}
Using the relation
\begin{equation}
 w^\Delta R_\alpha(kw)
 =
 C_\alpha k^{-\alpha}w^{d/2}J_\alpha(kw),
\end{equation}
we obtain
\begin{align}
 \cK_{\Delta,\rho}^{\mathrm{NN}}(x;w,y)
 &=
 \frac{w^{d/2}}
 {C_\alpha\sqrt{A_\Delta}}
 \int\frac{\dd^dp}{(2\pi)^d}
 \cos\bigl(p\cdot(x-y)\bigr)
 \int_0^\infty\dd k\,
 \rho(p,k)
 \frac{k^{\alpha+1}J_\alpha(kw)}
 {p^2+k^2}.
 \label{eq:neural-K-bessel}
\end{align}
We now evaluate the radial integral after removing the
spectral regulator in the appropriate sense. Using the
Schwinger representation
\begin{equation}
 \frac1{p^2+k^2}
 =
 \int_0^\infty\dd t\,e^{-t(p^2+k^2)}
\end{equation}
and the Bessel integral
\begin{equation}
 \int_0^\infty\dd k\,
 k^{\alpha+1}J_\alpha(kw)e^{-tk^2}
 =
 \frac{w^\alpha}{(2t)^{\alpha+1}}
 e^{-w^2/(4t)},
 \label{eq:neural-Bessel-Gaussian}
\end{equation}
we find
\begin{align}
 \int_0^\infty\dd k\,
 \frac{k^{\alpha+1}J_\alpha(kw)}{p^2+k^2}
 &=
 \frac{w^\alpha}{2^{\alpha+1}}
 \int_0^\infty\dd t\,
 t^{-\alpha-1}
 e^{-tp^2-w^2/(4t)}=
 p^\alpha K_\alpha(pw).
 \label{eq:neural-K-radial-integral}
\end{align}
Consequently, the neural mixed covariance becomes
\begin{equation}
 \cK_\Delta^{\mathrm{NN}}(x;w,y)
 =
 \frac{w^{d/2}}
 {C_\alpha\sqrt{A_\Delta}}
 \int\frac{\dd^dp}{(2\pi)^d}
 e^{ip\cdot(x-y)}
 p^\alpha K_\alpha(pw),
 \qquad p=|p|.
 \label{eq:neural-K-momentum}
\end{equation}
This is the momentum-space representation of the normalized
bulk-to-boundary propagator.

To return to position space, one can use the integral
representation of the modified Bessel function and the
Gaussian Fourier transform
\begin{equation}
 \int\frac{\dd^dp}{(2\pi)^d}
 e^{-tp^2+ip\cdot r}
 =
 \frac{1}{(4\pi t)^{d/2}}
 e^{-r^2/(4t)}.
 \label{eq:neural-K-Fourier}
\end{equation}
After the remaining parameter integration, one obtains
\begin{equation}
 \cK_\Delta^{\mathrm{NN}}(x;w,y)
 =
 \sqrt{A_\Delta}
 \left(
 \frac{w}{w^2+|x-y|^2}
 \right)^\Delta.
 \label{eq:neuralmixed}
\end{equation}
Comparing with \eqref{eq:unitK}, we arrive at the identity
\begin{equation}
 \cK_\Delta^{\mathrm{NN}}(x;Y)
 =
 \mathbb E_\eta[
 h_\eta(x)f_\eta(Y)]
 =
 \cK_\Delta(x;Y).
 \label{eq:neural-K-equality}
\end{equation}
The equality is understood after the prescribed spectral
limit. It is one of the central results of the neural
BDHM construction.

The bulk-to-boundary propagator has not been introduced
as a separate activation function or fitted to the
expected holographic kernel. It follows from averaging
the boundary and bulk neurons that already define the
free neural field. The phase integration produces the
boundary momentum dependence, while the Bessel spectral
integral determines the correct massive radial propagation.

For example, in $\AdS_4$ with $m^2L^2=-2$ and alternative
quantization $\Delta=1$, we have
\begin{equation}
 A_1=\frac{1}{2\pi^2},
\end{equation}
and hence
\begin{equation}
 \cK_1^{\mathrm{NN}}(x;w,y)
 =
 \frac{w}
 {\sqrt{2}\pi\bigl(w^2+|x-y|^2\bigr)}.
 \label{eq:neural-K-AdS4}
\end{equation}
This is consistent with the normalization and boundary limit of
the scalar BDHM kernel \eqref{eq:unitK}.

\subsection{The interacting boundary four-point function from NN--FT}
\label{sec:bdhm-neural-four}

We can now combine the bulk interaction with the boundary
neurons to obtain the connected boundary four-point
function entirely from neural parameter averages.

In Section~\ref{sec:interactions}, we considered several
ways to generate a quartic scalar interaction. For the
circular forward construction and the compensated
auxiliary-neuron construction, the resulting regulated
field interaction has the form
\begin{equation}
 V_4[\phi_N]
 =
 \frac{\lambda}{4!}
 \int_\Omega\dV_Y\,\phi_N(Y)^4,
 \label{eq:bdhm-neural-quartic}
\end{equation}
up to the chosen counterterms and regulator corrections.

The forward construction introduces this functional directly
in the neural parameter distribution. In the auxiliary-neuron
construction, the same quartic functional is obtained by
integrating out the additional binary variables in their
weak-coupling limit. Once the resulting field measures agree,
their perturbative bulk and boundary correlators agree as well.

To see the boundary four-point function explicitly, we use
the neural boundary operator
\begin{equation}
 \cO_N(x)
 =
 \frac{1}{\sqrt N}
 \sum_{i=1}^{N}a_i h_{\eta_i}(x).
 \label{eq:bdhm-four-boundary-field}
\end{equation}
For fixed hidden parameters, the quartic interaction induces
the leading connected cumulant of the output weights
\begin{equation}
 \kappa_4(a_i,a_j,a_k,a_l\mid\eta)
 =
 -\frac{\lambda}{N^2}V_{ijkl}
 +O(\lambda^2),
 \label{eq:bdhm-four-weight-cumulant}
\end{equation}
where
\begin{equation}
 V_{ijkl}
 =
 \int_\Omega\dV_Y\,
 f_{\eta_i}(Y)f_{\eta_j}(Y)
 f_{\eta_k}(Y)f_{\eta_l}(Y).
 \label{eq:bdhm-four-overlap}
\end{equation}
This is the same overlap tensor that appeared in the
bulk four-point calculation. The formula also follows
from the effective quartic measure obtained after
marginalizing the auxiliary neurons.

The connected boundary four-point function is
\begin{align}
 &\langle
 \cO_N(x_1)\cO_N(x_2)
 \cO_N(x_3)\cO_N(x_4)
 \rangle_{\mathrm c,\eta}^{(1)}
 \nonumber\\
 &\quad =
 \frac1{N^2}
 \sum_{i,j,k,l}
 h_{\eta_i}(x_1)h_{\eta_j}(x_2)
 h_{\eta_k}(x_3)h_{\eta_l}(x_4)
 \kappa_4(a_i,a_j,a_k,a_l\mid\eta).
 \label{eq:bdhm-neural-four-start}
\end{align}
Substituting \eqref{eq:bdhm-four-weight-cumulant}
and the overlap tensor, we obtain
\begin{align}
 &\langle
 \cO_N(x_1)\cdots\cO_N(x_4)
 \rangle_{\mathrm c,\eta}^{(1)}
 \nonumber\\
 &=
 -\frac{\lambda}{N^4}
 \sum_{i,j,k,l}
 \int_\Omega\dV_Y\,
 h_{\eta_i}(x_1)h_{\eta_j}(x_2)
 h_{\eta_k}(x_3)h_{\eta_l}(x_4) \times
 f_{\eta_i}(Y)f_{\eta_j}(Y)
 f_{\eta_k}(Y)f_{\eta_l}(Y).
 \label{eq:bdhm-neural-four-expand}
\end{align}
Each neuron index appears in one boundary feature and
one bulk feature. The four sums can therefore be
factorized
\begin{align}
 &\langle
 \cO_N(x_1)\cdots\cO_N(x_4)
 \rangle_{\mathrm c,\eta}^{(1)} =
 -\lambda\int_\Omega\dV_Y
 \prod_{a=1}^{4}
 \left[
 \frac1N\sum_{i=1}^{N}
 h_{\eta_i}(x_a)f_{\eta_i}(Y)
 \right].
 \label{eq:bdhm-neural-four-factorized}
\end{align}
The expression inside each bracket is the empirical
boundary--bulk covariance of the neural ensemble.
At fixed regulators, the law of large numbers gives
\begin{equation}
 \frac1N\sum_{i=1}^{N}
 h_{\eta_i}(x)f_{\eta_i}(Y)
 \longrightarrow
 \mathbb E_\eta[h_\eta(x)f_\eta(Y)]
 \qquad (N\to\infty).
 \label{eq:bdhm-empirical-mixed}
\end{equation}
Under suitable moment and integration bounds, we may
therefore take the infinite-width limit of
\eqref{eq:bdhm-neural-four-factorized}
\begin{align}
 &\lim_{N\to\infty}
 \langle
 \cO_N(x_1)\cdots\cO_N(x_4)
 \rangle_{\mathrm c}^{(1)}
 \nonumber\\
 &\quad =
 -\lambda\int_\Omega\dV_Y
 \prod_{a=1}^{4}
 \mathbb E_\eta[
 h_\eta(x_a)f_\eta(Y)].
 \label{eq:bdhm-neural-four-average}
\end{align}
We have already evaluated this average in
\eqref{eq:neural-K-equality}:
\begin{equation}
 \mathbb E_\eta[h_\eta(x_a)f_\eta(Y)]
 =
 \cK_\Delta(x_a;Y).
\end{equation}
Substituting it into the four-point function gives
\begin{equation}
 \langle
 \cO(x_1)\cdots\cO(x_4)
 \rangle_{\mathrm c}^{(1)}
 =
 -\lambda\int_{\AdS_{d+1}}\dV_Y
 \prod_{a=1}^{4}\cK_\Delta(x_a;Y),
 \label{eq:bdhm-neural-four-result}
\end{equation}
after the regulated limits have been taken.

In Poincar\'e coordinates, this is
\begin{equation}
 \langle
 \cO(x_1)\cdots\cO(x_4)
 \rangle_{\mathrm c}^{(1)}
 =
 -\lambda
 \int_0^\infty\frac{\dd w}{w^{d+1}}
 \int_{\RR^d}\dd^dy
 \prod_{a=1}^{4}
 \sqrt{A_\Delta}
 \left(
 \frac{w}{w^2+|x_a-y|^2}
 \right)^\Delta.
 \label{eq:bdhm-neural-four-coordinate}
\end{equation}
Equivalently,
\begin{equation}
 \langle
 \cO(x_1)\cdots\cO(x_4)
 \rangle_{\mathrm c}^{(1)}
 =
 -\lambda A_\Delta^2
 \cD_{\Delta\Delta\Delta\Delta}(x_i).
 \label{eq:bdhm-neural-D-result}
\end{equation}
This is exactly the leading BDHM boundary four-point function
derived independently in \eqref{eq:unitD}. The result
therefore follows directly from the explicit neural
parameter distribution, the boundary-neuron definition,
and the quartic interaction constructed in Section~2.

It is useful to check the normalization in a simple example.
Consider standard quantization in $\AdS_4$ with
\begin{equation}
 m^2L^2=-2,
 \qquad
 \Delta=2,
 \qquad
 A_2=\frac1{\pi^2}.
\end{equation}
The Fourier transform of the unit-normalized
bulk-to-boundary propagator is particularly simple:
\begin{equation}
 \widehat{\cK}_2(p;w)
 =
 \pi w e^{-|p|w}.
 \label{eq:bdhm-AdS4-K-momentum}
\end{equation}
Using the contact formula and removing the overall
momentum-conservation delta function, we obtain
\begin{align}
 &\langle
 \cO(p_1)\cO(p_2)\cO(p_3)\cO(p_4)
 \rangle_{\mathrm c}^{(1)\prime} =
 -\lambda
 \int_0^\infty\frac{\dd w}{w^4}
 \prod_{a=1}^{4}
 \left(\pi w e^{-|p_a|w}\right)
 \nonumber\\
 &\quad =
 -\lambda\pi^4
 \int_0^\infty\dd w\,
 e^{-w\sum_a|p_a|} =
 -\frac{\lambda\pi^4}
 {\sum_{a=1}^{4}|p_a|}.
 \label{eq:bdhm-AdS4-four-momentum}
\end{align}
For the unnormalized operator $\cO_{\mathrm{raw}}$,
the corresponding expression is
\begin{equation}
 \langle
 \cO_{\mathrm{raw}}(p_1)\cdots
 \cO_{\mathrm{raw}}(p_4)
 \rangle_{\mathrm c}^{(1)\prime}
 =
 -\frac{\lambda}
 {\sum_{a=1}^{4}|p_a|}.
 \label{eq:AdS4contact}
\end{equation}
The difference is entirely due to the normalization
of the four external boundary operators.

The critical binary-population construction described in
Appendix~\ref{app:critical-populations} can be treated in the same holographic framework
once its collective field has the required regulated
massive covariance and quartic measure. Its microscopic
degrees of freedom are different, so the explicit
Bessel-neuron factorization above does not apply to it
automatically. Nevertheless, if the limiting collective
field measure agrees with the same interacting scalar
theory, the leading BDHM boundary correlators agree.

At finite random-feature width, additional connected
correlations can arise from fluctuations of the empirical
free covariance. These are distinct from the interaction
contribution calculated above and vanish in the controlled
Gaussian-width limit. Likewise, taking the boundary limit
requires control of ultraviolet contact terms and the
normalization of the extrapolated field.

\paragraph{Non-uniqueness of bulk neural representations from boundary observables.}
The boundary results above do not uniquely determine the bulk free
field measure. An explicit counterexample is obtained by replacing
the coherent Bessel activation with independently sampled elementary
cosine neurons. We state the result for standard quantization,
$\alpha=\nu\geq0$, and for separated points after the prescribed
spectral limit. Define
\begin{equation}
 g_{\eta,u}(z,x)=B_\rho z^\Delta
 \cos(p\cdot x+kuz+b),\qquad
 u\sim\rho_\nu(u)\,\dd u.
 \label{eq:elementary-ridge-neuron}
\end{equation}
The Bessel neuron is the coherent average
$f_\eta(X)=\EE_u[g_{\eta,u}(X)\mid\eta]$. Replacing this average by
a single independently sampled $u$ changes the covariance because
\begin{equation}
 \EE_u[g_{\eta,u}(X)g_{\eta,u}(Y)]
 \ne \EE_u[g_{\eta,u}(X)]\EE_v[g_{\eta,v}(Y)].
 \label{eq:elementary-vs-coherent-covariance}
\end{equation}
The independently sampled architecture has the continuum covariance
\begin{equation}
 H_\Delta(X,Y)=A_\Delta
 \left[\frac{zw}{(z-w)^2+|x-y|^2}\right]^\Delta .
 \label{eq:elementary-cosine-kernel}
\end{equation}
It is AdS invariant, but is not the propagator of the selected
Klein--Gordon operator. With
$u_{XY}=((z-w)^2+|x-y|^2)/(2zw)=\cosh\ell-1$,
\begin{equation}
 P_{m,X}H_\Delta(X,Y)
 =-\frac{\Delta(2\nu+1)}{u_{XY}}H_\Delta(X,Y),
 \qquad X\ne Y.
 \label{eq:wrong-kernel-equation}
\end{equation}
Its singularity at small geodesic separation is
$H_\Delta\sim\ell^{-2\Delta}$, rather than the canonical
$G_\Delta\sim\ell^{-(d-1)}$ for $d+1>2$.

Nevertheless, the one-leg BDHM limit is identical
\begin{equation}
 A_\Delta^{-1/2}\lim_{z\to0}z^{-\Delta}
 H_\Delta((z,x),Y)
 =\cK_\Delta(x;Y).
 \label{eq:wrong-kernel-same-boundary}
\end{equation}
Consequently this incorrect bulk field reproduces both the
separated-point boundary two-point function and the leading quartic
contact integral \eqref{eq:unitD}. Neither is sufficient by itself
to establish equivalence of the bulk neural representation.

More precisely, for $\nu>0$ the separated-point kernel admits the
higher-mass decomposition
\begin{equation}
 H_\Delta(X,Y)=\sum_{n=0}^{\infty}b_n(\nu)G_{\Delta+n}(X,Y),
 \qquad b_n(\nu)=\frac{\nu+n}{\nu}\frac{(2\nu)_n}{n!},
 \quad b_0=1.
 \label{eq:wrong-kernel-tower}
\end{equation}
The coefficients are positive, and
$G_{\Delta+n}$ denotes the canonical propagator of a scalar with
$m_n^2=(\Delta+n)(\Delta+n-d)$. The additional bulk components
are invisible to this particular lowest-dimension boundary readout,
but contribute to diagrams with internal bulk propagators.
A finite coherent inner average over $M$ independent $u$-samples
has covariance
\begin{equation}
 G^{(M)}
 =\left(1-\frac1M\right)G_\Delta+\frac1M H_\Delta
 =G_\Delta+\frac1M(H_\Delta-G_\Delta).
 \label{eq:inner-width-contamination}
\end{equation}
Taking the outer Gaussian-process width $N\to\infty$ does not remove
this $1/M$ change of the limiting bulk covariance. An internal-line
exchange or loop diagram therefore tests the underlying neural
representation more strongly than the leading contact diagram.
The spectral evaluation of $H_\Delta$, its mass-tower expansion,
and the corresponding assumptions are given in
Appendix~\ref{app:coherent-counterexample}.

\section{Discussion and conclusions}
\label{sec:bdhm-summary}

In this work, we have developed a neural representation of
a massive scalar field theory on a fixed AdS background
and investigated its connection to holographic boundary
observables. Starting from the spectral decomposition of
the Klein--Gordon operator, we constructed a Bessel--cosine
neural field whose covariance reproduces the regulated
massive AdS propagator with the prescribed boundary
condition. We then introduced a local quartic interaction
through two complementary mechanisms: a direct deformation
of the neural parameter measure and an auxiliary-neuron
construction in which the interaction emerges from the
statistics of additional binary degrees of freedom.
Together, these constructions provide a framework for
representing both free and interacting scalar field
theories in terms of neural ensembles.

A central result of the holographic analysis is that
the boundary observables can be constructed directly
from the same neural degrees of freedom that define
the bulk field. Applying the BDHM prescription to the
individual Bessel--cosine neurons yields boundary neural
operators whose correlation functions reproduce those
of the corresponding scalar theory. In particular,
the free boundary two-point function takes the expected
conformal form
\begin{equation}
 \langle\cO(x)\cO(y)\rangle_0
 =
 \frac1{|x-y|^{2\Delta}},
\end{equation}
where $\Delta$ is determined by the scalar mass and
the chosen quantization.

The introduction of a local quartic interaction in
the bulk generates a nonvanishing connected boundary
four-point function. At leading order in the coupling,
we obtain
\begin{equation}
 \langle
 \cO(x_1)\cdots\cO(x_4)
 \rangle_{\mathrm c}^{(1)}
 =
 -\lambda A_\Delta^2
 \cD_{\Delta\Delta\Delta\Delta}(x_i),
\end{equation}
which coincides with the scalar contact Witten diagram.
In a renormalization scheme that preserves the free
two-point function through order $\lambda$, such as
the Wick-ordering prescription considered above,
the complete four-point function is
\begin{align}
 \langle\cO(x_1)\cO(x_2)\cO(x_3)\cO(x_4)\rangle
 &=
 \frac1{x_{12}^{2\Delta}x_{34}^{2\Delta}}
 +\frac1{x_{13}^{2\Delta}x_{24}^{2\Delta}}
 +\frac1{x_{14}^{2\Delta}x_{23}^{2\Delta}}
 \nonumber\\
 &\quad
 -\lambda A_\Delta^2
 \cD_{\Delta\Delta\Delta\Delta}(x_i)
 +O(\lambda^2),
 \label{eq:bdhm-full-boundary-four}
\end{align}
where $x_{ij}=|x_i-x_j|$.

The first three terms are the disconnected Gaussian
contractions characteristic of a generalized free
scalar primary, while the connected contribution
encodes the leading effect of the bulk interaction.
The resulting expression is crossing symmetric,
since the four external operators enter the contact
integral on equal footing. Its dependence on the
boundary conformal cross-ratios can be analyzed through
a conformal-block decomposition, providing a route
to extracting interaction-induced corrections to
operator dimensions and operator-product coefficients
of boundary composite operators.

An important feature of our construction is that
neither the bulk-to-boundary propagator nor the
boundary interaction kernel needs to be introduced
as an independent ingredient of the neural architecture.
The normalized boundary neurons follow directly from
the near-boundary limits of the bulk Bessel--cosine
features. Their mixed covariance with the bulk neurons
reproduces the massive bulk-to-boundary propagator,
which in turn determines the leading connected
boundary four-point function. Thus, the expected
holographic correlation functions emerge from the
same neural parameter distribution that defines
the underlying bulk field theory.

The neural representation of a given field theory is
not unique. As illustrated by the explicit construction
in Appendix~\ref{app:coherent-counterexample},
distinct neural architectures can reproduce the same
leading boundary observables while defining different
bulk field theories. In particular, agreement of the
boundary two-point function and the leading quartic
contact diagram does not guarantee equality of the
underlying bulk covariance or higher-order interaction
correlators. Establishing equivalence between neural
representations therefore requires matching the full
bulk field measures rather than a restricted set of
boundary correlation functions.

The precise interpretation of these results also
depends on the regularization and limiting procedures.
All exact probability-measure identities established
here are formulated at fixed spectral, bulk, and
finite-dimensional field regulators. The subsequent
limits depend on the construction under consideration
and must be taken in a controlled order. In particular,
the sequential procedures used in this work are
\begin{align}
 &\text{forward bulk construction:}
 &&N\to\infty
 \quad\text{at fixed }\Lambda\text{ and }\Omega,
 \notag\\
 &\text{auxiliary construction:}
 &&K\to\infty
 \quad\text{at fixed }N,\eta,\mathcal P,
 \quad\text{then }N\to\infty,
 \quad\text{then }\delta(\mathcal P)\to0,
 \notag\\
 &\text{boundary construction:}
 &&z\to0
 \quad\text{at fixed }\Lambda,
 \quad\text{then }N\to\infty,
 \quad\text{then }\Lambda\to\infty.
 \label{eq:orders-of-limits}
\end{align}
Here $N$ denotes the neural width, $K$ the auxiliary
population per bulk cell, $\mathcal P$ the bulk-cell
partition, and $\Lambda$ the spectral cutoff. These
orders specify the limits relevant to each construction;
no general interchangeability of the limits is assumed.

For the auxiliary-neuron construction, the displayed
population limit applies directly to the bare
compensated measure. If Wick subtraction is imposed
already at finite $K$, normalizability requires the
corresponding stability condition derived in
Appendix~\ref{sec:aux-wick-stability}. In particular,
the population limit must be taken through values
$K\geq K_{\mathrm{stab}}$ satisfying
\eqref{eq:aux-wick-K-stability}. Since this bound
depends on the regulated coincident covariance,
it need not remain uniform as the ultraviolet cutoff
is removed. Alternatively, one may first take the
bare $K\to\infty$ limit and subsequently Wick-order
the resulting quartic theory.

The removal of the spectral regulator requires
additional control. For the leading bulk contact
diagram, convergence to the unregulated expression
is ensured if, for separated external points,
the regulated propagators satisfy
\begin{align*}
 G_{\Delta,\Lambda}(X_r,Y)
 &\longrightarrow
 G_\Delta(X_r,Y),
 \qquad
 \left|G_{\Delta,\Lambda}(X_r,Y)\right|
 \leq M_r(Y),
 \qquad
 \prod_{r=1}^{4}M_r
 \in L^1(\AdS_{d+1},\dd V).
\end{align*}
Under these assumptions, dominated convergence
allows the spectral regulator to be removed from
the contact integral. Higher-order diagrams,
however, require the existence of the corresponding
renormalized limits of smeared Schwinger functions.
Such continuum limits are not established by the
finite-regulator neural construction itself and
remain subject to the renormalization properties
of the underlying scalar field theory.

These ultraviolet considerations also depend on
the bulk spacetime dimension $D=d+1$. The quartic
coupling has mass dimension
\begin{equation}
 [\lambda]=4-D=3-d.
\end{equation}
Accordingly, the scalar $\phi^4$ interaction is
super-renormalizable for $D<4$, perturbatively
renormalizable for $D=4$, and generally requires
an effective-field-theory interpretation for $D>4$.
Although Wick ordering removes the leading tadpole
contribution, higher-order corrections can require
additional mass, coupling, and other counterterms
consistent with the prescribed AdS boundary condition.
Moreover, the presence of the conformal boundary
may necessitate further boundary counterterms
beyond those suggested by ordinary bulk
power counting.

Several aspects of the construction therefore
remain to be developed. In particular, the present
analysis is formulated in Euclidean signature and
concerns regulated bulk and boundary correlation
functions of a scalar field on a fixed AdS background.
The covariance identities and perturbative matching
results do not, by themselves, establish reflection
positivity. A unitary Lorentzian interpretation would
require the appropriate positivity properties of
the Euclidean Schwinger functions, together with
a controlled reconstruction of the corresponding
quantum theory.

Likewise, reproducing the leading holographic
correlation functions does not by itself establish
an autonomous interacting boundary CFT or a
theory of dynamical bulk gravity. A more complete
holographic construction would require control
over the continuum and boundary limits, a consistent
renormalization prescription at higher orders,
and a well-defined spectrum of boundary operators
with their associated correlation functions.

Within the fixed-background scalar sector, however,
the results demonstrate that neural quantum field
theory provides a direct representation of both
the massive bulk propagator and the leading effects
of local interactions on holographic observables.
The Bessel--cosine architecture connects the spectral
properties of the Klein--Gordon operator to the
statistical structure of neural features, while
the direct and auxiliary interaction mechanisms
show how local bulk dynamics can be encoded in
neural probability measures. The BDHM construction
then relates these bulk neural fields to boundary
correlation functions without introducing an
independent boundary neural theory. This establishes
a concrete framework for investigating holographic
observables within neural representations of
interacting quantum field theories\footnote{Authors used AI tools during work on this manuscript. Author
reviewed and edited the content as needed and take full responsibility for the content of the publication.}.

\appendix

\section{Alternative representations of the free massive scalar}
\label{app:alternative-neurons}

The neural representation of a free Gaussian field
is not uniquely determined by the choice of activation
functions. Rather, the resulting field theory depends
on both the feature map and the statistical properties
of its coefficients. At a fixed finite-dimensional
regulator, consider a linear neural representation
$\phi=Fa$, where the coefficients are Gaussian with
covariance $\Sigma$. The prescribed massive scalar
theory is reproduced whenever
\begin{equation}
 F\Sigma F^T=G_\Delta=P_m^{-1},
 \label{eq:app-free-factorization}
\end{equation}
with the Klein--Gordon operator $P_m$ equipped with
the selected AdS boundary condition.

This factorization allows different neural architectures
to represent the same Gaussian field measure. The
alternative constructions discussed below illustrate
this freedom, differing primarily in their spatial
localization, spectral resolution, and the correlations
required among the neural coefficients.

\subsection{{Alternative Gaussian covariance factorizations}}
\label{app:covariance-factorizations}

A first alternative is to use deterministic spectral features. Let $P_mu_n=\varepsilon_nu_n$ be an orthonormal eigenbasis of a
positive regulated kinetic operator, with $\varepsilon_n>0$.
Truncating its inverse suggests
\begin{equation}
 \phi_M(X)=\sum_{n=1}^{M}\frac{a_nu_n(X)}{\sqrt{\varepsilon_n}},
 \qquad a_n\sim\mathcal N(0,1),
 \qquad
 G_M(X,Y)=\sum_{n=1}^{M}\frac{u_n(X)u_n(Y)}{\varepsilon_n}.
 \label{eq:app-spectral-neurons}
\end{equation}
The field is exactly Gaussian at finite $M$, although $G_M$ only
approximates the full propagator. Fourier--Bessel eigenmodes give an explicit
spectral basis in Poincar\'e AdS. Deterministic quadrature replaces random
feature fluctuations by spectral truncation and quadrature errors.

Another possibility is to construct features using the heat kernel or localized functions. For strictly positive self-adjoint $P_m$, let $H_t(X,Y)$ be the kernel
of $e^{-tP_m}$. The resolvent identity and heat-kernel semigroup law give
\begin{equation}
 G_\Delta(X,Y)
 =\int_0^\infty\dd t\int\dV_Z\,
 H_{t/2}(X,Z)H_{t/2}(Y,Z).
 \label{eq:app-heat-factorization}
\end{equation}
Thus, $F_{t,Z}(X)=H_{t/2}(X,Z)$ provides a continuous feature
factorization labeled by a center $Z$ and scale $t$. Such neurons,
as well as finite-element or wavelet features, are useful for localized
readouts and multiscale discretizations. Their sampling weights or
coefficient correlations must nevertheless reproduce the covariance,
locality of a feature alone does not guarantee the correct free theory.
The small-$t$ region requires ultraviolet regularization.

The Bessel activation also admits a coherent cosine representation. For $\alpha>-1/2$, the normalized radial Bessel activation satisfies
\begin{equation}
 R_\alpha(t)=\int_{-1}^{1}\dd u\,\rho_\alpha(u)\cos(ut),
 \qquad
 \rho_\alpha(u)=\frac{\Gamma(\alpha+1)}
 {\sqrt\pi\,\Gamma(\alpha+1/2)}(1-u^2)^{\alpha-1/2}.
 \label{eq:app-coherent-cosines}
\end{equation}
This represents each Bessel neuron as a coherent average of
ordinary cosines. The average over $u$ is taken inside the neuron,
before assigning its Gaussian output weight. Replacing it by separately
weighted, independently sampled cosines changes the covariance because,
in general,
\begin{equation}
 \mathbb E_u[\widetilde f_u(X)\widetilde f_u(Y)]
 \ne
 \mathbb E_u[\widetilde f_u(X)]\,
 \mathbb E_u[\widetilde f_u(Y)].
 \label{eq:app-coherent-distinction}
\end{equation}
This positive-density identity applies directly in standard quantization
and, for $\alpha=-\nu$, only when $0<\nu<1/2$.

All these factorizations may represent the same regulated Gaussian
field, but the Bessel--cosine representation is especially convenient
here: its radial normalization $R_\alpha(0)=1$ makes the BDHM limit
of an individual neuron elementary and permits direct evaluation
of mixed bulk--boundary averages.

\subsection{Independent inner cosines and higher-mass contamination}
\label{app:coherent-counterexample}

Here we derive the counterexample used in
Section~\ref{sec:bdhm}. We assume standard
quantization, $\alpha=\nu\geq0$, with $\Delta=d/2+\nu$.
All manipulations are first defined with spectral and bulk-volume
regulators; the displayed closed kernels refer to separated points
after the appropriate spectral limit.

For the elementary feature \eqref{eq:elementary-ridge-neuron},
averaging the shared phase and a \emph{single shared} $u$ gives
\begin{align}
 \EE_{b,u}[g_{\eta,u}(X)g_{\eta,u}(Y)]
 &=\frac{B_\rho^2}{2}(zw)^\Delta
 \cos[p\cdot(x-y)]R_\nu(k(z-w)),
 \label{eq:app-shared-inner-average}
\end{align}
By contrast, the covariance of the coherently averaged Bessel
features contains the product $R_\nu(kz)R_\nu(kw)$.
Using the density \eqref{eq:qneuron} and removing the Gaussian
spectral regulator at separated points gives
\begin{equation}
 H_\Delta(X,Y)
 =\frac{(zw)^\Delta}{C_\nu^2}
 \int\frac{\dd^dp}{(2\pi)^d}e^{ip\cdot(x-y)}
 \int_0^\infty\dd k\,
 \frac{k^{2\nu+1}R_\nu(k(z-w))}{p^2+k^2}.
 \label{eq:app-wrong-spectral-kernel}
\end{equation}
Introduce the Schwinger representation
$(p^2+k^2)^{-1}=\int_0^\infty\dd t\,e^{-t(p^2+k^2)}$.
The radial integration becomes
\begin{equation}
 \int_0^\infty\dd k\,k^{2\nu+1}e^{-tk^2}R_\nu(ks)
 =\frac{\Gamma(\nu+1)}{2t^{\nu+1}}
 e^{-s^2/(4t)},\quad t>0.
 \label{eq:app-wrong-radial-integral}
\end{equation}
The Gaussian momentum transform is
\begin{equation}
 \int\frac{\dd^dp}{(2\pi)^d}
 e^{-tp^2+ip\cdot r}
 =(4\pi t)^{-d/2}e^{-|r|^2/(4t)}.
 \label{eq:app-wrong-fourier-integral}
\end{equation}
Writing $\mathcal R^2=(z-w)^2+|x-y|^2$, the remaining integral is
\begin{equation}
 \int_0^\infty\dd t\,t^{-\Delta-1}
 e^{-\mathcal R^2/(4t)}
 =4^\Delta\Gamma(\Delta)\mathcal R^{-2\Delta},
 \label{eq:app-wrong-schwinger-integral}
\end{equation}
which establishes Eq.~\eqref{eq:elementary-cosine-kernel} with the
same $A_\Delta$ as Eq.~\eqref{eq:normalA}.

To test the bulk equation, write
$u_{XY}=\mathcal R^2/(2zw)$. For an AdS-invariant function $F(u)$,
\begin{equation}
 \nabla_g^2F(u)
 =u(u+2)F''(u)+(d+1)(u+1)F'(u).
 \label{eq:app-invariant-laplacian}
\end{equation}
Substituting $F(u)=u^{-\Delta}$ and
$m^2=\Delta(\Delta-d)$ gives
Eq.~\eqref{eq:wrong-kernel-equation}; the right-hand side is
nonzero away from coincidence.
Nevertheless,
\begin{equation}
 \lim_{z\to0}z^{-\Delta}H_\Delta((z,x),(w,y))
 =A_\Delta\left[\frac{w}{w^2+|x-y|^2}\right]^\Delta,
 \label{eq:app-wrong-boundary-limit}
\end{equation}
proving the equality of the mixed BDHM kernels and therefore of
the leading boundary quartic contact diagrams.

For $\nu>0$, the expansion
\eqref{eq:wrong-kernel-tower} can be obtained by comparing the
canonical separated-point Green kernels of dimensions $\Delta+n$.
The operator $P_{m}$ acts on those kernels as
\begin{equation}
 P_mG_{\Delta+n}
 =-n(2\nu+n)G_{\Delta+n}
 \qquad (X\ne Y),
 \label{eq:app-mass-shift}
\end{equation}
because the corresponding masses differ by
$n(2\nu+n)$. The explicit power-law kernel also satisfies
\begin{equation}
 P_mH_\Delta
 =-2(\nu+1)(2\nu+1)H_{\Delta+1}
 \qquad (X\ne Y).
 \label{eq:app-wrong-kernel-recurrence}
\end{equation}
Matching coefficients gives the recurrence
\begin{equation}
 n(2\nu+n)b_n(\nu)
 =2(\nu+1)(2\nu+1)b_{n-1}(\nu+1),
 \quad b_0=1,
 \label{eq:app-tower-coeff-recurrence}
\end{equation}
whose solution is
$b_n(\nu)=(\nu+n)(2\nu)_n/(\nu n!)$.
For example, $b_1=2(\nu+1)$ and
$b_2=(\nu+2)(2\nu+1)$. The infinite sum is interpreted at
separated points, with a regulator when required; it should not
be interchanged with coincident-point or loop limits without
additional estimates.

Finally, with
$f_\eta^{(M)}=M^{-1}\sum_{r=1}^Mg_{\eta,u_r}$ and independent
$u_r$ conditional on $\eta$, the $M(M-1)$ off-diagonal pairs in
$\EE[f_\eta^{(M)}(X)f_\eta^{(M)}(Y)]$ produce $G_\Delta$, while
the $M$ diagonal pairs produce $H_\Delta$. This proves
Eq.~\eqref{eq:inner-width-contamination}. In a quartic theory the
difference is detected by internal bulk lines, for instance by the
$O(\lambda^2/M)$ contribution obtained from replacing one factor of
$G_\Delta(Y,Z)$ in a one-loop four-point bubble by
$H_\Delta(Y,Z)-G_\Delta(Y,Z)$, before matching counterterms.

\section{Alternative approaches to interactions}
\label{app:alternative-interactions}

Besides action pullback and compensated auxiliary neurons, interactions
may arise from critical collective statistics or from more general
auxiliary probability laws. We present some alternative options below.

\subsection{Critical binary populations: a quartic interaction from entropy}
\label{app:critical-populations}
\label{sec:critical-populations}

Unlike Section~\ref{sec:aux}, which couples auxiliary neurons to an
existing Gaussian scalar, this construction builds the scalar field
from the collective fluctuations of binary neurons. At each of a
finite number of bulk cells $A$, introduce $n$ spins $s_{A,r}=\pm1$
and their magnetization
\begin{equation}
 M_A=\sum_{r=1}^{n}s_{A,r}.
 \label{eq:critical-magnetization}
\end{equation}
For a real symmetric matrix $Q$, choose the microscopic law
\begin{equation}
 \mathbb P_n(s)\propto
 \exp\left[\frac1{2n}\sum_A M_A^2
 -\frac1{2n^{3/2}}\sum_{A,B}M_AQ_{AB}M_B\right].
 \label{eq:critical-spin-law}
\end{equation}
Every term is pairwise in the elementary spins, since
$M_AM_B=\sum_{r,t}s_{A,r}s_{B,t}$. In particular, no
microscopic quartic spin coupling has been imposed.
The relation between critical Ising-type populations and scalar
$\phi^4$ measures, including the critical $n^{3/4}$ magnetization
scaling, has classical precedents in
Refs.~\cite{GriffithsSimon,EllisNewman1978};
we use this mechanism here at fixed regulated bulk-cell number.

For a fixed magnetization, the number of spin configurations is
$\Omega_n(M)=\binom{n}{(n+M)/2}$. Its large-deviation rate function is
\begin{equation}
 I(m)=\frac{1+m}{2}\log(1+m)
 +\frac{1-m}{2}\log(1-m)
 =\frac{m^2}{2}+\frac{m^4}{12}+\frac{m^6}{30}+O(m^8),
 \qquad m=\frac{M}{n}.
 \label{eq:critical-rate}
\end{equation}
The first term in \eqref{eq:critical-spin-law} cancels the
quadratic entropic cost $M_A^2/(2n)$. The leading surviving
local term is $M_A^4/(12n^3)$, which fixes the critical scaling
\begin{equation}
 x_A=\frac{M_A}{n^{3/4}}.
 \label{eq:critical-rescaling}
\end{equation}
At fixed bulk-cell number, the rescaled magnetizations converge
weakly to the collective-field measure
\begin{equation}
 \dd\mu_{\mathrm{crit}}(x)\propto
 \exp\left[-\frac12x^TQx-\frac1{12}\sum_Ax_A^4\right]
 \prod_A\dd x_A.
 \label{eq:critical-limit-measure}
\end{equation}
The convergence is weak, not total-variation convergence of the
lattice-valued magnetizations to a continuous density.

To represent the massive AdS scalar, let $Q_h(m)$ be a positive
regulated precision matrix for the chosen bulk geometry, mass,
and boundary condition, with invariant cell volumes $v_A$.
For $\lambda>0$, the field identification $x_A=c_A\phi_A$ with
\begin{equation}
 Q=C^{-1}Q_h(m)C^{-1},\qquad
 C=\operatorname{diag}(c_A),\qquad
 c_A^4=\frac{\lambda v_A}{2}
 \label{eq:critical-graph-map}
\end{equation}
turns the exponent in \eqref{eq:critical-limit-measure} into
\begin{equation}
 S_h[\phi]=\frac12\phi^TQ_h(m)\phi
 +\frac{\lambda}{4!}\sum_Av_A\phi_A^4.
 \label{eq:critical-target-graph}
\end{equation}
The quartic term follows from spin entropy, but the AdS quadratic
operator, geometry, and critical tuning are still prescribed.
This is a fixed-regulator representation, not a derivation of
AdS geometry from the spin model.

The first finite-population correction, on bounded collective-field
sets, is
\begin{equation}
 S_n(x)=\frac12x^TQx+\frac1{12}\sum_Ax_A^4
 +\frac1{\sqrt n}\sum_A\left(\frac{x_A^6}{30}
 -\frac{x_A^2}{2}\right)+O(n^{-1}).
 \label{eq:critical-finite-n}
\end{equation}
After compensating the quadratic correction, the sextic term is
positive, in contrast to the negative sextic correction
$-\gamma^6\sum_Av_A^{3/2}\phi_A^6/(45\sqrt K)$ from the
compensated binary auxiliaries in \eqref{eq:Klimit}. Their
leading quartic correlators can therefore agree while their
finite-population higher-point functions differ. Controlling
the simultaneous spin-population, bulk-continuum, and BDHM
boundary limits remains an open step.

\subsection{Generalized auxiliary distributions and higher-order interactions}
\label{app:general-interactions}

The binary auxiliary distribution is only one example. For a
centered, symmetric auxiliary variable $\sigma$ with unit variance,
write its cumulant-generating function as
\begin{equation}
 K_\sigma(t)=\log\mathbb E_\sigma[e^{t\sigma}],\qquad
 W_\sigma(t)=\frac{t^2}{2}-K_\sigma(t)
 =-\sum_{q\ge2}\frac{\kappa_{2q}(\sigma)}{(2q)!}t^{2q}.
 \label{eq:app-general-auxiliary}
\end{equation}
A negative fourth cumulant induces a positive quartic term. By
matching lower even moments to Gaussian ones, one can instead make
the first surviving interaction proportional to $\phi^{2q}$.
Replicating weak auxiliary units then gives finite even couplings,
provided the resulting full probability measure is normalizable.

\paragraph{A three-valued auxiliary with no sextic correction.}
\label{app:ternary-auxiliaries}

The auxiliary probability law can be chosen to suppress specific
unwanted higher-order interactions. As a simple example, let
\begin{equation}
 \sigma=
 \begin{cases}
  0,
  &\text{with probability }1-r,\\
  +r^{-1/2},
  &\text{with probability }r/2,\\
  -r^{-1/2},
  &\text{with probability }r/2.
 \end{cases}
 \label{eq:ternary-auxiliary-law}
\end{equation}
We use $r$ rather than $p$ here to avoid confusion with the boundary
momentum. This distribution is centered and has unit variance. Its
fourth and sixth cumulants are
\begin{equation}
 \kappa_4(\sigma)=\frac1r-3,
 \qquad
 \kappa_6(\sigma)=\frac1{r^2}-\frac{15}{r}+30.
 \label{eq:ternary-auxiliary-cumulants}
\end{equation}
The sixth cumulant vanishes when
\begin{equation}
 30r^2-15r+1=0.
\end{equation}
Of the two solutions, the choice
\begin{equation}
 r_*
 =
 \frac{15+\sqrt{105}}{60}
 \simeq0.4208
 \label{eq:ternary-optimal-probability}
\end{equation}
also gives
\begin{equation}
 \kappa_4(\sigma)
 =
 \frac{9-\sqrt{105}}{2}<0.
\end{equation}
The negative fourth cumulant therefore generates a positive quartic
interaction, while the vanishing sixth cumulant removes the leading
sextic correction.
Here ``no sextic correction'' refers to the direct sextic
coefficient in the local effective-potential expansion, not to all
six-point contributions: two quartic vertices and other finite-width
or higher-order effects can still generate connected six-point functions.

The compensated potential generated by one such auxiliary variable is
\begin{equation}
 W_r(t)
 =
 \frac{t^2}{2}
 -
 \log\left[
 1-r+r\cosh\left(\frac{t}{\sqrt r}\right)
 \right].
 \label{eq:ternary-compensated-potential}
\end{equation}
At $r=r_*$ its small-$t$ expansion is
\begin{equation}
 W_{r_*}(t)
 =
 -\frac{\kappa_4}{4!}t^4
 -\frac{\kappa_8}{8!}t^8
 +O(t^{10}),
 \qquad
 \kappa_8=165\sqrt{105}-1665>0,
 \label{eq:ternary-potential-expansion}
\end{equation}
with no $t^6$ term.

Using $K$ independent auxiliaries at cell $A$, choose
\begin{equation}
 g_A^{(K)}
 =
 \left[
 \frac{\lambda v_A}
 {K(-\kappa_4)}
 \right]^{1/4}.
 \label{eq:ternary-auxiliary-coupling}
\end{equation}
Then, on bounded field configurations,
\begin{align}
 \sum_AK\,W_{r_*}\!\left(g_A^{(K)}S_A\right)
 &=
 \frac{\lambda}{4!}
 \sum_Av_AS_A^4
 \nonumber\\
 &\quad
 -
 \frac{\kappa_8}
 {8!\,\kappa_4^2K}
 \sum_A(\lambda v_A)^2S_A^8
 +O(K^{-3/2}).
 \label{eq:ternary-effective-interaction}
\end{align}
Thus the leading finite-population correction is of order $K^{-1}$,
whereas for the binary auxiliary law it is the sextic term of order
$K^{-1/2}$. The auxiliary distribution therefore acts as a controllable
non-Gaussian interaction generator: its fourth cumulant fixes the
desired quartic term, while its higher cumulants determine the unwanted
finite-population operators.
Since $|\kappa_4(r_*)|<2$, matching the same $\lambda$ at fixed
$K$ and $v_A$ requires a larger microscopic coupling than the binary
construction. The missing $K^{-1/2}$ term improves the formal expansion
but does not by itself bound errors of every finite-$K$ observable.

The exact compensated potential remains stable. For $r\geq1/3$, the
even moments of $\sigma$ do not exceed the corresponding Gaussian
moments, which implies
\begin{equation}
 \mathbb E_\sigma[e^{t\sigma}]
 \leq e^{t^2/2},
 \qquad
 W_r(t)\geq0.
\end{equation}
Hence the negative octic coefficient in the local expansion does not
make the full bare auxiliary measure unstable. If Wick subtraction is
added already at finite $K$, the separate large-field stability
condition of Section~\ref{sec:aux-wick-stability} must still be checked.

There is a limitation on how literally such effective interactions can
be realized by a single auxiliary marginal. If a unit-variance real
variable produced an \emph{exact} compensated potential $W_\sigma(t)=at^4$
with $a>0$ for every real $t$, its cumulant generator would be
$K_\sigma(t)=t^2/2-at^4$. But every finite cumulant-generating
function is convex, whereas
\begin{equation}
 K_\sigma''(t)=1-12at^2<0
 \quad\text{for sufficiently large }|t|.
 \label{eq:app-aux-convexity-obstruction}
\end{equation}
Therefore a fixed auxiliary marginal cannot give a globally exact
pure positive quartic by quadratic compensation. This does not exclude
the weak-unit replicated limit in Section~\ref{sec:aux}.

\subsection{{Correlated and nonlocal readouts}}
\label{app:correlated-readouts}

Spatial locality depends separately on the field readouts. For the
smeared readout
\[
 S_A[\phi]=\int\dV_Y\,R_A(Y)\phi(Y),
\]
correlated auxiliary neurons with fourth cumulants
$\kappa^\sigma_{ABCD}$ induce
\begin{align}
 V_4[\phi]&=\frac1{4!}\int\prod_{r=1}^4\dV_{Y_r}\,
 \Lambda_4(Y_1,\ldots,Y_4)\prod_{r=1}^4\phi(Y_r),
 \nonumber\\
 \Lambda_4(Y_1,\ldots,Y_4)
 &=-\sum_{A,B,C,D}\kappa^\sigma_{ABCD}
 g_Ag_Bg_Cg_D\,
 R_A(Y_1)R_B(Y_2)R_C(Y_3)R_D(Y_4).
 \label{eq:app-interaction-kernel}
\end{align}
Here the full auxiliary covariance must also be compensated if the
leading quadratic term is to vanish. Point-local independent readouts
recover a contact vertex in the appropriate limit; finite-range
readouts can generate nonlocal and derivative interactions.
Consequently, the quartic power may follow from auxiliary statistics
while exact bulk locality remains a choice of architecture.

We next consider the stability conditions for correlated auxiliary variables. Negativity of the fourth cumulant of each auxiliary neuron does not
ensure that the full induced quartic form is nonnegative. For example,
consider two symmetric binary auxiliaries with
\begin{equation}
 \mathbb P(\sigma_1,\sigma_2)
 =\frac14(1+c\sigma_1\sigma_2),\qquad |c|<1.
 \label{eq:app-correlated-binary}
\end{equation}
Compensating their complete quadratic covariance and expanding the
joint cumulant generator gives
\begin{equation}
 W_{c,4}(t_1,t_2)
 =\frac{t_1^4+t_2^4}{12}
  +\frac c3(t_1^3t_2+t_1t_2^3)
  +\frac{c^2}{2}t_1^2t_2^2.
 \label{eq:app-correlated-quartic}
\end{equation}
Along the antisymmetric direction,
\begin{equation}
 W_{c,4}(t,-t)=\frac{(1-c)(1-3c)}6\,t^4.
 \label{eq:app-correlated-negative-direction}
\end{equation}
The quartic form is nonnegative for every real $(t_1,t_2)$ if and
only if $|c|\leq1/3$. More generally, the required condition for a
nonnegative compensated quartic in every collective direction is
$\kappa_4(\sum_Au_A\sigma_A)\leq0$ for all real $u_A$.
Failure of this test concerns the quartic truncation; it need not
invalidate the exact, bounded-auxiliary probability law.

Some smeared interaction channels are redundant after imposing the equations of motion. Although finite-range readouts generate derivative terms in the
action, not every such term supplies a new tree-level boundary
four-point structure. To see this, take a differential readout
$R=1+a\ell^2\nabla_g^2+O(\ell^4)$ and a vertex
$(\lambda/4!)\int\dV\,(R\phi)^4$. A free bulk-to-boundary leg obeys
$(-\nabla_g^2+m^2)K_\Delta=0$, and hence
\begin{equation}
 RK_\Delta=(1+a\ell^2m^2)K_\Delta+O(\ell^4).
 \label{eq:app-smear-on-shell}
\end{equation}
At this order the corresponding contact diagram is simply rescaled
by $(1+a\ell^2m^2)^4$; no new cross-ratio dependence follows.
Boundary contributions and genuinely nonlocal operators require
separate treatment. A composite-pair readout, on the other hand,
can generate independent interactions. For example, with
$E=\phi^2$ and
$R=1+\ell^2\nabla_g^2+(\ell^4/2)\nabla_g^4+\cdots$,
\begin{equation}
 \int\dV\,E RE
 =\int\dV\left[\phi^4-\ell^2(\nabla\phi^2)^2
 +\frac{\ell^4}{2}(\nabla_g^2\phi^2)^2+\cdots\right],
 \label{eq:app-composite-pair-expansion}
\end{equation}
up to boundary terms. Since
$\nabla_g^2\phi^2=2(\nabla\phi)^2+2\phi\nabla_g^2\phi$,
the four-derivative term contains
$4[(\nabla\phi)^2]^2$, which can provide independent quartic
kinematic dependence after removing equations-of-motion-redundant
operators. Here $(\nabla\phi)^2=g^{AB}\partial_A\phi\partial_B\phi$.

\subsection{Finite-population stability and interaction diagnostics}
\label{app:population-diagnostics}
\label{sec:aux-wick-stability}

The convergence argument above was formulated for the bare compensated
auxiliary potential $V_K$. In that case the exact inequality
$W(t)\geq0$ immediately guarantees that the interaction cannot
destabilize the underlying Gaussian measure. A small additional
qualification is needed if Wick ordering is imposed already at finite
auxiliary population $K$.

The reason is simple. Wick ordering of a quartic interaction contains
a negative quadratic contribution. For the limiting $\phi^4$ theory
this causes no stability problem, because the positive quartic term
dominates at large field amplitude. At finite $K$, however, the exact
auxiliary potential is not globally quartic. Indeed,
\begin{equation}
 W(t)=\frac{t^2}{2}-\log\cosh t
 =\frac{t^2}{2}-|t|+\log2+o(1),
 \qquad |t|\to\infty.
 \label{eq:aux-W-large-field}
\end{equation}
Thus the exact finite-$K$ interaction grows only quadratically in the
far tails. A sufficiently large negative quadratic Wick counterterm
could therefore overcome this growth and make the finite-$K$
probability measure non-normalizable. We now state explicitly the
condition that prevents this from happening.

Let $\mathcal C_A$ denote the regulated coincident covariance used for
normal ordering at the bulk cell $A$. If normal ordering is defined
relative to the conditional Gaussian covariance, then
$\mathcal C_A=C_{AA}$. If instead it is defined relative to the
deterministic target covariance, then
$\mathcal C_A=G_{\Delta,\rho}(Y_A,Y_A)$. At a fixed regulator,
\begin{equation}
 :\!S_A^4\!:\,_{\mathcal C_A}
 =
 S_A^4-6\mathcal C_A S_A^2+3\mathcal C_A^2.
 \label{eq:aux-cell-wick-polynomial}
\end{equation}
The term relevant for the large-field behavior is therefore
\begin{equation}
 -\frac{\lambda}{4}v_A\mathcal C_A S_A^2.
 \label{eq:aux-wick-negative-quadratic}
\end{equation}

Recall that the local readout is linear in the Gaussian coefficients,
\begin{equation}
 S_A(a)=F_Aa,
 \qquad
 F_A=(F_{Ai})_i.
\end{equation}
After adding the Wick subtraction, the exact finite-$K$ action becomes
\begin{align}
 S_{K,\mathrm W}(a)
 &=
 \frac12 a^Ta
 +
 \sum_AK
 W\!\left(g_A^{(K)}F_Aa\right)
 -
 \frac{\lambda}{4}
 \sum_Av_A\mathcal C_A(F_Aa)^2
 +\text{const.},
 \label{eq:aux-finite-K-wick-action}\\
 g_A^{(K)}
 &=
 \gamma v_A^{1/4}K^{-1/4}.
 \notag
\end{align}

Using the exact identity
$W(t)=t^2/2-\log\cosh t$, we can collect all quadratic
contributions and write
\begin{equation}
 S_{K,\mathrm W}(a)
 =
 \frac12a^TQ_Ka
 -
 \sum_AK\log\cosh\!\left(g_A^{(K)}F_Aa\right)
 +\text{const.},
 \label{eq:aux-finite-K-wick-exact}
\end{equation}
where
\begin{equation}
 Q_K
 =
 I+
 \sum_A
 \left[
 K\bigl(g_A^{(K)}\bigr)^2
 -
 \frac{\lambda}{2}v_A\mathcal C_A
 \right]
 F_A^TF_A .
 \label{eq:aux-wick-stability-matrix}
\end{equation}
Here $I$ is the identity matrix in the finite-dimensional space of
Gaussian coefficients $a_i$.

Equation~\eqref{eq:aux-wick-stability-matrix} has a direct
interpretation. The identity matrix $I$ comes from the original
Gaussian measure and provides positive quadratic suppression in every
direction of parameter space. The term proportional to
$K(g_A^{(K)})^2$ is the positive large-field contribution generated
by the exact auxiliary potential. The term proportional to
$-\lambda v_A\mathcal C_A/2$ is the negative contribution introduced
by Wick subtraction. The finite-$K$ measure remains normalizable only
when the sum of these effects is still positive in every direction.

Concretely, the required condition is
\begin{equation}
 a^TQ_Ka>0
 \qquad
 \text{for every nonzero coefficient vector }a.
 \label{eq:aux-wick-positive-definite}
\end{equation}
Equivalently, all eigenvalues of $Q_K$ must be strictly positive.

To see why this condition controls the probability measure, suppose
that the smallest eigenvalue of $Q_K$ is $q_K>0$. Then
\begin{equation}
 a^TQ_Ka\geq q_K\|a\|^2.
\end{equation}
Moreover,
\begin{equation}
 \log\cosh x\leq |x|,
 \qquad
 |F_Aa|\leq\|F_A\|\,\|a\|.
\end{equation}
The large-field action is therefore bounded from below by
\begin{equation}
 S_{K,\mathrm W}(a)
 \geq
 \frac{q_K}{2}\|a\|^2
 -
 \left[
 \sum_AK|g_A^{(K)}|\|F_A\|
 \right]\|a\|
 -\text{const.}
\end{equation}
At sufficiently large $\|a\|$, the positive quadratic term dominates
the term linear in $\|a\|$. The probability density consequently
decays at least as fast as a Gaussian in the far tails, and the
partition function is finite.

The opposite situation explains why the condition is necessary. If
$Q_K$ has a negative eigenvalue, there is a direction in parameter
space along which the quadratic part of the action decreases as
$-\|a\|^2$. The remaining logarithmic term in
\eqref{eq:aux-finite-K-wick-exact} cannot compensate for this
decrease, and the action runs to $-\infty$. If $Q_K$ has a zero
eigenvalue, the quadratic suppression disappears along the
corresponding direction; the remaining $-\log\cosh$ contribution
either drives the action to $-\infty$ or leaves an unsuppressed flat
direction. Thus strict positivity of $Q_K$ is the appropriate
finite-regulator stability condition.

In practice, one does not need to diagonalize the full matrix in
order to obtain a simple sufficient condition. Since every matrix
$F_A^TF_A$ is positive semidefinite, it is enough to require
\begin{equation}
 K\bigl(g_A^{(K)}\bigr)^2
 \geq
 \frac{\lambda}{2}v_A\mathcal C_A
 \qquad
 \text{for every }A.
 \label{eq:aux-wick-cellwise-stability}
\end{equation}
Under this stronger condition all additional quadratic contributions
in \eqref{eq:aux-wick-stability-matrix} are nonnegative, and therefore
$Q_K$ is automatically positive.

Using
\begin{equation}
 g_A^{(K)}
 =
 \gamma v_A^{1/4}K^{-1/4},
 \qquad
 \lambda=2\gamma^4,
\end{equation}
the condition becomes
\begin{equation}
 K
 \geq
 K_{\mathrm{stab}}
 :=
 \max_A
 \left\{
 \frac{\lambda}{2}v_A\mathcal C_A^2
 \right\}.
 \label{eq:aux-wick-K-stability}
\end{equation}
This is only a convenient sufficient bound; the exact matrix
condition \eqref{eq:aux-wick-positive-definite} may already be
satisfied for smaller values of $K$.

It is important that all these statements are made at fixed field and
ultraviolet regulators. The coincident covariance $\mathcal C_A$
depends on the regulator and generally grows when the ultraviolet
cutoff is removed. Consequently, the lower bound on $K$ need not be
uniform in the continuum limit. This is one reason for keeping the
order of limits explicit: the population limit $K\to\infty$ is taken
while the field regulator and the bulk-cell partition are held fixed.

For randomly sampled neural features, $F_A$ and, in the conditional
normal-ordering scheme, $\mathcal C_A$ depend on the feature
realization $\eta$. The stability condition should then first be
understood at fixed $\eta$. The quenched construction is well defined
when it holds for almost every regulated feature realization. A
single feature-independent value of $K$ may be used whenever the
regulated coincident covariances are bounded uniformly over the
feature ensemble.

Finally, the stability requirement is compatible with the desired
large-$K$ quartic limit. For every fixed field amplitude $x$,
\begin{equation}
 K\,W\!\left(xK^{-1/4}\right)
 \longrightarrow
 \frac{x^4}{12},
 \qquad K\to\infty.
 \label{eq:aux-W-large-K}
\end{equation}
In fact, the left-hand side increases with $K$ toward the quartic
limit.
To see this, use $W''(t)=\tanh^2t$ and $W(0)=W'(0)=0$ to write,
for $t>0$,
\begin{equation}
 \frac{W(t)}{t^4}
 =\int_0^1(1-u)u^2
 \left[\frac{\tanh(tu)}{tu}\right]^2\,\dd u.
 \label{eq:aux-W-monotonic-proof}
\end{equation}
Because $\tanh v/v$ decreases for $v>0$, this ratio decreases with
$t$; substituting $t=|x|K^{-1/4}$ proves the claimed monotonicity
for fixed $x$ (with equality at $x=0$). At the same time,
\begin{equation}
 K\bigl(g_A^{(K)}\bigr)^2
 =
 \gamma^2v_A^{1/2}\sqrt K,
\end{equation}
so the positive quadratic suppression governing the far tails becomes
stronger as $K$ increases. Hence, once a sufficiently large value
$K_0$ satisfies the stability condition, all larger populations are
also stable. The finite-$K$ Wick-subtracted measures can then be
taken smoothly to the same regulated Wick-ordered quartic measure.

There is also a simpler alternative prescription. One may keep the
exact finite-$K$ auxiliary measure in its bare compensated form, for
which $W(t)\geq0$ guarantees normalizability automatically, first take
the limit $K\to\infty$, and only then introduce the Wick subtraction
in the resulting quartic theory. In that order of limits no additional
finite-$K$ stability condition is required. The discussion above is
needed only when one wishes to define a Wick-subtracted auxiliary
probability measure already at finite population.

\paragraph{A finite-population six-point discriminator.}
The agreement of the leading quartic contact functions need not extend to
finite-population higher-point corrections. For the replicated auxiliaries,
Eq.~\eqref{eq:Klimit} contains the local sextic term
\begin{equation}
  \delta V_{6,\mathrm{aux}}
  =\sum_A u_A^{\mathrm{aux}}\phi_A^6,
  \qquad
  u_A^{\mathrm{aux}}=-\frac{\gamma^6v_A^{3/2}}{45\sqrt K}.
  \label{eq:sextic-aux-coefficient}
\end{equation}
The critical binary populations of
Appendix~\ref{app:critical-populations}, after adjustment of their
$O(n^{-1/2})$ quadratic correction, instead contain
\begin{equation}
  \delta V_{6,\mathrm{crit}}
  =\sum_A u_A^{\mathrm{crit}}\phi_A^6,
  \qquad
  u_A^{\mathrm{crit}}
  =\frac{c_A^6}{30\sqrt n}
  =\frac{\gamma^6v_A^{3/2}}{30\sqrt n}.
  \label{eq:sextic-critical-coefficient}
\end{equation}
The last equality uses the matched leading coupling
$\lambda=2\gamma^4$ and the positive choice
$c_A^4=\lambda v_A/2$. Thus the sextic coefficients have opposite signs.

To isolate their implication, write a small sextic perturbation of the
matched regulated quartic theory as $\delta V_6=\sum_Au_A\phi_A^6$.
At lowest order about the common Gaussian reference measure, its
connected six-point contact contribution is
\begin{equation}
 \delta G_{\mathrm c,6}^{(6)}(X_1,\ldots,X_6)
 =-6!\sum_Au_A\prod_{i=1}^6 C(X_i,Y_A)+O(\lambda u,u^2),
 \label{eq:sextic-sixpoint-contact}
\end{equation}
where the displayed term is the sextic contact contribution only.
The complete six-point function also receives quartic exchange terms
beginning at order $\lambda^2$ and, under finite-population matching,
possible parameter and counterterm corrections. Therefore the sign
comparison in \eqref{eq:sextic-aux-coefficient}--\eqref{eq:sextic-critical-coefficient}
is a distinction between microscopic corrections to the interaction,
not by itself a sign theorem for the complete six-point correlator.
Both sextic coefficients vanish in the respective large-population
limits at fixed bulk regulator; whether a residual distinction
survives a correlated continuum limit requires separate analysis.

\paragraph{A mass-response consistency identity.}
The mass dependence of the bulk correlators provides a test that does not
require fitting another interaction vertex. At a fixed finite regulator,
assume that the boundary domain and the remaining bare couplings are held
fixed when $m^2$ is varied, and write
\begin{equation}
 S_m[\phi]=S_{\mathrm{rest}}[\phi]
   +\frac{m^2}{2}\int_\Omega\dV_Y\,\phi(Y)^2.
 \label{eq:mass-response-action}
\end{equation}
For an observable $\mathcal A[\phi]$ with no explicit $m^2$
dependence, differentiation of the normalized measure gives
\begin{equation}
 \frac{\partial}{\partial m^2}\langle\mathcal A\rangle_m
 =-\frac12\int_\Omega\dV_Y\,
  \Cov_m\!\bigl(\mathcal A,\phi(Y)^2\bigr).
 \label{eq:mass-response-general}
\end{equation}
For a centered scalar measure, its connected-cumulant decomposition
yields
\begin{align}
 \frac{\partial}{\partial m^2}G_{2,m}(X_1,X_2)
 &=-\int_\Omega\dV_Y\,
   G_{2,m}(X_1,Y)G_{2,m}(Y,X_2)\nonumber\\
 &\quad-\frac12\int_\Omega\dV_Y\,
   G_{4,\mathrm c,m}(X_1,X_2,Y,Y).
 \label{eq:mass-response-two-point}
\end{align}
At the Gaussian point $G_{4,\mathrm c,m}=0$ and the equation reduces
to the resolvent identity
\begin{equation}
 \partial_{m^2}G_m=-G_m^2,
 \label{eq:mass-response-resolvent}
\end{equation}
where the operator product includes integration over the intermediate
bulk coordinate. For a lattice regulator, all bulk integrals in these
identities are replaced by the corresponding weighted cell sums.
If the renormalization counterterms or the boundary domain change with
$m^2$, their derivatives must also be included.
In particular, the source-free standard and alternative domains of
the continuum inverse-square radial operator have mass-dependent
asymptotic exponents. Hence the simple mass derivative applies directly
to a fixed finite-dimensional field space with a mass-independent
domain and boundary action; varying the continuum AdS mass while
retuning its boundary condition requires accounting for that variation. Thus
\eqref{eq:mass-response-two-point} is a consistency condition for a
single family of measures with the stated fixed-parameter convention,
not an automatic identity for arbitrarily retuned neural models at
different masses.

\subsection{Distinguishing non-Gaussianity from the desired interaction}
\label{app:gaussian-obstruction}

Random hidden features can make a neural field non-Gaussian without
producing the desired repulsive quartic interaction. Indeed, suppose
$\phi\mid\eta$ is centered Gaussian with random covariance $C_\eta$.
For a real smeared field $\Phi_J=\int\dV\,J\phi$ and conditional
variance $V_\eta=(J,C_\eta J)$, Gaussian contractions give
\begin{equation}
 \kappa_4(\Phi_J)=3\operatorname{Var}_\eta(V_\eta)\ge0.
 \label{eq:app-gaussian-mixture-sign}
\end{equation}
By contrast, a positive local $\lambda\phi^4$ interaction has
\begin{equation}
 \kappa_4(\Phi_J)
 =-\lambda\int\dV_Y\,[(G_\Delta J)(Y)]^4
 +O(\lambda^2),
 \label{eq:app-repulsive-sign}
\end{equation}
which is negative at leading order for nontrivial $J$.
Thus fluctuations of a centered Gaussian covariance alone cannot
replace the interaction mechanisms used in the main text.

Other possibilities include nonlinear transport of Gaussian neural
parameters and stochastic sampling dynamics. Such methods can
represent interacting measures, but may simply encode an action
chosen beforehand. To test a more general neural architecture,
one may reconstruct its leading amputated vertex from the connected
four-point function,
\begin{equation}
 \Lambda_4=-(G_h^{-1})^{\otimes4}G_{\mathrm c}^{(4)}
 +O(\lambda^2),
 \label{eq:app-interaction-amputation}
\end{equation}
and check whether it approaches the local quartic contact kernel
rather than an extended nonlocal interaction.
Here $G_h^{-1}$ is taken only on the nondegenerate regulated
field space (or on its specified support); a finite-rank empirical
feature covariance need not be invertible on a larger field space. These finite-regulator
tests do not by themselves establish continuum or boundary convergence.

\bibliographystyle{JHEP} 
\bibliography{bibliography}

\end{document}